\documentclass[11pt, a4paper]{article}
\usepackage{amsfonts,amsmath,color,graphicx,stmaryrd,amssymb,mathdots,esint,tikz}
\usepackage{mathrsfs, enumerate}
\usepackage{amscd, lieart}
\usepackage{pdfcomment}
\usepackage{xypic}
\usepackage{pgfplots}
\DeclareMathAlphabet{\mathpzc}{OT1}{pzc}{m}{it}

\makeatletter
\@addtoreset{equation}{section}
\makeatother
\newtheorem{theorem}{Theorem}[section]
\newtheorem{conjecture}[theorem]{Conjecture}
\newtheorem{proposition}[theorem]{Proposition}
\newtheorem{lemma}[theorem]{Lemma}
\newtheorem{corollary}[theorem]{Corollary}

\newtheorem{example}{Example}[section]
\newtheorem{definition}[example]{Definition}

\newtheorem{remark}[example]{Remark}
\newtheorem{hypothesis}[example]{Hypothesis}
\def\br{\begin{remark}\rm\small}
\def\er{\end{remark}}
\def\bt{\begin{theorem}}
\def\et{\end{theorem}}
\def\bcj{\begin{conjecture}}
\def\ecj{\end{conjecture}}
\def\bd{\begin{definition}}
\def\ed{\end{definition}}
\def\bp{\begin{proposition}}
\def\ep{\end{proposition}}
\def\bl{\begin{lemma}}
\def\el{\end{lemma}}
\def\bc{\begin{corollary}}
\def\ec{\end{corollary}}
\def\bh{\begin{hypothesis}}
\def\eh{\end{hypothesis}}
\def\beaq{\begin{eqnarray}}
\def\eeaq{\end{eqnarray}}

\newcommand{\beq}{\begin{equation}}
\newcommand{\eeq}{\end{equation}}
\newcommand{\bea}{\begin{eqnarray}}
\newcommand{\eea}{\end{eqnarray}}

\newcommand{\dd}{\mathrm{d}}

\newcommand{\bra}{\left\langle}
\newcommand{\ket}{\right\rangle}

\newcommand{\re}{{\rm e}}

\definecolor{rouge}{rgb}{0.84,0.18,0.07}
\definecolor{bleu}{rgb}{0.22,0.41,0.74}
\definecolor{vertf}{rgb}{0.08,0.46,0.07}
\usepackage[pdftex]{hyperref}
\hypersetup{colorlinks,urlcolor=vertf,citecolor=rouge,linkcolor=bleu,filecolor=black}

\newcommand{\de}{{\partial}}

\newcommand{\rd}{\mathrm{d}}
\newcommand{\ri}{\mathrm{i}}

\newcommand{\fU}{\mathfrak{U}}

\renewcommand{\Re}{\mathfrak{Re}}

\newcommand{\bbS}{\mathbb{S}}
\newcommand{\bbN}{\mathbb{N}}

\newcommand{\bbZ}{\mathbb{Z}}
\newcommand{\bbR}{\mathbb{R}}
\newcommand{\bbC}{\mathbb{C}}

\newcommand{\bbP}{\mathbb{P}}

\newcommand{\bbQ}{\mathbb{Q}}

\def\bary{\begin{array}} 
\def\eary{\end{array}} 
\def\ben{\begin{enumerate}} 
\def\een{\end{enumerate}}
\def\bit{\begin{itemize}} 
\def\eit{\end{itemize}}
\def\nn{\nonumber} 

\renewcommand{\cR}{\mathcal{R}}

\newcommand{\cO}{\mathcal{O}}
\newcommand{\cT}{\mathcal{T}}

\newcommand{\cP}{\mathcal{P}}
\newcommand{\cC}{\mathcal{C}}
\newcommand{\DD}{\mathcal{D}}
\newcommand{\LL}{\mathcal{L}}
\newcommand{\cS}{\mathcal{S}}

\newcommand{\cN}{\mathcal{N}}

\newcommand{\cG}{\mathcal{G}}
\newcommand{\cA}{\mathcal{A}}

\newcommand{\cM}{\mathcal M}

\renewcommand{\l}{\left}
\renewcommand{\r}{\right}

\begin{document}

\sloppy

\pagestyle{plain}
\addtolength{\baselineskip}{0.10\baselineskip}
\begin{center}

\vspace{26pt}

{\Large \textbf{Chern--Simons theory on spherical Seifert manifolds, topological strings and integrable systems}}

\vspace{26pt}

\textsl{Ga\"etan Borot}\footnote{Max Planck Institut f\"ur Mathematik,
  Vivatsgasse 7, 53111 Bonn,
  Germany. E-mail: \texttt{gborot@mpim-bonn.mpg.de}} and
\textsl{Andrea Brini}\footnote{UMR 5149 du CNRS, Institut Montpelli\'erain Alexander
  Grothendieck, Universit\'e de Montpellier, case courrier 51, 34095
  Montpellier Cedex 5, France. E-mail: \texttt{andrea.brini@univ-montp2.fr }}
\end{center}

\vspace{20pt}

\begin{center}
\textbf{Abstract}
\end{center}

\noindent 

We consider the Gopakumar--Ooguri--Vafa correspondence, relating ${\rm U}(N)$ Chern--Simons
theory at large~$N$ to topological strings, in the 
context of spherical Seifert \mbox{3-manifolds}. These are quotients  $\mathbb{S}^{\Gamma} = \Gamma\backslash\bbS^3$
of the three-sphere by the free action of a finite isometry group. Guided by
string theory dualities, we propose a large~$N$ dual description in terms
of both A- and B-twisted topological strings on (in general non-toric) local Calabi--Yau threefolds. The
target space of the B-model theory is obtained from the spectral curve of 
Toda-type integrable systems constructed on the double Bruhat cells of the simply-laced
group identified by the ADE label of $\Gamma$. Its mirror A-model
theory is realized as the local Gromov--Witten theory of suitable ALE
fibrations on $\mathbb{P}^1$, generalizing the results known for
lens spaces. We propose an explicit construction of the
family of target manifolds relevant for the correspondence, which we verify through a large $N$ analysis of the matrix model that
expresses the contribution of the trivial flat connection to the Chern--Simons
partition function. Mathematically, our results put forward an identification between the
$1/N$ expansion of the $\mathrm{sl}_{N + 1}$ LMO invariant of $\mathbb{S}^\Gamma$ and
a suitably restricted Gromov--Witten/Donaldson--Thomas partition
function on the A-model dual Calabi--Yau. This $1/N$ expansion, as well as
that of suitable generating series of perturbative quantum invariants of fiber knots in $\mathbb{S}^\Gamma$, is computed by the Eynard--Orantin topological recursion.

%





\vspace{26pt}
\pagestyle{plain}
\setcounter{page}{1}


\section{Introduction}

 In a series of celebrated works \cite{Gopakumar:1998ki, Ooguri:1999bv},
 Gopakumar, Ooguri and Vafa (GOV) proposed the existence of a
 duality between 
${\rm U}(N)$ Chern--Simons theory at level $k$ on $\mathbb{S}^3$
 \cite{Witten:1988hf}
and the topological A-model on the resolved conifold $Y= \mathrm{Tot}[\cO(-1)
  \oplus \cO(-1) \rightarrow \mathbb{P}^1]$. 
From a physical perspective, this identification provides a concrete
instance, and one where exact computations can be performed in detail,  of 't~Hooft's idea that the $1/N$ expansion of a gauge
theory with adjoint fields in the strong $g_{{\rm YM}}^2 N$ limit should be amenable
to a dual description in terms of a first quantized string theory. Originally
restricted to the partition function and closed string observables
\cite{Gopakumar:1998ki}, the correspondence was later extended to incorporate
Wilson loops along the unknot \cite{Ooguri:1999bv} and topological branes; progress in open/closed mirror symmetry
\cite{Hori:2000ck, Aganagic:2000gs, BKMP} has further allowed to
rephrase the correspondence in terms of the topological B-model on the
smoothing of the conifold singularity.\\

Mathematically, the main consequence of this physics-inspired duality is a
striking connection of theories of invariants from two domains of mathematics that
are {\it a priori} quite separated. On the one hand, Witten's heuristic approach to Chern--Simons invariants can be
recast in the context of quantum groups and modular tensor categories to yield {\it
  bona fide} invariants of links in 3-manifolds
\cite{Reshetikhin:1990pr,Reshetikhin:1991tc}; when the Chern--Simons gauge
group is ${\rm U}(N)$ or ${\rm SO}/{\rm Sp}(N)$, this leads respectively to the HOMFLY and
Kauffman invariants of links. Furthermore,
the perturbative expansion of the Chern--Simons functional integral around the
trivial flat connection leads to the theory of finite type invariants \cite{BNfinite}, via the Kontsevich integral and L\^{e}--Murakami--Ohtsuki (LMO) invariants. On the flip side, the topological A-model on a
Calabi--Yau $3$-fold $X$ is mathematically defined in terms of suitable theories of
moduli of curves in $X$, either via
stable maps \cite{Kontsevich:1994na} or ideal sheaves \cite{MR1634503}. In
particular, for the case of the unknot the Gopakumar--Ooguri--Vafa correspondence asserts that
Chern--Simons knot invariants should be identified with suitable virtual counts
of open Riemann surfaces on the dual Calabi--Yau $3$-fold $X$. By mirror
symmetry and the remodeling formalism \cite{BKMP,EOBKMP}, this can be recast
in the form of the topological recursion of \cite{EOFg} on the mirror curve of $X$.\\

As a detailed instance of the gauge/string correspondence, and because of
its far-reaching implications in geometry and topology, the GOV correspondence
has been the subject of intense study both in the physics and mathematics
communities. After the relation between Gromov--Witten invariants of the
resolved conifold and the $\mathfrak{sl}_{N + 1}$ quantum invariant of the unknot in $\mathbb{S}^3$ had been proved \cite{MR1954265, Katz:2001vm}, a
natural question was whether the correspondence could be extended so as to
encompass other classical gauge groups \cite{Bouchard:2004iu,Bouchard:2004ri},
knots\footnote{In an allied context, a vast program of computation of HOMFLY
  invariants, exploring also possible new relations with matrix models, has recently been undertaken by
  the mathematical physicists at ITEP; see in particular \cite{Alexandrov:2014nla,Mironov:2015aia} and references therein.} \cite{Labastida:2000yw,BEMknots,Diaconescu:2011xr,Aganagic:2013jpa},
and 3-manifolds. The generalization to manifolds beyond $\mathbb{S}^3$ is
perhaps the least studied, with all results to date confined to the case of lens spaces \cite{Aganagic:2002wv,
Halmagyi:2003ze,Brini:2008ik}.

\subsection{Scope of the paper}

The purpose of this paper is to propose an extension of the GOV correspondence
to the case of spherical Seifert manifolds. Our objects of study will be quotients $\mathbb{S}^{\Gamma} = \Gamma\backslash
\mathbb{S}^3$ by the free isometric action of a cyclic or binary
polyhedral group $\Gamma\subset {\rm SU}(2)$; one notable example is
the Poincar\'e homology sphere, corresponding to $\Gamma={\rm P}_{120}$ being the
binary icosahedral group. We offer here a conjectural dual description of the $1/N$ expansion
in terms of both A- and B-type topological strings, together with a precision
check for the contribution of the trivial flat connection, as follows. \\

On one hand, we associate to each $\Gamma$ a local Calabi--Yau $3$-fold
$Y^{\Gamma}$, serving as the A-model target space; this is constructed in Section~\ref{sec:geotrans} by 
a natural $\Gamma$-equivariant generalization of the conifold transition of
\cite{Gopakumar:1998ki} for $T^*\bbS^3$. When $\Gamma$ is non-abelian,
the $\Gamma$-action has the effect of reducing the rank of the
automorphism group of $Y^\Gamma$ to two, so that
$Y^{\Gamma}$ is non-toric. At first sight
this may be a hindrance towards finding a mirror B-model picture, as in particular
there is no explicit Hori--Vafa mirror here. However, the M-theory uplift of the
Katz--Klemm--Vafa geometric engineering to five compactified dimensions
\cite{Katz:1996fh,Lawrence:1997jr} suggests that the planar part of the topological
string free energy should be governed by special geometry on a family of
$5d$ Seiberg--Witten curves (Section~\ref{sec:geotrans2}), with gauge group $\cG_\Gamma$  
specified by the ADE label of $\Gamma$ via the McKay correspondence. Furthermore,  in light of the
connection of $4d$ pure $\cN=2$ Yang--Mills theory with the classical ADE Toda chain, it is natural to speculate that the $5d$ curves
should arise as the spectral curves of some relativistic deformation of the Toda
chain, as has been known for a long time for the case $\cG=\mathrm{SU}(p)$ \cite{MR1090424, Nekrasov:1996cz}. We will then be compelled to propose
that the B-model target space will be given by the family of spectral curves
of the Toda-type classical integrable system recently constructed in
\cite{Williams:2012fz,Fock:2014ifa} on the double
Bruhat cells of the loop group $\hat{\cG}$, as we recall in
Section~\ref{sec:Todain}. Concretely, the Toda spectral curves take the form
\beq
\mathcal{P}^{{\rm Toda}}_{\mathcal{G}^{\#}} (X,Y;u) = \det\big[Y\mathbf{1} -
  \rho_{\rm min}(L_{\mathsf{w}}^{\cG^{\#}})\big] = 0\,,
\label{eq:Todasc}
\eeq
where $L_{\mathsf{w}}^{\cG^{\#}}$ is the Lax matrix of the Toda
system on a suitable cell $\mathsf{w}$ of the affine co-extended group
$\cG^{\#}$, $\rho_{\rm min}$ is an irreducible representation of $\cG$ of minimal dimension, and $X\in\bbC^*$ is the
spectral parameter of the Lax matrix. The right-hand side expands in the spectral invariants of
the Lax matrix, which are encoded in $R = {\rm rank}(\cG)$ independent
parameters $u = (u_1,\ldots,u_R)$ -- these are the Hamiltonians for the
Toda classical integrable system on $\cG$, and they correspond to
the classical Weyl-invariant order parameters of the gauge theory vacua. We
also have one additional parameter $u_0$ associated to the affine root of $\cG^{\#}$, which plays the role
of the speed of light in the mechanical system, and is related to the exponentiated
volume of the base $\mathbb{P}^1$ in the mirror A-model. \\

\begin{figure}[h]
\centering
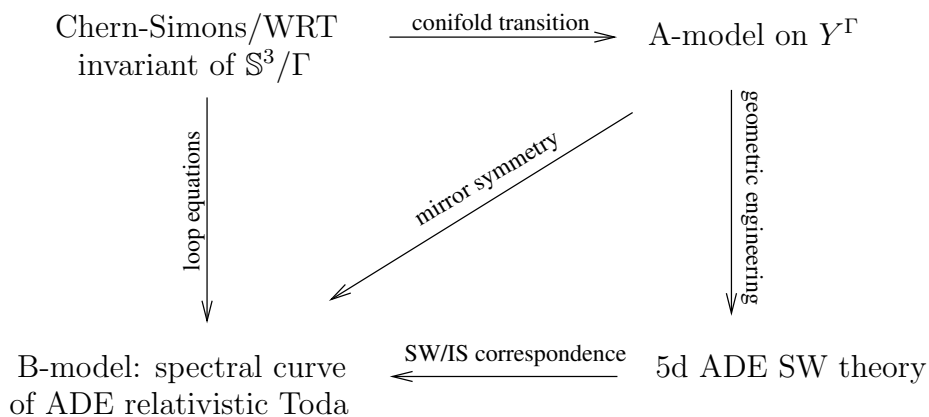
\caption{The chain of dualities behind our proposal.}
\label{fig:dualities}
\end{figure}

On the other hand, the $\mathfrak{sl}_{N + 1}$ evaluation of the LMO invariant of
the Seifert space $\mathbb{S}^{\Gamma} = \mathbb{S}^3/\Gamma$  is given by the
partition function of a random matrix model, and observables in this matrix
model encode perturbative quantum invariants of fiber knots; to disambiguate
notations, $\mathcal{D}_\Gamma$ in the following denotes a Dynkin diagram of type
$A$, $D$ or $E$, bijectively associated with the
cyclic or binary polyhedral group $\Gamma$ (A-model), and also with the compact, simply connected, simply-laced Lie group $\mathcal{G}_{\Gamma}$ (B-model). This matrix model has a spectral curve:
\beq
\mathcal{P}_{\mathcal{D}_{\Gamma}}^{{\rm LMO}}(X,Y;\hat{\lambda}) = 0,\qquad  \lambda \triangleq \hat{\lambda}/\sigma = N\hbar/\sigma\,,
\eeq
which depends on the three-dimensional geometry only via $\Gamma$ and the Seifert invariant $\sigma$ defined in \eqref{sigmde}. The latter only appears in the definition of the renormalized 't~Hooft parameter $\lambda$. The all-order asymptotic expansion of the partition function and observables can be obtained from the topological recursion of Eynard--Orantin \cite{EOFg} applied to this curve -- this is a B-model computation, in view of the remodeling proposal \cite{BKMP}.\\

We propose that, for a suitable restriction $u = u(\lambda)$ of parameters on the Toda side, the curves $\mathcal{P}^{{\rm LMO}}_{\mathcal{D}_{\Gamma}}$ and $\mathcal{P}_{\mathcal{G}_{\Gamma}}^{{\rm Toda}}$ agree up to an abelian factor $(Y - 1)^{\bullet}$.
Therefore, the generating functions of LMO invariants and perturbative quantum invariants
of fiber knots in $\mathbb{S}^{\Gamma}$ receive an interpretation as suitably
restricted Gromov--Witten/Donaldson--Thomas partition functions of
$Y^{\Gamma}$. Our proposal passes several non-trivial tests, and automatically retrieves the known results for the case of lens spaces $L(1,p)$, where $\Gamma=\bbZ/p\mathbb{Z}$ is a cyclic group, $Y^\Gamma$ is a toric
variety, and $\cG=\mathrm{SU}(p)$. The non-toric cases have so far remained unexplored, and they are the main focus of this paper.

\subsection{Summary of results and organization of the article}

We now describe more precisely our results and their mathematical status. They can be grouped in three main strands.\\

Firstly, we construct the A-model geometries $Y^\Gamma$ by a direct
generalization of the geometric transition for the case of spherical Seifert
spaces (Section~\ref{sec:geotrans}). We also highlight an extension of the
holomorphic disk counting of \cite{Aganagic:2000gs,Katz:2001vm,MR2861610} to
the non-abelian orbifold case, and introduce the generating functions of
open/closed Gromov--Witten invariants that are relevant for the
discussion. Secondly, we propose (Section~\ref{sec:geotrans2}) and carry out the detailed construction of the spectral
curves \eqref{eq:Todasc} for $\cG=A,D,E_6,E_7$; this requires substantial work
and occupies the bulk of Section~\ref{Stringside}. For $\cG=E_8$,
computational complexity restricts the amount of data we can extract, while
still allowing us to make some universal predictions on the form of
\eqref{eq:Todasc}, as well as a complete derivation of the spectral curve at the
special point in moduli space corresponding to the $\Gamma$-orbifold of the conifold. Thirdly, for all $\cG \neq E_8$, combining these results with \cite{BESeifert}, we can establish that the $\mathfrak{sl}_{N + 1}$ LMO invariants of $\mathbb{S}^\Gamma$ are computed by the
Eynard--Orantin invariants of the Toda curves, which
may be regarded as a restricted, B-model version of the GOV
correspondence; for $\cG = E_8$, a complete
proof is out of reach of our methods, but we propose it as a conjecture
passing non-trivial checks.\\

The strategy we employ in our proof runs as follows: the LMO invariant on any Seifert space has been
computed in \cite{BNrational, MarinoCSM}, and for weight system $\mathfrak{sl}_{N + 1}$ it takes
the form of the partition function for a random $N \times N$ hermitian matrix
model. The authors of \cite{BESeifert} rely on \cite{BGK} to prove the
existence of an asymptotic expansion when $N \rightarrow \infty$, and on
\cite{BEO} to show that the latter is computed by the topological recursion
applied to the spectral curve $\mathcal{P}_{\mathcal{D}_{\Gamma}}^{{\rm LMO}}$
of the matrix model. This material is reviewed in Section~\ref{CSsec}. The
computation of the LMO spectral curve occupies Section~\ref{S4LMO}, and is
completed for $A$, $D$ and $E_6$, while the result for $E_7$ and $E_8$
involves a number of parameters, in principle fixed by algebraic constraints
that we could not solve. We however point out that, compared to \cite{BESeifert}, the complete expression of the LMO curve for $E_6$ is new (Section~\ref{E6geom}).\\

Our comparison statement is presented in Section~\ref{conjs} and implemented in
Section~\ref{Stringside}. It boils down to a general recipe to identify the function
$u(\lambda)$ such that the LMO spectral curve and the Toda spectral curve
specialized to $u \leftarrow u(\lambda)$ coincide
(Conjecture~\ref{conj:spec}). We give the expression of $u(\lambda)$ for $\mathcal{D} \in
\{A,D,E_6\}$, thus proving the conjecture. Algebraic complexity challenges the computation for $E_7$
and $E_8$, but we are however able to prove that the specialization exists for
$E_7$, and we find exact agreement of the Toda/LMO curves at the conifold
point (i.e. $\lambda = 0$ on the LMO side) for $E_8$, as well as a more general equality of their vertical slope polynomial. This comparison pertains to the left vertical arrow in
Figure~\ref{fig:dualities}, and can be formulated as follows (see
Section~\ref{Sback} for the relevant notation):

\begin{proposition}
\label{propsimple}Let $(E_j[X;u])$ be the eigenvalues of the Lax matrix of the
Toda integrable system for the affine co-extended group of type ADE, specified
by fundamental character values $u_1,\ldots,u_{R}$, Casimir $u_0 = -\exp(-\chi_{{\rm orb}} \lambda/2)$ and spectral parameter $X$. There exist a specialization $(u_i(\lambda))_i$ and an explicit vector $\hat{v}_j \in \mathbb{Z}^a$ such that the Taylor expansion of $E_j[X;u(\lambda)]$ near $X \rightarrow \infty$ is equal to:
\beq
\mathcal{Y}_{\hat{v}_j}(X) = \prod_{\ell = 0}^{a - 1} \big[\mathcal{Y}(e^{2{\rm i}\pi \ell/a}X^{1/a})\big]^{\hat{v}_j(\ell)},
\label{eq:ylmo}
\eeq
with:
\beq
\mathcal{Y}(X) = -X^{1/a}\,c\,\exp\Big(\frac{\chi_{{\rm orb}}\,\lambda}{a}
\sum_{k \geq 0} X^{-k/a}\,\langle \mathrm{Tr}\,\,\mathcal{U}^{k}
\rangle^{(0)}\Big)\,,
\label{eq:trsmm}
\eeq
and where $\langle \mathrm{Tr}\,\,\mathcal{U}^{k} \rangle^{(0)}$ is the large
$N$ limit of the moments of the random matrix $\mathcal{U}$ in the Seifert
matrix model. Furthermore, the full $1/N$ asymptotic expansion of $\bra
\mathrm{Tr}\,\,\mathcal{U}^{k_1} \dots
\mathrm{Tr}\,\,\mathcal{U}^{k_n}\ket_{\rm conn.}$ is computed from
\eqref{eq:ylmo}--\eqref{eq:trsmm} by the Eynard--Orantin recursion \cite{EOFg}. For $k_i \in (a/a_m)\mathbb{Z}$, this is identified with the
$1/N$ expansion of the perturbative quantum invariants (in virtual $k$-th power sum representation) of the knot going along the fiber of order $a_m$ in the Seifert manifold.
\end{proposition}

\br Combining the results of \cite{Hansen} and \cite{MarinoCSM}, one sees that
the matrix model observables described in Section~\ref{CSsec} appear as one
term in the expression of the $\mathfrak{sl}_{N+1}$ quantum invariants of fiber knots in Seifert manifolds produced by
the Witten--Reshetikhin--Turaev--Wenzl TQFT at roots of unity; in Chern-Simons theory, localization
heuristics identifies it with the contribution of the trivial flat connection
to the Chern--Simons path integral. Throughout the paper, we will use the name
``perturbative quantum invariants'' to refer to these quantities. Whenever the
trivial connection is isolated, i.e. for lens spaces and the Poincar\'e
sphere, these should coincide with the dominant contribution to the saddle-point asymptotics of
Wilson loops in Chern--Simons theory.
\er

For the $A$-series, this correspondence has been known to extend to the perturbative
expansion in Chern--Simons theory around a general flat connection \cite{Halmagyi:2003ze}. Its formal
analogy with the general simply-laced case cries out for generalization to the
$D$- and $E$-series, and we speculate in
Conjecture~\ref{conj:gen} on extending our statements to an arbitrary flat
background. \\

The link between the A- and the B- model geometry -- i.e. the diagonal arrow
in Figure~\ref{fig:dualities} -- will be explored
in a subsequent publication \cite{GWADE}, where more details can be found on the computations leading
to the results of Section~\ref{Stringside}.

\subsubsection*{Acknowledgements}

We are particularly grateful to Albrecht Klemm for numerous discussions and
his participation at an early
stage of this work. The article contains numerical simulations by Alexander
Wei\ss{}e and we also thank him for his help on some heavy computations we have
done. This project was initiated following discussions at the workshop
``Hamiltonian PDEs, Frobenius manifolds and Deligne--Mumford moduli spaces"
hosted by SISSA in September 2013, and the authors wish to express their thanks to the organizers
for their invitation. We would also like to thank D.~Zagier for his answers on
factorization of polynomials, C.~Bonnaf\'e, S.~Gukov, W.~Lerche, D.~Pei, G.~Thompson and T.~Weigel for
discussions and/or correspondence, as well as the Theory and Geometry groups at Caltech, Imperial
College, the University of Milano--Bicocca, the
KdV Institut in Amsterdam, MPIM in Bonn and the Isaac Newton Institute in Cambridge for their kind hospitality while this
work was carried out. The work of G.B. benefits from the support
of the Max-Planck Gesellschaft. The work of A.B. is partially supported by the 
GNFM--INdAM.

\section{Chern-Simons theory and Seifert spaces}
\label{CSSeis}

\label{Sback}

This section reviews the main characters in our play, starting from the LMO
invariants and Chern--Simons theory of Seifert $3$-manifolds
(Section~\ref{CSSeis}), and in particular the spherical ones. We
also discuss rigorous aspects of the matrix model approach. Then, we argue on
physical grounds using large $N$ dualities, geometric transitions
(Section~\ref{sec:geotrans}) and geometric engineering
(Section~\ref{sec:geotrans2}), how Chern--Simons theory on $\mathbb{S}^{{\rm ADE}}$
relates to $d=5$, $\cN=1$ pure Yang--Mills theory with ADE gauge group, and
in turn to the classical integrable systems that govern its effective action
up to two derivatives (Section~\ref{sec:Todain}). This is the necessary material to present our two main conjectures in Section~\ref{conjs}.

\subsection{Geometry of Seifert 3-manifolds}

Seifert fibered spaces are manifolds $M^3$ that are $\mathbb{S}^{1}$-bundles over orbifold surfaces
\cite{Seifertbook}. When the base surface is the sphere $\mathbb{S}^2$ with
$r$ orbifold points of order
$a_1,\ldots,a_r$, $M^3$ can be realized by rational surgery on the link 
in $\mathbb{S}^3$, consisting of one main component passing through $r$
meridians. The surgery slopes are $1/b$ on the main component, and $a_m/b_m$ on the $m$-th meridian. Here, $a_m > 0$ and $0 \leq b_m < a_m$ is coprime to $a_m$. There exist moves changing the surgery data but giving isomorphic Seifert spaces. Nevertheless, the uple $(a_1,\ldots,a_r)$ and
\beq
\label{sigmde}\sigma \triangleq b + \sum_{m = 1}^r \frac{b_{m}}{a_m}\,
\eeq
are invariants of Seifert fibered spaces. For $r \geq 3$, $(a_1,\ldots,a_r)$ is a topological invariant of $M^3$, whereas the cases $r = 1$ or $2$ realize lens spaces in several inequivalent ways as Seifert fibered spaces. Two quantities are particularly important:
\beq
a \triangleq {\rm lcm}(a_1,\ldots,a_r),\qquad \chi_{{\rm orb}} \triangleq 2 - r + \sum_{m = 1}^r \frac{1}{a_m}.
\eeq
A presentation of the fundamental groups of Seifert spaces was described in
\cite{Seifertbook} and the fundamental groups identified in \cite{Orlik}: we
remind this in Appendix~\ref{pi1list}. The key fact is that $\pi_1(M^3)$ is
finite iff $\chi_{{\rm orb}} > 0$ and $\sigma \neq 0$;
 this occurs for lens spaces
or for $r = 3$ exceptional fibers of order $(2,2,p)$, $(2,3,3)$, $(2,3,4)$,
$(2,3,5)$. Then, the orbifold  fundamental groups of the $2d$-base of the Seifert fibration is the
  spherical triangle group $\Gamma=(a_1,a_2,a_3)$. 
The resulting 3-manifolds $\mathbb{S}^\Gamma \triangleq \Gamma\backslash\bbS^3$ are {\it
  spherical Seifert spaces}: these are quotients of the $3$-sphere by a finite
group of isometries acting smoothly, linearly and freely. Up to central
extension, as reviewed in Appendix~\ref{pi1list}-\ref{gpact}, the list of possible groups
is exhausted by $\Gamma\subset {\rm SL}(2,\bbC)$ being a cyclic or binary polyhedral group. By the
McKay correspondence \cite{McKay}, these have an ADE classification given in Table~\ref{ADElabe}.
Throughout the text, we will employ the labeling by ADE Dynkin diagrams $\DD_\Gamma$
to refer to the corresponding Seifert geometry. \\

\begin{table}[!h]
\centering
\begin{minipage}{12cm}
\centering
\begin{tabular}{|c|c|c|}
\hline
{\rule{0pt}{2.5ex}}{\rule[-1.4ex]{0pt}{0pt}} Exceptional fibers & $\Gamma$ & $\mathcal{\DD}_\Gamma$ \\
\hline
{\rule{0pt}{2.5ex}}{\rule[-1.4ex]{0pt}{0pt}} $(p)$ & $\bbZ/p\bbZ$ & $A_{p-1}$ \\
\hline
{\rule{0pt}{2.5ex}}{\rule[-1.4ex]{0pt}{0pt}} $(2,2,p)$ &  ${\rm Q}_{4(p + 2)}$ & $D_{p+2}$ \\
\hline
{\rule{0pt}{2.5ex}}{\rule[-1.4ex]{0pt}{0pt}} $(2,3,3)$ & ${\rm P}_{24}$ & $E_6$ \\
\hline
{\rule{0pt}{2.5ex}}{\rule[-1.4ex]{0pt}{0pt}} $(2,3,4)$ & ${\rm P}_{48}$ & $E_7$ \\
\hline
{\rule{0pt}{2.5ex}}{\rule[-1.4ex]{0pt}{0pt}} $(2,3,5)$ & ${\rm P}_{120}$ & $E_8$ \\
\hline
\end{tabular}
\caption{\label{ADElabe}
 ADE labeling of spherical Seifert manifolds. ${\rm Q}_{4p}$ is the binary 
  dihedral group, of order $4p$; ${\rm P}_{24}$, ${\rm P}_{48}$ and ${\rm P}_{120}$ denote the binary
  tetra-, octa-, and icosa-hedral groups respectively. 
 }
\end{minipage}
\end{table}

As $\pi_1(\mathbb{S}^\Gamma)=\Gamma$ is finite, $H_1(\mathbb{S}^\Gamma,\mathbb{Z})$ is purely torsion and $\mathbb{S}^\Gamma$ is always a rational homology sphere ($\mathbb{Q}$HS). In our list, the only case where we obtain an integer homology sphere is the $E_8$ case with $b_1 = b_2 = b_3 = -b = 1$: this is the Poincar\'e sphere.

\subsection{LMO invariant}
\label{LMOmatrix}

Before getting to Chern--Simons theory in Section~\ref{CSsec}, we first present the mathematical avatar about which this article is mainly concerned: the LMO invariant \cite{LMO}. It is a graph-valued formal series associated to any rational homology sphere. The choice of a simple Lie algebra $\mathfrak{g}$ gives an evaluation of the graphs, and converts this series into a formal series with rational coefficients:
\beq
\ln \mathcal{Z}_{{\rm LMO}}(M^3) = \sum_{g \in \mathbb{N}/2} \hbar^{2g - 2}\,\mathcal{F}_{g}(M^3) \in \hbar^{-2}\mathbb{Q}[[\hbar]]\,.
\eeq
Bar-Natan and Lawrence \cite{BNrational} obtained a surgery formula allowing them to compute the LMO invariant of Seifert manifolds which are $\mathbb{Q}$HS, and after picking up a simple Lie algebra, the result takes the form:
\beq
\label{SeiLMO}\mathcal{Z}_{{\rm LMO}}^{\mathfrak{g}}(M^3) = C_{\hbar}^{\mathfrak{g}}(M^3) \int_{\mathfrak{h}} \dd\mathbf{\phi}\,\prod_{\alpha > 0} \big({\rm sinh}[(\alpha\cdot \phi)/2]\big)^{2 - r} \prod_{m = 1}^r {\rm sinh}[(\alpha\cdot \phi)/2a_{m}]\big)\,e^{-\phi^2 /(2\sigma\hbar)}\,.
\eeq
$\mathfrak{h}$ is the (real) Cartan subalgebra of $\mathfrak{g}$, the product ranges over all positive roots, and $(x,y) \mapsto x\cdot y$ is the Killing bilinear form, and $\dd\mathbf{\phi}$ the corresponding Riemannian volume. $C_{\hbar}^{\mathfrak{g}}(M^3)$ is an explicit prefactor involving $a_m$, $\sigma$ and the Casson--Walker invariant of $M^3$ \cite{WalkerCasson}. Apart from this contribution, the only dependence on $b_m$ is hidden in the parameter $\sigma$ defined in \eqref{sigmde}. \\

We will be mainly interested in the weight system of the Lie algebra $\mathfrak{sl}_{N+1}$. In this case, elementary combinatorics shows that the LMO invariant can be repackaged by setting $\hat{\lambda} = N\hbar$ into a well-defined formal series:
\beq
\label{znfom}\ln \mathcal{Z}_{{\rm LMO}}^{\mathfrak{sl}(N+1)}(M^3) = \sum_{g \in \mathbb{N}} N^{2g - 2}\,\mathcal{F}_{g}(\hat{\lambda};M^3),\qquad \mathcal{F}_{g}(M^3;\hat{\lambda}) \in \mathbb{Q}[[\hat{\lambda}]]\,.
\eeq
$\mathcal{F}_{h}$ are called the free energies. In the case of Seifert manifolds, we prefer to define:
\beq
\lambda \triangleq  N\hbar/\sigma = \hat\lambda/\sigma\,,
\eeq
and \eqref{SeiLMO} for Seifert spaces becomes:
\beq
\label{MMmodel}\mathcal{Z}_{{\rm LMO}}^{\mathfrak{sl}_{N+1}}(M^3) = C_{\hbar}^{\mathfrak{sl}_{N+1}}(M^3)\!\!\int_{\mathbb{R}^N}\!\! \prod_{1 \leq i < j \leq N} \!\!\! \big({\rm sinh}[(\phi_i - \phi_j)/2]\big)^{2 - r}\!\prod_{m = 1}^r {\rm sinh}[(\phi_i - \phi_j)/2a_m]\,\prod_{i = 1}^N \re^{-N\phi_i^2/2\lambda}\dd\phi_i\,.
\eeq

\subsection{The matrix model approach}

The right-hand side of \eqref{MMmodel} provides a definition for a function of
an integer $N$ and a positive parameter $\hat\lambda$, that we denote
$Z_N(M^3;\hat\lambda)$. This is a convergent matrix integral, and its large $N$ asymptotic
behavior for a fixed $\hat\lambda > 0$ can be studied rigorously with the
techniques recently developed in \cite{BGK}. The main result of \cite{BGK}
relies on an assumption of strict convexity, which is here satisfied when
$\chi_{{\rm orb}} > 0$ and 
$\hat\lambda > 0$ is small enough. One then obtains, for any $g_0 \geq 0$, an asymptotic expansion of the form:
\bea
\label{zndef} Z_N(M^3;\lambda) & \triangleq & \int_{\mathbb{R}^N} \prod_{1 \leq i < j \leq N} \big({\rm sinh}[(\phi_i - \phi_j)/2]\big)^{2 - r}\,\prod_{m = 1}^r {\rm sinh}[(\phi_i - \phi_j)/2a_m]\,\prod_{i = 1}^N \re^{-N\phi_i^2/2\lambda}\dd\phi_i
 \nonumber \\
 \label{znexp}& = & N^{N + 5/12}\,\exp\Big(\sum_{g = 0}^{g_0} N^{2 - 2g}\,F_g(\lambda;M^3) + \cO(N^{2 - 2g_0})\Big)\,,
 \eea
and $F_g(M^3;\lambda)$ extends as an analytic function of $\lambda$ in a vicinity of $0$. It was proved in \cite{BEO} that the $F_g$ are computed by the topological recursion of \cite{EOFg}. This requires only the knowledge of the spectral curve of the matrix model, here conveniently defined as:
\beq
\label{Wsu} W_{0,1}(x) \triangleq \lim_{N \rightarrow \infty} \frac{1}{N}\Big\langle \sum_{i = 1}^N \frac{x}{x - e^{\phi_i/a}}\Big\rangle\,,
\eeq
and the knowledge of the two-point function:
\beq
W_{0,2}(x_1,x_2) \triangleq \lim_{N \rightarrow \infty} \bigg\{\Big\langle \sum_{i_1,i_2 = 1}^N \frac{x_1x_2}{(x_1 - \re^{\phi_{i_1}/a})(x_2 - \re^{\phi_{i_2}/a})}\Big\rangle - \Big\langle \sum_{i_1 = 1}^N \frac{x_1}{x_1 - \re^{\phi_{i_1}/a}}\Big\rangle\Big\langle \sum_{i_2 = 1}^N \frac{x_2}{x_2 - \re^{\phi_{i_2}/a}}\Big\rangle\bigg\}.
\eeq
It turns out that $W_{0,2}(x_1,x_2)$ can be analytically continued as a meromorphic
function of $2$ variables in the same curve, i.e. on $\{(x_1,y_1,x_2,y_2) \in \mathbb{C}^4,\quad y_i = W_{0,1}(x_i)\}$. The topological recursion then provides a universal algorithm to
compute the whole $1/N$ asymptotic expansion of correlation functions, and then  $F_g$ for $g \geq 2$. Beyond computations which are anyway rather heavy to perform explicitly, we learn that, to understand the  singularities of the continuation of $(F_{g})_{g \geq 2}$, $\partial_{\lambda} F_{1}$ and $\partial_{\lambda}^2 F_{0}$ as an analytic function of $\lambda$ in the complex plane, it is enough to understand the singularities of the analytic family of curves $\{y = W_{0,1}(x)\}_{\lambda}$. \\

\br
One may ask what these analytic functions
$F_g(\lambda)$ in \eqref{znexp} have to do with the formal series
$\mathcal{F}_{g}(\lambda)$ in \eqref{znfom}. It can be proved that the Taylor series of $F_g(\lambda)$ at $\lambda \rightarrow
0$ gives $\mathcal{F}_{g}(M^3;\lambda)$. Indeed, it is easy to show that the
formal series $\mathcal{F}_g(M^3;\lambda)$ satisfy some loop equations (let us
call them formal), expressing them as generating series of a certain set of
ribbon graphs with Boltzmann weights prescribed by \eqref{SeiLMO}, and these
equations have a unique solution (see e.g. \cite{Bstuff}).
 It is also well-known that $Z_N(M^3;\lambda)$ satisfies a set of loop equations, obtained for instance by integration by parts in the matrix model. Inserting the form of the asymptotic expansion \eqref{znexp} in these equations, collecting the powers of $N$, and collecting order by order in the Taylor expansion when $\lambda \rightarrow 0$, we obtain the same formal loop equations that were satisfied by $\mathcal{F}_{g}(\lambda)$. We can then conclude by uniqueness of the solution of the formal loop equations.
\er

To recap, the matrix model and the study of $F_g(\lambda)$ give a method to
compute and establish convergence properties and analytic continuation of the
formal series $\mathcal{F}_{g}(\lambda)$. The main task lies in the
computation of the spectral curve, which was mainly addressed in
\cite{BESeifert} by one of the authors. It turns out that among Seifert
spaces, only the ADE geometries have an algebraic spectral curve, with a
subtlety that will be explained in Section~\ref{S41}. In Section~\ref{S42} we
review the construction of the matrix model spectral curves, which consists in
describing the monodromy group of $W(x)$, and exhibiting the unique function
that admits the singular behavior and branchcuts required by the problem.

\subsection{Chern--Simons theory}
\label{CSsec}
In physics, the LMO invariant captures the $\hbar \rightarrow 0$, perturbative
expansion of the Chern--Simons functional integral on $M^3$ with compact, simply-connected gauge group $G =
\exp(\mathfrak{g})$,
\bea
\label{eq:ZCS}
Z_{\rm CS}^{\mathfrak{g}}(k, M^3) &=& \int_{\mathscr{A}/\mathscr{G}} [\DD\cA]
\exp\l(\frac{\ri k}{2\pi} \mathrm{CS}[\cA]\r), \\
\mathrm{CS}[\cA] &=& \int_{M^3} \l(\mathcal{A} \wedge \rd \mathcal{A}+\frac{2}{3} \mathcal{A}^3\r),
\eea
around the trivial flat connection, $\mathcal{A}=g \rd g^{-1}$; here $k \in \mathbb{N}^*$ is the
Chern--Simons level, and the LMO variables are identified as $\hbar = 2{\rm
  i}\pi/(k + h^{\vee})$, $\hat{\lambda}=h^{\vee}\cdot \hbar$ with $h^{\vee}$ the dual
Coxeter number of $\mathfrak{g}$. The full partition function $Z_{{\rm
    CS}}^{\mathfrak{g}}$ of Chern--Simons of Seifert manifolds that are
$\mathbb{Q}$HS can be found in various ways, depending on the mathematical
starting point one chooses for Chern--Simons theory -- which morally realize
the path integral with Chern--Simons action. They all lead to the same answer
for Seifert spaces, and $Z_{{\rm LMO}}^{\mathfrak{g}}$ appears as one term
within $Z_{{\rm CS}}^{\mathfrak{g}}$.\\

In a Hamiltonian context, Mari{\~n{o}} \cite{MarinoCSM} cleverly used the gluing rules of the
Wess--Zumino--Witten TQFT, the Kac--Peterson formula for the $S$- and $T$-matrices, and the
surgery presentation of Seifert spaces to derive the
formula \eqref{SeiLMO} for $Z_{{\rm LMO}}^{\mathfrak{g}}$. His work generalized to all simply-laced Lie algebra an
observation of Lawrence and Rozansky \cite{Law} for $\mathfrak{sl}_{2}$, and
can be seen as the TQFT analogue of the surgery approach of
\cite{BNrational}. His matrix model representation has then been rederived
via functional localization, either by exploiting the $\bbS^1$-action of the
Seifert fibration to reduce \eqref{eq:ZCS} to a discrete sum over flat
connections over the orbifold sphere \cite{BT1,BT2}, or by taking a choice of
a contact structure on $M^3$ and resorting to
non-abelian localization \cite{BeasWitten,Beasley:2009mb} to single out the
contribution of isolated flat connections\footnote{It should be stressed that
  the trivial flat connection is isolated only in the case of lens spaces and
  the $E_8$ Seifert geometry. Therefore, identifying $Z_{{\rm
      LMO}}^{\mathfrak{g}}$ with the trivial connection is only legitimate in
  those cases.}, or yet again \cite{Kallen} by employing localization in a topological twist
of a parent supersymmetric theory \cite{Kapustin}. The authors of
\cite{Beasley:2009mb,Kallen,BT2} 
also show that the insertion of a Wilson line $\mathcal{W}_{\cR}(K_{a_m})$
along the exceptional fiber of order $a_m$ decorated with a representation
$\mathcal{R}$ is represented in terms of $\phi \in \mathfrak{h}$ as an
insertion of the character ${\rm
  ch}_{\mathcal{R}}(e^{\phi_1/a_m},\ldots,e^{\phi_N/a_m})$. For $\mathfrak{g}
= \mathfrak{sl}$, the characters are the symmetric functions, and the
definition of $\mathcal{W}_{\mathcal{R}}$ can be extended by linearity to the
whole character ring. If we restrict to the contribution of the trivial flat connection, a good way to encode all of them at the same time is to define the correlators of the matrix model. The latter are defined, for $n \geq 1$, as:
\beq
W_n(x_1,\ldots,x_n) \triangleq \Big\langle \prod_{j = 1}^n \sum_{i_j = 1}^N \frac{x}{x - e^{\phi_{i_j}/a}}\Big\rangle_{{\rm conn.}}
\eeq
with respect to the measure in \eqref{zndef}, and they depend implicitly on
$\lambda$. For our purposes, it is helpful to work with connected observables,
as they enjoy a well-defined $1/N$ expansion. For $k$ an integer, let $p_k$ be the $k$-th power sum character. Then, we have:
\beq
\big\langle \prod_{j = 1}^n \mathcal{W}_{p_{k_j}}(K_{a_{m_{j}}}) \big\rangle_{{\rm conn.}} = \Big[\prod_{j = 1}^n x^{-k_j (a/a_{m_j})}\Big]\,\,W_n(x_1,\ldots,x_n)\,.
\eeq
We can read off invariants of knots going along the various exceptional fibers $K_{a_m}$ by looking at the coefficients of expansion of the correlators when $x_i \rightarrow \infty$ (or $x_i \rightarrow 0$) for orders that are multiples of $a/a_m$. \\

The discussion of Section~\ref{LMOmatrix} applies to the $W_n$ as well. For the spherical Seifert geometries, the work of \cite{BGK} establishes an asymptotic expansion when $N \rightarrow \infty$:
\beq
\label{connW}W_n(x_1,\ldots,x_n) = \sum_{g \geq 0} N^{2 - 2g-n}\,W_{g,n}(x_1,\ldots,x_n)
\eeq
at least for $\lambda > 0$ small enough. The coefficient of $x^{-k(a/a_m)}$ in
the Laurent expansion at infinity of the function:
\beq
W(x) \triangleq W_{0,1}(x)
\eeq
defining the spectral curve computes the planar limit of the HOMFLY invariant of $K_{a_{m}}$ colored with
the virtual character $p_k$. The other coefficients do not seem to have an
interpretation in terms of 3d topology, but they do influence the monodromy of
the spectral curve\footnote{In the case of lens spaces, invariants of fiber
  knots are related to invariants of torus knots in $\mathbb{S}^3$. We point out that \cite{KJ} defines and compute a new spectral curve that only contains the
  physical part of the information (i.e. the planar limit of HOMFLY's of the
  torus knots) skipping the other coefficients. They are able to find a (very
  complicated) 2-point function which, after applying topological recursion, still
  gives the ``physical part'' of the correct higher genus expansion. From a
  conceptual point of view, it is simpler to keep on with spectral curves that
  may contain knot-theoretic irrelevant information, which are used to get the
  higher genus corrections, and only then discard coefficients which do not have a knot-theoretic interpretation. The equivalence between the two approaches is guaranteed by a property of commutation with ``forgetting information'' enjoyed by the topological recursion, see \cite{BEO}.}.

\section{Construction of $Y^{\Gamma}$ and topological string dualities}

\subsection{Topological large $N$ duality} 

\label{mimi}

As for any quantum gauge theory with gauge group ${\rm U}(N)$ and fields in the adjoint representation,
the formal perturbative expansion of the Chern--Simons path integral can be
formulated as an expansion in ribbon graphs $\mathsf{G}$, 
 whose dual graphs are
triangulations of a closed oriented topological 2-manifold $S_{\mathsf{G}}$. Elementary
combinatorics then shows that each loop in the diagram contributes a factor of
$\hat \lambda= g_{{\rm YM}}^2 N$, and the overall topology contributes a factor of $g_{{\rm YM}}^{-2\chi(C_{\mathsf{G}})}$ \cite{tHooft}. In
particular, the perturbative free energy takes the form
\beq
\mathcal{F}^{\mathfrak{sl}_{N+1}}_{{\rm CS}}(M^3;g_{{\rm YM}})=\sum_{g,n\geq 0} \mathcal{F}_{g,n}(M^3)\hat{\lambda}^n g_{{\rm YM}}^{4g-4} \in g_{{\rm YM}}^{-4}\mathbb{Q}[[\hat{\lambda},g_{{\rm YM}}^4]].
\eeq
For the case of ${\rm U}(N)$ Chern--Simons theory on a closed oriented $3-$manifold
$M^3$, Witten showed \cite{WittenCS} that this can be reinterpreted as the
target string field theory of the open topological A-model on the cotangent bundle $T^*M^3$, with $N$ Lagrangian
A-branes wrapping the image of the zero section (see \cite{Marino:2004uf}
for a review). Here, the string coupling constant should be identified with
$g_s=g_{{\rm YM}}^2$; in particular, the ribbon graph expansion
translates into 
a virtual count of open holomorphic worldsheet instantons with A-type Dirichlet
boundary condition on $M^3$. A formal resummation of the contribution of the connected contribution of the boundary --~the ``holes'' in the worldsheet~--
gives rise to a formal closed string expansion,
\beq
\mathcal{F}^{\mathfrak{sl}_{N + 1}}_{\rm CS}(M^3,g_{{\rm YM}})=\sum_{g\geq
  0}g_s^{2g-2}\cdot \hat{\lambda}^{-(2g - 2)}\,\mathcal{F}_{g}(M^3;\hat \lambda),
\eeq
When $M^3=\bbS^3$, Gopakumar and Vafa identified the closed string model as the
closed topological A-model on the resolved conifold $\mathrm{Tot}[\cO(-1) \oplus \cO(-1)
\rightarrow \mathbb{P}^1]$: here $g_s$ is the closed string coupling constant,
and $\hat \lambda$ is identified with the
K\"ahler parameter of the base $\mathbb{P}^1$. Geometrically, this target space is
obtained from $T^*\bbS^3$ by a complex degeneration to a normal singular variety (the singular
conifold) obtained by contracting the base $\mathbb{S}^3$, followed by a
minimal crepant resolution of the resulting singularity  with
a $\mathbb{P}^1$ as its exceptional locus. While there are obstructions to extend this circle of ideas to more general
3-manifolds \cite{Brini:2008ik}, it is still natural to conjecture, in view of
the positive results of 
\cite{Halmagyi:2003ze}, that the same scenario could apply to the case of
spherical Seifert manifolds and $\Gamma \subset \rm SU(2)$ quotients of the conifold, as we now
describe.

\label{sec:geotrans}

\subsection{Geometric transition for $\mathbb{S}^3$}

Let us review the conifold transition for the simplest case of $\bbS^3$ with unit
radius. Since $\bbS^3\simeq \mathrm{SU}(2)$ is a Lie group, $T^*\mathbb{S}^3$ is a trivial bundle; its
fiber at identity is the space ${\rm i}\mathcal{H}_0(2,\mathbb{C})$ of
traceless anti-hermitian $2\times 2$ matrices. Any matrix $A \in
\mathrm{GL}(2,\mathbb{C})$ can be written uniquely by polar decomposition $M =
Ue^{H}$ where $U \in \mathrm{U}(2)$ and $H \in \mathcal{H}(2,\mathbb{C})$
definite positive, and if we restrict to $\mathrm{det}(A) = 1$, we must have $\mathrm{det}(U) = 1$ and $\mathrm{tr}(H) = 0$. Therefore, the polar decomposition gives an isomorphism:
\beq
T^*\mathbb{S}^3 \mathop{\simeq}^{\rho} \mathrm{SL}(2,\mathbb{C})
\eeq
This description can be fit into a flat family $\psi: X = {\rm GL}(2,\mathbb{C}) \to \bbC^*$ given by the determinant map. Then the fiber $X_{[\mu]}$ at a point $\mu$ such that ${\rm Im}\,\mu =0$ and ${\rm Re}\,\mu>0$ is isomorphic to the cotangent bundle $T^*\mathbb{S}^3_{[\mu]}$ of a sphere
with radius $\mu$. Explicitly, writing 
\beq
\rho(A) = w_0 + {\rm i}\vec{w}\cdot\vec{\sigma},\qquad w_j = p_j + {\rm i}q_j
\label{eq:rhoA}
\eeq
realizes $X_{[\mu]}$ as the real complete intersection in $T^*\mathbb{R}^4$
cut out by $\sum_{j=1}^4 q_j^2 - p_j^2=\mu,$ $\sum_{j=1}^4 q_j p_j=0$. \\

 Let us add the locus of non-invertible matrices to form:
\beq
\tilde \psi\,:\, {\rm Mat}(2,\mathbb{C}) \longrightarrow \bbC\,.
\eeq
The fiber $X_{[0]}$ above $\mu = 0$ is the singular quadric $\det A=0$. It admits a canonical minimal resolution 
\beq
\pi\,:\,\widehat{X} \longrightarrow X_{[0]},\qquad  \widehat{X} \triangleq \big\{(\rho(A),v) \in X_{[0]} \times \mathbb{P}^1,\quad \rho(A) v =0\big\},\qquad 
\label{eq:rescon} 
\eeq
where $\pi$ is the projection to the first factor. The point $A = 0$ is
singular in $X_{[0]}$, and its fiber is a complex projective line with
$[v_1:v_2]$ as homogeneous coordinates. Using coordinate charts on
$\mathbb{P}^1$ exhibits $\widehat{X}$ as the total space of $\cO(-1)\oplus
\cO(-1)\to \mathbb{P}^1$, i.e. the resolved conifold. As a symplectic
manifold, it supports a one-dimensional family of complexified K\"ahler forms coming from
its presentation in \eqref{eq:rescon}, namely
%
\beq
\omega_{t_{ \rm B}}=i_1^* \omega_{\bbC^4} + t_{\rm B} i_2^* \omega_{\rm FS},
\label{eq:kform}
\eeq
where $i=(i_1,i_2)$ is the factorized form of the embedding $i : X \hookrightarrow
\mathrm{Mat}(2,\bbC) \times \bbP^1$ from \eqref{eq:rescon}, and
$\omega_{\bbC^4}$ and $\omega_{\rm FS}$
 are respectively the canonical K\"ahler form on
$\mathrm{Mat}(2,\bbC)\simeq \bbC^4 \simeq T^* \bbR^4$ and the Fubini--Study form on $\bbP^1$.

\subsection{Geometric transition for $\mathbb{S}^{\Gamma}$}
\label{YGamma}

We now consider the action of finite groups of isometries of $\mathbb{S}^3$, reviewed in Appendix~\ref{gpact}. The morphism $\rho$ is compatible with the isometric action of left and right multiplication on $\mathbb{S}^3 \simeq {\rm SU}(2)$. This means that, if we denote $\tilde{\Phi}_4$ the lift of this action to an action by symplectomorphisms on $T^*\mathbb{S}^3$, we have for any $(q_1,q_2) \in \mathrm{SU}(2)\times \mathrm{SU}(2)$ and any $A \in T^*\mathbb{S}^3$,
\beq
\rho(\tilde{\Phi}_{4}(q_1,q_2)\cdot A) = q_1\rho(A)q_2^{-1}.
\eeq

Let us focus on the left action by a finite
subgroup $\Gamma\subset\mathrm{SU}(2)$. This is a fiberwise action on
$\psi\,:\,X \rightarrow \mathbb{C}^*$, which is free on each fiber
$X_{[\mu]}$. When $\mu>0$, we claim that the set of equivalence classes is
just isomorphic to $T^*\mathbb{S}^\Gamma$. Indeed, consider the local diffeomorphism on
$\bbR^8$ given by 
\beq
\bary{ccccccc}
\tilde p_1 &=& q_1 p_1 + q_2 p_2+q_3 p_3 + q_4 p_4, & \qquad & \tilde p_2 &=& q_1 p_2 - q_2 p_1+q_4 p_3 - q_3 p_4, \\
\tilde p_3 &=& q_3 p_1 + q_4 p_2-q_1 p_3 - q_2 p_4, & \qquad & \tilde p_4 &=& q_3 p_2 - q_4 p_1-q_2 p_3 + q_1 p_4,
\eary 
\eeq
and 
\beq
\tilde q_i = \frac{q_i}{\sqrt{\mu+\sum_{j = 1}^4 p_j^2}}.
\eeq
It is non-singular everywhere for $\mu >0$, and the resulting real sixfold is just
$\bbR^3 \times \bbS^3$, cut out in $\mathbb{R}^8$ by:
\beq
\label{eq:invts3}
\tilde p_1 = 0,\qquad \sum_{i=1}^4 \tilde q_i^2 = 1.
\eeq
Using the generators of $\Gamma$ given in Appendix~\ref{gpact}, it can be checked that the coordinates $\tilde p_i$ are $\Gamma$-invariant so that the quotient is:
\beq
X^\Gamma_{[\mu]} = 
\frac{\mathrm{Spec}\,\mathbb{C}[A]^{\Gamma}}{\langle \det A = \mu \rangle} \simeq \mathbb{R}^3 \times \mathbb{S}^\Gamma
\eeq
which is isomorphic to $T^*\mathbb{S}^{\Gamma}$ by Stiefel's theorem. \\

\label{mcka}
Now, let us look at the $\Gamma$-action on the resolution $\widehat{X}$. It only acts on the first factor of
\eqref{eq:rescon}, and hence this is a fiberwise action on $p\,:\,\widehat{X}
\rightarrow \mathbb{P}^1$ (the second factor in \eqref{eq:rescon}). The fiber
at a point $z \in \mathbb{P}^1$ is isomorphic to the du Val singularity
$\Gamma\backslash\mathbb{C}^2$, and the resulting target geometry can be
studied in two distinguished chambers of the stringy K\"ahler moduli space.  Let
\beq
R \triangleq {\rm rank}(\mathcal{G}_{\Gamma})\,.
\eeq
In
the orbifold chamber, we are looking at the orbifold A-model on $Y^\Gamma_{\rm
  orb}\triangleq [\Gamma \backslash \cO^{\oplus 2}_{\bbP^1}(-1)]$. Its degree two orbifold quantum cohomology -- i.e. the space of marginal deformations of
the A-model chiral ring -- is generated by classes $(\delta,
(\xi_j)_{j=1}^R)$; here $\delta$ is the class of the base of $[Y^\Gamma]_{[0]} \to
\bbP^1$, where $[]_{[0]}$ denotes the untwisted sector, and $\xi_j$ are
twisted orbifold cohomology classes of Chen--Ruan degree two. In the large
radius chamber, we take a crepant resolution $Y^\Gamma_{\rm res}$ of the singularities of
$Y^\Gamma_{\rm orb}$ obtained by canonically resolving the surface singularity
$\Gamma\backslash \bbC^2$ fiberwise. The resulting Calabi--Yau
threefold $Y^\Gamma$ is thus an ALE fibration
over $\mathbb{P}^1$, with fibers given by configurations of rational curves
having normal bundle $(0,-2)$, and whose
intersection matrix equates the negative of the Cartan matrix of
$\cG_\Gamma$ \cite{reid2012val}. Then $H^2(Y^\Gamma_{\rm res})$ is generated as a
vector space by the base class $\delta$ above, plus classes $(\gamma_j)_{j = 1}^{R}$ representing the nodes in the chain of exceptional fiber
$\mathbb{P}^1$'s. In the following we will often write
$Y^\Gamma$ to refer to either of the two
chambers whenever the context applies to both of them. \\

\subsection{A-model: Gromov--Witten theory on $Y^{\Gamma}$}
\label{sec:amodel}

In terms of the coordinates $\{a_{ij}=\rho(A)_{ij}\}_{i,j=1,2}$ and $[v_1:v_2]$ of
\eqref{eq:rhoA} and \eqref{eq:rescon}, $Y^\Gamma$ supports a natural
$T\simeq \bbC^*$-action given by
\beq
\l(a_{11},a_{12},a_{21},a_{22}; [v_1: v_2]\r) \longrightarrow 
\l(
\mu a_{11},  a_{12}, \mu a_{21} , a_{22}\,;\, [\mu^{-1}v_1: v_2]\r)\,,
\label{eq:torus}
\eeq
Here $T$ acts trivially on the canonical bundle: on the full resolution
$Y^\Gamma_{\rm res}$, it has a compact fixed locus $Y^\Gamma_{{\rm res},T}$
consisting of two fibers above $[0:1]$ and $[1:0]$, each isomorphic to a disjoint union of a $\bbP^1$ with $(R - 2)$ points; 
likewise, its fixed locus on $Y^\Gamma_{\rm
  orb}$ is the union of two $\Gamma$-orbifold points, i.e. $Y^\Gamma_{{\rm orb},T} \simeq B\Gamma \sqcup B\Gamma$.
The A-model/Gromov--Witten closed free
energy of $Y^\Gamma$ is then defined/computed by localization \cite{MR1666787}:
\bea
\label{eq:GW1}
F^{\rm GW}(Y^\Gamma) & \triangleq & \sum_{g\geq 0} g_s^{2g-2}F^{\rm
  GW}_g(Y^\Gamma, t), \\
\label{eq:GW2}
F_{g}^{{\rm GW}}(Y^\Gamma, t) & \triangleq & \sum_{n=0}^\infty\sum_{\beta \in
  H^2(Y^\Gamma,\bbZ)} \frac{\bra \Phi(t), \dots, \Phi(t) \ket_{g,\beta}^{Y^\Gamma}}{n!},  \\
\bra \varphi_1, \dots, \varphi_n \ket & \triangleq & \int_{[\overline{\cM_{g,0}}(Y_T^\Gamma,\beta)]^{\rm
  virt}} \mathrm{ev}_1^*\varphi_1 \cup \dots \cup \mathrm{ev}_n^* \varphi_n  \in \bbQ(\mu),
\label{eq:GW3}
\eea
where $\mu=c_1(\cO_{BT}(1))$ denotes the equivariant parameter of $T$ and
$\Phi$ is a cohomology class specified by linear
coordinates $t$ on $H^\bullet(Y^\Gamma)$. In
fact, as the torus action is Calabi--Yau (i.e. it preserves the holomorphic volume form), Gromov--Witten
invariants in positive degree \eqref{eq:GW3} do not depend on $\mu$ \cite{moop}, nor do the
higher genus invariants for $g \geq 2$ and all
$\beta$. Equations \eqref{eq:GW1}-\eqref{eq:GW3} will be our candidate for the
A-model dual of the Chern--Simons free energy at large $N$.\\

\subsubsection*{$A$-branes}
\label{sec:abranes}
The geometry of $Y^\Gamma$ offers also a natural candidate for an A-model
description of the large $N$ expansion of the Wilson loops along fiber knots,
\eqref{connW}, in terms of open Gromov--Witten invariants
\cite{Katz:2001vm,MR2861610}. On the resolved conifold $Y = Y^{\Gamma = \{1\}}$, consider the anti-holomorphic involution
$\sigma: Y \to Y$ induced by $\sigma(a_{22})=\overline{a_{11}}$, 
$\sigma(a_{21})=\overline{a_{12}}$. Equivalently, this means $\sigma(w_{0,3})=\overline{w_{0,3}}$,
$\sigma(w_{1,2})=-\overline{w_{1,2}}$ in \eqref{eq:rhoA}) and $v_i\to 
\overline{v_{3-i}}$. Its fixed locus is thus isomorphic to $\bbR^2 \times
\mathbb{S}^1$, where the circle is given by the equator of the base $\bbP^1$, and it is
Lagrangian with respect to the canonical K\"ahler form \eqref{eq:kform}, as
the first summand in \eqref{eq:kform} changes sign under $\sigma$, and the
second vanishes as $Y_{\sigma} \cap \bbP^1$ has non-vanishing codimension.\\

When $\Gamma\subset \mathrm{SU}(2)$ is cyclic, the $\Gamma$-action descends to a free action on the fixed
locus $Y_{\sigma}$: this simultaneously defines Lagrangian
branes  on
$Y^\Gamma_{\rm orb}$ and $Y^\Gamma_{\rm res}$ by respectively taking the orbit
space $Y_{\sigma}^{\Gamma} \triangleq \mathcal{L}_{\rm orb}^\Gamma$ for
$Y^\Gamma_{\rm orb}$, and the
transform $\mathcal{L}_{\rm res}^\Gamma$ of this condition under the
resolution map for $Y^\Gamma_{\rm res}$\footnote{We again omit
  the subscript from $\mathcal{L}_{\rm res}^\Gamma$ and $\mathcal{L}_{\rm
    orb}^\Gamma$ whenever the statements apply to both.}. When $\Gamma$ is non-abelian,
on the other hand, the $\Gamma$ action does not descend to an action on the
$\sigma$-fixed locus, as can be checked directly on the generators
\eqref{eq:iota}--\eqref{eq:kappa}. However, $\Gamma$ preserves the symplectic form \eqref{eq:kform} on $Y$ (see Appendix~\ref{gpact}) , and one can check that the images of $Y_\sigma$ under the degree 3 (resp. 2) generator $\jmath$
(resp. $\kappa$) are Lagrangians having empty intersection with $Y_\sigma$. Then, defining
\beq
Y^{\Gamma}_\sigma = 
\l\{
\bary{rl}
Y_\sigma & \DD_\Gamma=A,\\
Y_\sigma \sqcup \iota(Y_\sigma) & \DD_\Gamma=D,\\
\bigsqcup_{\phi={\rm id}, \jmath, \jmath^2} \phi(Y_\sigma) &
\DD_\Gamma=E_6, E_7,\\
\bigsqcup_{\phi={\rm id}, \jmath, \jmath^2, \kappa} \phi(Y_\sigma) &
\DD_\Gamma=E_8,\\
\eary
\r.
\eeq
the $\Gamma$-action descends on $Y^{\Gamma}_\sigma$ to give Lagrangian branes
$\mathcal{L}_{\rm orb}^\Gamma$ and $\mathcal{L}_{\rm res}^\Gamma$ as before. These branes have topology $\bbR^2/(\bbZ/q_\Gamma \bbZ) \times \mathbb{S}^1$, where $q_\Gamma$
is tabulated in Table~\ref{tab:qgamma}; notice that the $\Gamma$-action leaves
the base $\mathbb{P}^1$ unaffected (hence the $\mathbb{S}^1$ factor) and that
$Y_{\sigma}^{\Gamma}$ is constructed from Lagrangian copies of $Y_{\sigma}$ in
the orbit of ``non-cyclic'' generators $\iota$, $\jmath$ and $\kappa$, hence the $\Gamma$-action factors through a residual cyclic action on $Y_{\sigma}$, giving rise to a cyclic quotient of $\mathbb{R}^2$.
\begin{table}[!h]
\beq
\bary{|c|c|}
\hline
\DD_\Gamma & q_\Gamma \\
\hline
A_{p-1} & p \\
\hline
D_{p+2} & 2p+4 \\
\hline
E_{6,8} & 4 \\
\hline
E_7 & 8\\
\hline
\eary
\eeq
\caption{Orders of the residual cyclic group action on $Y_\sigma$ for
  $\DD_\Gamma=A_n, D_n, E_n$.}
\label{tab:qgamma}
\end{table}

As for
the usual toric case, the Calabi--Yau torus action \eqref{eq:torus} allows
then to
define a virtual counting theory of stable open maps \cite{MR2861610,Katz:2001vm,Brini:2011ij} to $Y^\Gamma$ having
Dirichlet boundary conditions on $\mathcal{L}^\Gamma$  via equivariant residues on
$\overline{\cM}_{g,n}$ and $\overline{\cM}_{g,n}(\bbP^1,\beta)$ (for
$Y^\Gamma_{\rm res}$) or $\overline{\cM}_{g,n}(B\Gamma)$ (for $Y^\Gamma_{\rm orb}$):
\bea
\bra \varphi_1, \dots, \varphi_n\ket^{Y^\Gamma,\LL^\Gamma}_{g,n,\zeta, \vec d}
& \triangleq & \int_{[{\overline{\cM}_{g,n}}(Y^\Gamma,\mathcal{L}^\Gamma,\zeta,\vec
    d)_T]^{\rm virt}}\frac{\mathrm{ev}_1^*\varphi_1 \cup \dots \cup \mathrm{ev}_n^* \varphi_n}{\re_T\l( N^{\rm
    virt}_{{\overline{\cM}_{g,n}}(Y^\Gamma,\mathcal{L}^\Gamma,\zeta,\vec d)_T}
  \r)}.
\eea
Here, $n$ is the number of connected components of the boundary of the source curve,  $\vec d = (d_1, \dots, d_n)$ with
$d_i\in H_1(\mathcal{L}^\Gamma)$ describe their winding around the equator, and $\zeta\in H_2(Y^\Gamma,\mathcal{L}^\Gamma)$ is the relative homology class representing the image of the open worldsheet
in $Y^\Gamma$. This can be packaged into formal generating series:

\beq
W_{g,n}^{{\rm GW}}(Y^\Gamma,\mathcal{L}^\Gamma;t, w)\triangleq
\sum_{n,\zeta,\vec d}
\frac{\bra \Phi(t), \dots, \Phi(t) \ket_{g,n,\zeta, \vec d}^{Y^\Gamma, \mathcal{L}^\Gamma}}{n!}
\prod_{i=1}^n \frac{w_i^{d_i}}{d_i!}.
\label{eq:wghgw}
\eeq
where $t$ are again quantum cohomology parameters accounting for localized
primary insertions. On the resolution, the divisor equation puts
\eqref{eq:GW2} and \eqref{eq:wghgw} in the form of the familiar worldsheet instanton
expansion\footnote{The class $\beta \in H_2(Y^\Gamma)$ here is retrieved as the image of $\zeta$
under the connecting morphism in the relative homology exact
sequence for $(Y^\Gamma,\mathcal{L}^\Gamma)$. As the constraint $\de \zeta=\sum_i
d_i$ for the moduli space to be non-empty singles out a unique pre-image
$\zeta$ for $\beta$, we slightly abuse notation and switch $\zeta
\leftrightarrow \beta$ to emphasize the dependence of $W_{g,n}^{{\rm GW}}$ on the
bulk/boundary moduli.} 

\bea
\label{eq:FgGW}
F_{g}^{{\rm GW}}(Y^\Gamma_{\rm res}, t) &=& \sum_{\beta \in
  H^2(Y^\Gamma_{\rm res},\bbZ)} \bra 1 \ket_{g,\beta}^{Y^\Gamma_{\rm res}} \re^{\beta \cdot t} \\
W_{g,n}^{{\rm GW}}(Y^\Gamma_{\rm res},\mathcal{L}^\Gamma_{\rm res};t, w) &=& \sum_{\beta,\vec d} 
 \bra 1 \ket_{g,\beta,\vec d}^{Y^\Gamma_{\rm res},\mathcal{L}^\Gamma_{\rm res}} \re^{\beta \cdot t}
\prod_{i=1}^n \frac{w_i^{d_i}}{d_i!}.
\label{eq:WgnGW}
\eea

\subsection{Geometric engineering and mirror symmetry}
\label{sec:geotrans2}
When $\Gamma$ is a cyclic group, $Y^\Gamma$ is a toric variety
and it admits a family
of mirror spectral curves $(\mathcal{C}^{\Gamma},\Omega^{\Gamma})$ described by Hori--Iqbal--Vafa
\cite{Hori:2000ck,Brini:2008ik}. When $\Gamma$ is non-abelian, $Y^\Gamma$ is not toric anymore as
the fibers in the ALE fibration only possess the one-dimensional torus action
\eqref{eq:torus}, which is the the lift of the scalar action on
$\Gamma\backslash\bbC^2$; as a result the standard toric methods used to deduce an explicit
picture in terms of mirror Calabi--Yau $3$-folds (let alone mirror curves) do
not apply here. However, at least in some special limits
it has been argued in the physics literature that the genus zero A-topological
string on $Y^\Gamma$ 
should be governed by special geometry on a family of curves. 
Denoting by $t_{\rm B}$ and $t_j$ the K\"ahler
parameters of $\delta$, $\gamma_j \in H^2(Y^\Gamma_{\rm res},\bbZ)$, it was proposed in a series of papers \cite{Kachru:1995fv, Klemm:1996bj,
  Katz:1996fh} that the $g = 0$ free energy of the type A--topological
string on $Y^\Gamma_{\rm res}$ should coincide with the prepotential of $\cN=2$, $d=4$
pure super Yang--Mills with gauge group $\cG_\Gamma$ upon identifying the quantum
Coulomb moduli as $a_j=t_j/\epsilon$, the holomorphic scale as
$\Lambda=\re^{-t_{\rm B}/4}/\epsilon$, and taking the limit $\epsilon\to 0$. This
limit
corresponds to a type IIA compactification on a $K3$ where we
``zoom'' around an ADE singularity by sending the Planck mass to
infinity. The overall effect is to decouple the gravitational modes and give
rise at the same time to enhanced ADE gauge symmetry. Further fibering that
over a $\mathbb{P}^1$ yields a pure gauge field theory in $d=4$ with eight
supercharges and no hypermultiplets as the effective four-dimensional theory. As a
result, in this degenerate situation we {\it do} expect a spectral curve mirror: this
is the Seiberg--Witten curve of the geometrically engineered gauge theory.\\

What about the case of finite $\epsilon$? When $\Gamma = \mathbb{Z}/p\mathbb{Z}$, i.e.
$\cG_\Gamma= A_{p-1}$, it was argued in \cite{Lawrence:1997jr} that uplifting
the reasoning above to M-theory compactified on a circle gives rise to {\it exactly}
the same type of identification, where now the UV scale $1/\epsilon$ is identified with
 the inverse of the radius of the eleventh dimensional circle. This
gives an exact identification of the gauge theory prepotential of the
resulting $\cN=1$, $d=5$ field theory with the topological string free energy: 
the ``field theory limit'' of \cite{Kachru:1995fv, Klemm:1996bj,Katz:1996fh}
becomes here just the limit from five to four dimensions. 
The upshot is that the sought-for mirror of $Y^{\mathbb{Z}/p\mathbb{Z}}$ should take the
form of a 
$d=5$ Seiberg--Witten curve for the pure gauge theory with group
$\cG_\Gamma$. When $\cG_\Gamma=A_{p-1}$, this was obtained by Nekrasov in
\cite{Nekrasov:1996cz}, and the resulting geometry is the spectral curve
$\mathcal{C}_{A_{p-1}}^{\rm SW}$ of the
periodic relativistic Toda chain with $p$-particles \cite{MR1090424}:
\beq
\mathcal{C}^{{\rm SW}}_{A_{p-1}} =\Big\{(X,Y) \in \bbC^*\times\bbC^*,\qquad 
\re^{-t_{\rm B}/2}\l(X+ X^{-1}Y^p\r) = Y^p+\sum_{k=1}^{p-1} u_{p-k} (-Y)^k +1  \Big\}, 
\label{eq:CA}
\eeq
equipped with the $1$-form:
\beq
\Omega^{{\rm SW}}_{A_{p-1}} =\log{Y} \frac{\rd X}{X}.
\eeq
Unsurprisingly, this coincides with the Hori--Iqbal--Vafa mirror of $Y^{\mathbb{Z}/p\mathbb{Z}}$. Using brane constructions,
Nekrasov's result has been generalized to arbitrary classical groups, and
in particular $\cG=D_{p + 2}$ in \cite{Thei}: 
\beq
\mathcal{C}^{{\rm SW}}_{D_{p + 2}} =\big\{ (X,Y) \in \bbC^*\times\bbC^*,\quad 
\re^{-t_{\rm B}/2}\l(X+ X^{-1}\r)(Y^2-1)^2 Y^{p}=(-1)^p 2^{-2p}\prod_{j=1}^{p + 2} (Y-r_j)(Y-r_j^{-1})
  \big\},
\label{eq:CD}
\eeq
again with the canonical Seiberg--Witten differential $\Omega^{{\rm SW}}_{D_{p + 2}}
= \log{Y}\rd X /X$. \\

No results are available in the literature for the exceptional cases away from
the $4d$ limit
(see however \cite{Lerche:1996an,Eguchi:2001fm} for the $E_6$ and $E_7$
cases when $\epsilon\to 0$). However, Nekrasov's original insight \cite{Nekrasov:1996cz} naturally suggests
that the resulting geometry should be in all cases the spectral curve of a
relativistic deformation of the Lie-algebraic Toda systems relevant for
the four-dimensional limit \cite{Martinec:1995by}. Fortunately, the relevant technology
for the construction of the spectral curves has recently become available
since the work of Williams \cite{Williams:2012fz} and
Fock--Marshakov \cite{Fock:1997aia,Fock:2014ifa}, as we now turn to review.

\subsection{B-model: the classical affine co-extended ADE Toda chain}
\label{sec:Todain}

A simple, simply-laced Lie group $\cG$ of rank $R$, with maximal torus $\cT$, can be endowed with a canonical
Drinfeld--Jimbo Poisson structure 
\beq
\l\{g \stackrel{\otimes}{,} g\r\}=-\frac{1}{2}[r,g\otimes g],
\label{eq:DJ}
\eeq
where
\beq
r=\sum_{\alpha \in \Delta^+} e_\alpha \otimes e_{\overline \alpha}+\frac{1}{2}\sum_{i=1}^{R} h_i \otimes h_i
\eeq
is the canonical solution of the classical Yang--Baxter equation on $\cG$
\cite{MR1636293}; here $\Delta_+$ is the set of positive roots, and $(h_i,e_\alpha, e_{\overline{\alpha}})$ is a
Chevalley basis of generators of $\mathrm{Lie}(\cG)$. We choose a labeling of
the nodes of the Dynkin diagram of $\cG$ by $i = 1,\ldots,R$, which leads in
turn to a labeling of the Cartan generators. $\cG$ has a cell-decomposition  
\beq
\cG=\coprod_{\mathsf{w} \in \mathfrak{W}_{\cG} \times \mathfrak{W}_{\cG}}
\cG_{\mathsf{w}}, 
\eeq
where $\mathfrak{W}_\cG$ is the Weyl group of $\cG$ and the double Bruhat cells $\cG_{\mathsf{w}}$ are themselves Poisson
manifolds. As $\cT\subset \cG$ is a trivial Poisson subgroup of $\cG$, the Poisson
structure \eqref{eq:DJ} descends to Poisson structures on $\cG/\cT$ and $\cG_{\mathsf{w}}/\cT$,
where the quotient is taken by the adjoint action of the torus. Given a
standard decomposition of a word $\mathsf{w}\in \mathfrak{W}_{\cG}\times \mathfrak{W}_{\cG}$ of length $l$ into reflections $\mathsf{w}= \psi_{i_{1}} \circ \cdots \circ \psi_{i_{l}}$ with respect to the simple roots $\alpha_{i_j}$ labeled by the nodes $i_j$ of the Dynkin diagram, the map
\beq
\bary{cccl}
L_{\mathsf{w}}^\cG : & (\bbC^*)^{l} & \longrightarrow & \cG_{\mathsf{w}}/\cT \\
& \{\varkappa_m\}_{m=1}^{l} & \longrightarrow & \prod_{m=1}^{l} H_{i_m}(\varkappa_m) E_{i_m}
\eary
\label{eq:lax1} 
\eeq
is a Poisson morphism with respect to the logarithmically constant Poisson structure on $(\bbC^*)^l$
determined by the exchange matrix $\epsilon$ on the corresponding Poisson
quiver (see \cite{Kruglinskaya:2014pza}):
\beq
\{\varkappa_i,\varkappa_j\}=\epsilon_{ij} \varkappa_i \varkappa_j.
\eeq
In \eqref{eq:lax1}, $H_{i}(\varkappa) = \exp(\varkappa h_i)$ and $E_{i} = \exp(e_{i})$ are elements of $\mathcal{G}$ obtained by exponentiating the Chevalley generators.  The operator $L^\cG_{\mathsf{w}}$ is the Lax matrix of a classical integrable system on $\cG_{\mathsf{w}}/\cT$: the
coefficients of its characteristic polynomial give then a set of independent
Ad-invariant (hence Poisson commuting) functions on $\cG_{\mathsf{w}}/\cT$. \\ 

When $\cG=\mathrm{SL}(p + 1)$, the resulting mechanical system is the open
relativistic Toda chain with $p$ sites \cite{MR1090424}. As was the case for
the Lie-algebraic version of the non-relativistic Toda system, generalizing this picture
to the periodic case relevant for the discussion of the previous section
amounts to extending the construction above to the case of affine Lie groups. It was
proposed in \cite{Fock:2014ifa} that the relevant Poisson
submanifolds in this case should be constructed on the co-extended loop group $\cG^\#\simeq
{\rm Loop}(\cG) \rtimes \bbC^*$, upon projecting onto elements having trivial
co-extension. In particular, we focus on the double Bruhat cell labeled by the cyclically
irreducible word:
\beq
\mathsf{w} \triangleq 1\bar1 \dots R \overline{R}\,.
\eeq
The corresponding Lax matrix $L_{\mathsf{w}}^{\cG^\#}$ is obtained from
$L_{\mathsf{w}}^\cG$ by adjoining a 
spectral parameter-dependent contribution by the affine root of ${\rm Loop}(
\cG)$ \cite{Kruglinskaya:2014pza}, as
\beq
L_{\mathsf{w}}^{\cG^\#}(\varkappa_1,  \varkappa_{\overline{1}}, \ldots, \varkappa_{R},\varkappa_{\overline{R}}\,;\,X) \triangleq
\prod_{i=1}^{R} H_i(\varkappa_i) E_i H_{i}(\varkappa_{\overline{i}}) E_{\overline{i}} E_0(X/\varkappa_0)
E_{\overline{0}}(X^{-1})\,, 
\label{eq:lax2}
\eeq
with $\varkappa_0, X\in\bbC^*$ and the product is done starting from $i = 1$ on the left and ending at $i = R$ on the right. Denote by $(\chi_{\omega_i})_{i = 1}^{R}$ the characters of
the fundamental representation with highest weight $\omega_i$, where
$\omega_i(\alpha_j) = \delta_{ij}$. We have a map
\bea
u: (\bbC^*)^{2R} \times \bbC^*_{\varkappa_0} \times \bbC^*_X & \longrightarrow & \bbC_u^{R}
\nn\\
L_{\mathsf{w}}^{\cG^\#} & \longmapsto & \chi_{\omega_i}(L_{
   \mathsf{w}}^{\cG^\#,[0]})
\eea
obtained by taking the constant term $L_{\mathsf{w}}^{\cG^\#,[0]}$ in the
Laurent expansion of $L_{\mathsf{w}}^{\cG^\#} \in \cG[X,X^{-1}]$ and then
evaluating its fundamental characters. This is a
submersion of $(\bbC^*)^{2R+2}$ onto a Zariski open subset $\mathfrak{U}_\cG$ of 
$\bbC^{R}$ with the linear coordinates:
\beq
u_i=\chi_{\omega_i}(L_{\mathsf{w}}^{\cG^\#,[0]})
\eeq
giving a complete set of hamiltonians in involution. Furthermore, let $l_i\in
\bbN$
be the coefficients of the highest positive root in the $\alpha$-basis for
$\cG$. Then, upon projecting to trivial co-extension,
\beq
u_0\triangleq \varkappa_{0}^{1/2}\prod_{i=1}^R \varkappa_i^{l_i}= \varkappa_0^{-1/2}\prod_{i=1}^R \varkappa_{\overline{i}}^{-l_i}
\label{eq:casimir}
\eeq
gives a Casimir for the Poisson bracket on $\cG^\#_{\mathsf{w}}$. Fix now an arbitrary irreducible representation $\rho \in
{\rm Rep}(\cG)$. The characteristic polynomial of
$\rho(L_{\mathsf{w}}^{\cG^\#})$  then gives a 
family of plane curves $\mathcal{C}^{\rm Toda}_{\cG^{\#}}\subset (\bbC^*)^2$ over
$\hat{\mathfrak{U}}_{\cG} \triangleq \bbC^*_{u_0}\times \mathfrak{U}_\cG$. The
curve above a point 
$u=(u_0,\dots, u_{R})$ given by 
\beq
\mathcal{C}^{\rm Toda}_{\cG^{\#},\rho} = \big\{ (X,Y) \in \bbC^*\times\bbC^*,\qquad  \det\big[Y
\mathbf{1} - \rho(L_{\mathsf{w}}^\cG(\varkappa; X))\big]=0\big\}. 
\label{eq:sctoda}
\eeq
We further equip $\mathcal{C}^{\rm Toda}_{\cG^{\#},\rho}$ with the $1$-form:
\beq
\Omega_{\cG^{\#},\rho}^{{\rm Toda}} = \log{Y} \frac{\rd X}{X}\,.
\eeq
When $\cG = A_1 = \mathrm{SL}(2)$ and $\rho = \Box$ is the fundamental representation, this is just the
holomorphic Poincar\'e $1$-form on the phase space of the relativistic Toda particle.

\section{The two main conjectures}
\label{S4}
\label{conjs}
It can easily be shown that, upon specializing \eqref{eq:sctoda} to
$(\cG=A_{p}, \rho = \Box)$ and $(\cG=D_{p + 2}, 
\rho =\mathbf{2(p + 2)}_{\bf v})$, we obtain
\cite{Nekrasov:1996cz,Kruglinskaya:2014pza} that: 
\beq
\mathcal{C}^{\rm Toda}_{A_p,\Box}= \mathcal{C}^{\rm
 SW}_{A_p},\qquad  {\rm and} \qquad \mathcal{C}^{\rm Toda}_{D_{p +
    2},\mathbf{2(p + 2)}_{\bf v}} = C^{\rm SW}_{D_{p + 2}}
\eeq
after suitably identifying the action variables $u = (u_0,\ldots,u_{R})$ in
\eqref{eq:sctoda} with the classical Coulomb vacuum expectation values in 
 \eqref{eq:CA}-\eqref{eq:CD}. This compels us to
formulate the two following conjectures. \\

Let $\Gamma \subset \mathrm{SL}(2,\bbC)$ be an isometry group of
$\mathbb{S}^3$ isomorphic to a cyclic or binary
polyhedral group $\Gamma$. Let $\mathcal{D}_{\Gamma}$ its Dynkin diagram
determined by McKay correspondence (Table~\ref{ADElabe}), and $\cG_\Gamma$ the
associated simply connected, simply-laced Lie group. We specialize $\rho_{\rm min}$ to be an irreducible
$\cG_\Gamma$-module of minimal dimension, as in the following table (we will
comment on non-minimal representation at the end of this Section). We can thus
abbreviate $\mathcal{C}^{{\rm Toda}}_{\cG^{\#},\rho_{\rm min}} \triangleq \mathcal{C}^{{\rm Toda}}_{\cG^{\#}}$, and denote the family $\psi\,:\,\mathcal{C}^{{\rm Toda}}_{\cG^{\#}} \rightarrow \mathfrak{U}_{\cG}$.
\begin{table}[!h]
\begin{center}
\begin{tabular}{|c|c|}
\hline
{\rule{0pt}{2.5ex}}{\rule[-1.4ex]{0pt}{0pt}} $\cG_\Gamma$ & $\rho_{\rm min}$ \\
\hline
{\rule{0pt}{2.5ex}}{\rule[-1.4ex]{0pt}{0pt}} $A_{p-1}$ & $\square$, $\overline\square$ \\
\hline
{\rule{0pt}{2.5ex}}{\rule[-1.4ex]{0pt}{0pt}} $D_4$ & $\mathbf{8}_{\bf v}$,
$\mathbf{8}_{\bf s}$,$\mathbf{8}_{\bf c}$\\
\hline
{\rule{0pt}{2.5ex}}{\rule[-1.4ex]{0pt}{0pt}} $D_{p + 2}$ & $\mathbf{2(p +
  2)}_{\bf v}$, $p > 2$ \\
\hline
{\rule{0pt}{2.5ex}}{\rule[-1.4ex]{0pt}{0pt}} $E_6$ & $\mathbf{27}$, $\mathbf{\overline{27}}$ \\
\hline
{\rule{0pt}{2.5ex}}{\rule[-1.4ex]{0pt}{0pt}} $E_7$ & $\mathbf{56}$ \\
\hline
{\rule{0pt}{2.5ex}}{\rule[-1.4ex]{0pt}{0pt}} $E_8$ & $\mathbf{248}$\\
\hline
\end{tabular}
\caption{\label{tab:rmin}Minimal irreducible modules for the ADE Lie groups.}

\end{center}
\end{table}

The first conjecture states that, upon suitable restriction of the action
variables in $\mathfrak{U}_{\cG_\Gamma}$ and quantum cohomology parameters
  of $Y^\Gamma$, the (affine co-extended) Toda spectral curves are a subfamily of mirror curves of $Y^\Gamma$ that coincides with the
LMO spectral curves of $\mathbb{S}^\Gamma$. Here, the only place where the Seifert
invariant $\sigma$ appears is in the rescaling $\lambda = \hat\lambda/\sigma$
of the string coupling constant. Recall that $c = \exp(\chi_{{\rm orb}}\lambda/2a)$.
\begin{conjecture} \label{conj:spec}
\ben[(a)]
\item There exists a family of curves $\phi\,:\,\widetilde{\mathcal{C}}^{{\rm LMO}}_{\mathcal{D}_{\Gamma}} \rightarrow \mathfrak{T}^{{\rm LMO}}_{\mathcal{D}_{\Gamma}}$ over a $1$-dimensional base, and a finite surjective map $\kappa\,:\,\mathfrak{T}_{\mathcal{D}_{\Gamma}}^{{\rm LMO}} \rightarrow \mathbb{C}^*_{c}$, such that the germs at $c = 1$ of the LMO spectral curve and of $\kappa \circ \phi$ are canonically isomorphic.
\item The base $\mathfrak{T}^{{\rm LMO}}_{\mathcal{D}_{\Gamma}}$ is isomorphic to $\mathbb{A}^1$.
\item We have a commutative diagram:
\beq
\begin{xy}
(0,20)*+{\widetilde{\mathcal{C}}^{{\rm LMO}}_{\mathcal{D}_{\Gamma}}} ="a"; (30,20)*+{\mathcal{C}_{\cG_\Gamma^{\#}}^{\rm Toda}}="b";
(0,0)*+{\mathfrak{T}_{\mathcal{D}_{\Gamma}}^{{\rm LMO}}} ="c"; (30,0)*+{\mathfrak{U}_{\cG_{\Gamma}}}="d";
{\ar_{\vartheta} "a";"b"};
{\ar_{\phi} "a";"c"};{\ar^{\psi} "b";"d"};
{\ar_\theta "c";"d"}; \nonumber
\end{xy}
\eeq
where $\theta$ is a finite immersion and $\vartheta$ restricted to any fiber is an isomorphism.

\item There exists a choice $t \leftarrow t(\lambda)$ of quantum cohomology parameters such
  that the generating series $F_g$ (resp. $W_{g,n}$) computed by the
  topological recursion to the restricted subfamily $\cC^{{\rm Toda}}_{\cG_\Gamma^{\#}}|_{{\rm Im}\,\theta}$ 
  coincide with the genus-$g$ closed (resp. $n$-holes, open)  Gromov--Witten potential of the $3$-fold geometry $(Y^\Gamma,\mathcal{L}^\Gamma)$ described in \eqref{eq:GW1}, \eqref{eq:wghgw}. Up to symplectic transformations of $(X,Y)$ and overall multiplication by a constant, the $1$-form to use as input of the recursion is $\Omega^{{\rm Toda}}_{\cG^{\#}} = \ln Y\,\dd X/X$ restricted to ${\rm Im}\,\theta$.
\een
\end{conjecture}

We formulate a second conjecture, extending the previous one to generic action variables/generic vacua in Chern--Simons theory. Since $\pi_1(\mathbb{S}^{\Gamma}) = \Gamma$, the set of critical points of the Chern--Simons action is:
\beq
\mathfrak{V}_{\Gamma,N} \triangleq \big\{{\rm flat}\,\,{\rm U}(N)\,\,{\rm connections}\,\,{\rm on}\,\,\mathbb{S}^{\Gamma}\,\,{\rm modulo}\,\,{\rm gauge}\big\} \simeq {\rm Hom}(\Gamma,{\rm U}(N))/{\rm U}(N)\,.
\eeq
and we let $\mathfrak{V}_{\Gamma} = \lim_{N \rightarrow \infty}
\mathfrak{V}_{\Gamma,N}$ be its direct limit with respect to the
composition of morphisms given by the embedding ${\rm U}(N)
\hookrightarrow {\rm U}(N + 1)$. 
By the McKay correspondence \cite{McKay}, irreducible representations of $\Gamma$ are labeled by the nodes of the extended Dynkin diagram $\tilde{\mathcal{D}}_{\Gamma}$. The affine node labels the trivial representation, and for $i \geq 1$, these dimensions coincide with the components of the highest root of $\cG_{\Gamma}$ in the basis of simple roots. We can then describe:
\beq
\mathfrak{V}_{\Gamma} = \mathbb{N}^{R + 1},\qquad \mathfrak{V}_{\Gamma,N} = \l\{(N_0,\ldots,N_R) \in \mathbb{N}^{R + 1},\quad N_0 + \sum_{i = 1}^r D_iN_i = N\r\}.
\eeq
When $N \rightarrow \infty$, we consider a background
$[\mathcal{A}]_{\mathtt{t}}$ parametrized by $\mathtt{t}_i \triangleq
N_i\hbar$ for $i \in \llbracket 0,R \rrbracket$, and in particular the rank is
encoded in $\hat\lambda = \mathtt{t}_{0} = N\hbar$. We also define
$\mathtt{c}_i = \exp(\chi_{{\rm orb}}\mathtt{t}_i/2a)$. Let now
$\mathcal{F}_g(\mathbb{S}^{\Gamma},\mathtt{t})$ and
$\mathcal{W}_{g,n}(\mathbb{S}^{\Gamma},\mathtt{t};\vec{x})$ be the perturbative free
energies and correlators of ${\rm U}(N)$ Chern--Simons theory expanded around
the background $[\mathcal{A}]_{\mathtt{t}}$, which is defined at least
formally as a series in $\mathtt{t}$ by the ribbon graph expansion of
Section~\ref{mimi}. While it is not clear to us if this can be given a matrix
model-like expression beyond the A-series, e.g. by collecting certain terms in the exact
Chern--Simons partition functions derived in \cite{MarinoCSM,BT2}, the
spectral curve in the background $[\mathcal{A}]_{\mathtt{t}}$ can nevertheless
be defined as in \eqref{Wsu} from $\mathcal{W}_{0,1}$, and it yields a family
of curves $\phi_{0}\,:\,\mathcal{C}^{{\rm CS}}_{\mathcal{D}_{\Gamma}}
\longrightarrow (\mathbb{C}^{R + 1}_{\mathtt{t}})_{{\rm formal}}$ where the
notation for the base means that it is a priori a formal neighborhood of $0$
in $\mathbb{C}^{R + 1}$. In light of the previous remark, we are unable to
propose an independent computation for this Chern--Simons spectral curve in a general background, but we speculate:

\begin{conjecture}
\label{conj:gen}
\ben[(a)]
\item There exists a family of curves $\phi\,:\,\widetilde{\mathcal{C}}^{{\rm CS}}_{\mathcal{D}_{\Gamma}} \rightarrow \mathfrak{T}$ over an $(R + 1)$-dimensional base, and a finite surjective map $\kappa\,:\,\mathfrak{T}_{\mathcal{D}_{\Gamma}} \rightarrow \prod_{i = 0}^r \mathbb{C}^*_{\mathtt{c}_i}$, such that $\phi_0$ and the germ at $c = 1$ (i.e. $\mathtt{t} = 0$) of $\kappa\circ \phi$ are canonically isomorphic.

\item We have a commutative diagram:
\beq
\begin{xy}
(0,20)*+{\widetilde{\mathcal{C}}^{{\rm CS}}_{\mathcal{D}_{\Gamma}}} ="a"; (30,20)*+{\mathcal{C}_{\cG_\Gamma^{\#}}^{\rm Toda}}="b";
(0,0)*+{\mathfrak{T}_{\mathcal{D}_{\Gamma}}}="c"; (30,0)*+{\mathfrak{U}_{\cG_{\Gamma}}}="d";
{\ar_\vartheta "a";"b"};
{\ar_{\phi} "a";"c"};{\ar^{\psi} "b";"d"};
{\ar_\theta "c";"d"}; \nonumber
\end{xy}
\eeq
where $\theta$ is a finite map and $\vartheta$ is a fiberwise isomorphism.

\item There exists a section $\tilde{\mathtt{t}}$ (the {\rm orbifold mirror
  map}) of $\theta$ such that the topological recursion applied to $\mathcal{C}^{{\rm Toda}}_{\mathcal{G}_{\Gamma}^{\#}}$ and the $1$-form $\ln Y\,\dd X/X$ (maybe up to rescaling by a constant) computes, above the point $u$, the free energy $\mathcal{F}_g(\mathbb{S}^{\Gamma},\tilde{\mathtt{t}}(u))$ and the correlators $W_{g,n}(\mathbb{S}^{\Gamma},\tilde{\mathtt{t}}(u);\vec{x})$, with $x = X^{1/a}$. 

\item There exists an affine automorphism $\ell_\Gamma\in
  \bbC^{r_{\cG_\Gamma}} \rtimes
  \mathrm{End}(\bbC^{r_{\cG_\Gamma}}) $ such that 
\beq
\mathcal{F}^{\rm CS}_g(\mathbb{S}^\Gamma,\mathtt{t}) = F^{\rm GW}_g(Y^\Gamma_{\rm orb},\ell_\Gamma(t)),\qquad \mathcal{W}^{\rm CS}_{g,h}(\mathbb{S}^\Gamma,\mathtt{t},\vec x) = W^{\rm
  GW}_{g,n}(Y^\Gamma_{\rm orb},\mathcal{L}^{\Gamma}_{\rm orb},\ell_\Gamma(t);\vec x).
  \eeq
Furthermore, there exists a unique change of normalization of the periods of
$\mathcal{W}^{\rm CS}_{0,2}$ (call it $\widetilde{\mathcal{W}}^{\rm
  CS}_{0,2}$) such that the ensuing topological recursion on
$\widetilde{\mathcal{C}}^{{\rm CS}}_{\mathcal{D}_{\Gamma}}$ gives generating
  functions $\widetilde{\mathcal{F}}^{\rm CS}_g$,
  $\widetilde{\mathcal{W}}^{\rm CS}_{g,n}$ with
\beq
\widetilde{\mathcal{F}}^{\rm CS}_g(\mathbb{S}^\Gamma,\mathtt{t}) = F^{\rm GW}_g(Y^\Gamma_{\rm res},\ell_\Gamma(t)),\qquad \widetilde{\mathcal{W}}^{\rm CS}_{g,n}(\mathbb{S}^\Gamma,\mathtt{t},\vec x) = W^{\rm
  GW}_{g,n}(Y^\Gamma_{\rm res},\mathcal{L}^{\Gamma}_{\rm res},\ell_\Gamma(t);\vec x).
\eeq
\een
\end{conjecture}

\br (On minimal orbits and minimal irreps).
The construction of the Toda spectral curve involves
the choice of a minimal-dimensional representation $\rho_{\rm min}$ of
$\cG_\Gamma$; picking up a different representation leads to a curve, which cannot
be simply reconstructed from the minimal one, but that should however lead to the same
free energies \cite{Martinec:1995by}. On the other hand, we will see in Section~\ref{S41} that the
construction of the LMO spectral curve likewise depends on the choice of a vector $v \in
\mathbb{Z}^a$ with finite orbit under a monodromy group $\mathfrak{W}' = {\rm
  Weyl}(\mathcal{D}_\Gamma')$ for a certain $\mathcal{D}_\Gamma' \subseteq
\mathcal{D}_\Gamma$: different choices of $v$ contain equivalent information
which is just repackaged differently, though in a non-trivial way, since the degree of the
curves is related to the size of the orbit of $v$. One may wonder if there
is a set-theoretic injection of the set of finite monodromy orbits into
$\mathrm{Rep}(\cG_\Gamma)$, and whether the
higher degree curves on the LMO side should be obtained from (suitable
restrictions of) non-minimal Toda spectral curves. 
\er

\br (Central extensions of $\Gamma$). A finite isometry subgroup $\tilde \Gamma \subset {\rm SO}(4)$ of $\bbS^3$ is generically a non-trivial central
extension of one of the finite groups $\Gamma$ of Table~\ref{ADElabe} (see
Appendix~\ref{pi1list} for more details). The
reasoning of Section~\ref{sec:amodel} would lead us to consider now ALE
fibrations over the weighted projective line, as in this case the
$\tilde\Gamma$-action on the resolved conifold acts effectively on the base
$\bbP^1$. It was shown by one of the authors in \cite{Brini:2008ik} for the $A$-series that the
geometric transition argument cannot be applied verbatim in this setting. We leave this question to future investigations.
\er

\br ($\Gamma$-action and orientifolds).~
In our brane construction of Section~\ref{sec:amodel}, if we instead chose $\sigma(a_{11})=\overline{a_{22}}$,
$\sigma(a_{12})=-\overline{a_{21}}$ as our anti-holomorphic involution, we would have that $Y_\sigma=\emptyset$: this would
correspond to the orientifold of the resolved conifold considered by Sinha and
Vafa in \cite{Sinha:2000ap}, and in turn to
Chern--Simons theory on $\mathbb{S}^3$ with SO/Sp gauge group at large $N$. In contrast
with the discussion of Section~\ref{sec:abranes}, it is straightforward to check that in this case the $\Gamma$-action commutes
with the real involution for all finite $\Gamma \subset {\rm SU}(2)$. In particular,
open and closed {\it real} versions of the Gromov--Witten potentials
\eqref{eq:FgGW} and \eqref{eq:WgnGW} can be defined by unoriented
localization, as in \cite{Bouchard:2004ri,Diaconescu:2003dq}. On the other
hand, SO/Sp Chern-Simons invariants of $\mathbb{S}^{\Gamma}$ can also be
computed from a matrix model analysis and the topological recursion \cite[Section
  8]{BESeifert}: the spectral curve and two-point function is the same as for
${\rm SU}$ up to a renormalization $\lambda \rightarrow \lambda/2$, but the
initial data is enriched by an (explicit) $1$-point genus $1/2$ function. It
is possible to formulate the analogue of Conjectures
\ref{conj:spec}-\ref{conj:gen} in this context, but we will not venture in
collecting supporting evidence here.
\er

\section{Computations I: the LMO curves}
\label{S4LMO}
\subsection{LMO spectral curves}
\label{S41}

The LMO spectral curve is characterized as a solution of a maximization problem, which can be presented in several ways. In terms of the large $N$ spectral density $\varrho(\phi)$ for the $\phi_i$'s, we have the saddle point equation:
\beq
\label{lineae}\fint \varrho(\phi')\Big\{(2 - r)\ln{\rm sinh}[(\phi - \phi')/2] + \sum_{m = 1}^{r} \ln {\rm sinh}[(\phi - \phi')/2a_m]\Big\} \leq \frac{\phi^2}{2\lambda}\,,
\eeq
with equality on the support of $\varrho$, and $\varrho \geq 0$ with total
mass $\int \varrho(\phi)\dd\phi = 1$. When $\chi_{{\rm orb}} > 0$, one can
show that the solution of this problem is unique, the support is a segment
$S$, and $\varrho(\phi)$ is of the form $\mathbf{1}_{S}(\phi)\sqrt{Q(\phi)}$
with $Q$ a positive, real-analytic function vanishing at the endpoints of
$S$. Given the symmetry $\{\phi_i \rightarrow -\phi_i,\quad 1 \leq i \leq N\}$
of the model \eqref{zndef}, $S$ must be symmetric around $0$. Its
determination is part of the problem. The usual method is to solve the linear
equation \eqref{lineae} for a fixed arbitrary segment, then list the possible
segments compatible with the other constraints (total mass $1$, vanishing of the $\varrho$ at the edges). This list is usually finite, and if there is not already a unique solution, the correct one is singled out by the positivity constraint $\varrho \geq 0$. However, it is by no means easy to solve explicitly singular integral equations of the form \eqref{lineae} on a segment.\\

The linear equation \eqref{lineae} can be rewritten in several equivalent forms. In terms of the resolvent:
\beq
W(x) \triangleq \int \frac{x\,\varrho(\phi)\dd\phi}{x - e^{\phi/a}},\qquad \varrho(\phi) \triangleq \frac{W(e^{\phi/a} - {\rm i}0) - W(e^{\phi/a} + {\rm i}0)}{2{\rm i}\pi\,e^{\phi/a}}\,,
\eeq
it becomes, for all $x \in S$:
\beq
W(x + {\rm i}0) + W(x - {\rm i}0) + (2 - r)\sum_{\ell =
  1}^{a - 1} W(\zeta_{a}^{\ell}x) + \sum_{m = 1}^r \sum_{\ell_m = 1}^{a/a_m - 1}
   W(\zeta_{a/a_m}^{\ell_m}x) = (a^2/\lambda)\ln x + (a/2)\chi_{{\rm orb}}
\eeq
where $\zeta_{k}$ is a primitive $k$-th root of unity. The symmetry $\{\phi_i \rightarrow -\phi_i\,\,\,1 \leq i \leq N\}$ implies:
\beq
\label{symW}W(x) + W(1/x) = 1\,. 
\eeq
We can get rid of the right-hand side and of log-singularities by defining:
\beq
\mathcal{Y}(x) \triangleq -cx\exp\big[(\chi_{{\rm orb}}\lambda/a)W(x)\big],\qquad c \triangleq \exp(\chi_{{\rm orb}}\lambda/2a)\,.
\eeq
By construction, $\mathcal{Y}(x)$ is a holomorphic function on $\mathbb{C}\setminus S$, with behavior
\beq
\label{growth}\mathcal{Y}(x) \mathop{\sim}_{x \rightarrow 0} -cx,\qquad \mathcal{Y}(x) \mathop{\sim}_{x \rightarrow \infty} -c^{-1}x
\eeq
and satisfying:
\beq
\label{mono} \forall x \in S,\qquad \mathcal{Y}(x + {\rm i}0)\mathcal{Y}(x -
      {\rm i}0)\Big[\prod_{\ell = 1}^{a - 1}
          \mathcal{Y}(\zeta_{a}^{\ell}x)\Big]^{2 - r} 
          \cdot\prod_{m = 1}^{r} \Big[\prod_{\ell_m = 1}^{a/a_m - 1} \mathcal{Y}(\zeta_{a/a_m}^{\ell_m})\Big] = 1\,.
\eeq
The symmetry \eqref{symW} becomes:
\beq
\label{Ysym} \mathcal{Y}(x)\mathcal{Y}(1/x) = 1\,.
\eeq
Equation \eqref{growth} can be seen as a description of generators for the
monodromy group $\mathfrak{G}$ of the analytic function $\mathcal{Y}(x)$, and
\eqref{growth} are constraints imposed on the singularities of the solution
away from the branchcuts (here meromorphic singularities at $0$ and
$\infty$). \cite{BESeifert} presented a general strategy to solve a class of
monodromy equations including \eqref{mono}. It leads at least to a
partially explicit solution when certain finite subgroups of $\mathfrak{G}$
are identified, as one can then express the solution in terms of an algebraic function. Among the Seifert matrix model, the ADE cases turned out to be very special, because they are the ones that can be obtained from algebraic curves.

\begin{theorem}\cite{BESeifert}
\label{Lcal}Consider the equation \eqref{mono} with $(a_1,\ldots,a_r)$ arbitrary positive integers and $\chi_{{\rm orb}} \neq 0$. $\mathfrak{G}$ is infinite, except in the case $(2,2,2p')$ where $\mathfrak{G}$ is the symmetric group in $2p' + 1$ elements. Besides, the two following points are equivalent:
\begin{itemize}
\item[$(i)$] $\chi_{{\rm orb}} > 0$.
\item[$(ii)$] There exists a non-zero $v \in \mathbb{Z}^{a}$ such that \beq
\label{Ydefff}\mathcal{Y}_{v}(x) \triangleq \prod_{j = 0}^{a - 1} [\mathcal{Y}(\zeta_{a}^{j}x)]^{v_j}
\eeq has a finite monodromy group. So, there is a polynomial $\mathcal{P}_{v}$ in two variables, depending on $\lambda$, such that $\mathcal{P}_{v}(x,\mathcal{Y}_{v}(x)) = 0$.
\end{itemize}
\end{theorem}

While $\mathcal{Y}(x)$ has a cut on $S$ only, $\mathcal{Y}_{v}(x)$ has a branchcut on $S_j = \zeta_{a}^{-j}S$ whenever $v_j \neq 0$. Knowing $\mathcal{Y}_v(x)$ is enough to retrieve $W(x)$ since:
\beq
v_j\,(W(\zeta_{a}^jx) - 1) = \frac{a}{\chi_{{\rm orb}}\lambda} \oint_{S_j} \frac{\dd \xi}{2{\rm i}\pi}\,\frac{\ln\big(\mathcal{Y}_{v}(\xi)/(-c\xi)\big)}{x - \xi}\,.
\eeq
A simple computation from \eqref{mono} shows that there exists linear involutions $T_j \in {\rm GL}(a,\mathbb{Z})$ describing the monodromy of these new functions:
\beq
\forall v \in \mathbb{Z}^{a},\,\,\forall x \in S_j,\qquad \mathcal{Y}_{v}(x + {\rm i}0) = \mathcal{Y}_{T_j(v)}(x - {\rm i}0)\,.
\eeq
The monodromy group $\mathfrak{G}$ is isomorphic to the linear subgroup
generated by the $T_j$ for $j = 0,\ldots,(a - 1)$. Lemma~\ref{Lcal} is then an
answer to the question: does there exist a non-zero vector $v$ with finite
$\mathfrak{G}$-orbit? The answer is positive only for the ADE cases, and we
can actually be more precise: there exists a decomposition in two lattices
$\mathbb{Z}^a = E_0 \oplus E$, where $E$ is stable under $\mathfrak{G}$, and
the group generated by $T_j|_{E}$ for $j = 0,\ldots,(a - 1)$ is conjugate to
the Weyl group $\mathfrak{W}' $ of a finite root system. The latter are also
classified by Dynkin diagrams $\mathcal{D}'$ of ADE type, and it turns out
that $\mathcal{D}'$ is always a sub-diagram of the Dynkin diagram
$\mathcal{D}$ attached to the Seifert geometry (see Table~\ref{proporPv}),
with equality only in the $E_8$ case and certain lens spaces. Then, one can
show that $v$ has finite $\mathfrak{G}$-orbit iff $v \in E$. In that case,
describing the monodromy of $\mathcal{Y}_{v}(x)$ reduces to the
well-known classification of the orbits of the Weyl group 
$\mathfrak{W}'$ \cite{GPbook}, which are in correspondence with the parabolic subgroups of $\mathfrak{W}'$, themselves described as the reflection groups attached to the (possibly disconnected) sub-diagrams $\mathcal{D}''$ strictly included in $\mathcal{D}'$.\\

\subsection{Definition of $\mathcal{P}^{{\rm LMO}}_{\mathcal{D}}$}

For computational purposes, it is natural to choose $v$ in an orbit of minimal
size. If $v$ lies in a minimal orbit, all the other minimal orbits are 
  obtained -- up to rescaling -- by shifting with
  \beq
  \varepsilon\,:\,(v_j)_{j} \mapsto  (v_{j + 1\,\,({\rm mod}\,\,a)})_{j}.
  \eeq
This shift amounts to replacing $x$ with $\zeta_{a}^{-1}x$. Minimal orbits are given in Section~\ref{Dcases} for $D$ cases, in Section~\ref{E6geom} for $E_6$ and in Appendices~\ref{appE7orb} and \ref{appE8orb} for $E_7$ and $E_8$.  They happen to be stable under some power of the shift $\varepsilon^{a/a'}$, i.e. $\mathcal{P}_{v}(x,y)$ is a polynomial in $x^{a'}$, for some $a'$ dividing $a$. 

\begin{itemize}
\item[$\bullet$] In the Seifert geometry $\mathcal{D} = D_{2p' + 2}$, $v$ and $\varepsilon[v]$ generated two disjoint minimal orbits, and $P_{v}(x,y)$ is a actually a polynomial in $x^{p'}$. Then:
\beq
\mathcal{P}^{{\rm LMO}}_{D_{2p' + 2}}(X,Y) \triangleq \mathcal{P}_{v}(x,Y)\mathcal{P}_{v}(-x,Y),\qquad X = x^{2p'}.
\eeq
\item[$\bullet$] For $\mathcal{D} = D_{2p' + 3}$, there is a unique minimal orbit (all the shifts are in the same orbit), so $P_{v}(x,Y)$ is a polynomial in $X = x^{4p'}$ that we denote $\mathcal{P}^{{\rm LMO}}_{D_{2p' + 3}}(X,Y)$.
\item[$\bullet$] For $\mathcal{D} = E_6$, the triality of $\mathcal{D}' = D_4$
  is responsible for the existence of $3$ minimal orbits, generated by $v$,
  $\varepsilon[v]$ and $\varepsilon^2[v]$, and $\mathcal{P}_{v}(x,y)$ is a polynomial in $x^2$. Then, we introduce the polynomial: 
\beq
\mathcal{P}^{{\rm LMO}}_{E_6}(X,Y)\triangleq
\mathcal{P}_{v}(x,Y)\mathcal{P}_{v}(\zeta_{3}x,Y)\mathcal{P}_{v}(\zeta_{3}^{-1}x,Y),\qquad X = x^6.
\label{eq:PmmE6}
\eeq
\item[$\bullet$] Similarly, for $\mathcal{D} = E_7$, the duality of
  $\mathcal{D}' = E_6$ results in the existence of $2$ minimal orbits
  generated by $v$ and $\varepsilon[v]$, and $P_{v}(x,Y)$ is a polynomial in $x^6$. Then, we introduce the polynomial:
\beq
\mathcal{P}_{E_7}^{{\rm LMO}}(X,Y)\triangleq \mathcal{P}_{v}(-x,Y)\mathcal{P}_{v}(x,Y),\qquad X = x^{12}.
\label{eq:PmmE7}
\eeq
\item[$\bullet$] For $\mathcal{D} = E_8$, there is a unique minimal orbit, so $\mathcal{P}_{v}(x,Y)$ is a polynomial in $X = x^a = x^{30}$, that we denote $\mathcal{P}^{{\rm LMO}}_{E_8}(X,Y)$.
\end{itemize}

In all cases, we have set $X = x^{a}$, and our definition for the LMO spectral curve is:
\beq
\mathcal{C}^{{\rm LMO}}_{\mathcal{D}} = \big\{(X,Y) \in \mathbb{C}^*\times \mathbb{C}^*,\qquad \mathcal{P}^{{\rm LMO}}_{\mathcal{D}}(X,Y) = 0\big\}\,.
\eeq
Equivalently, the ideal $\mathcal{P}^{{\rm LMO}}_{\mathcal{D}}(X,Y) = 0$ is obtained by elimination of $x$ in the equations $\{\mathcal{P}_{v}(x,Y) = 0,\,\,X = x^a\}$. Considering $\mathcal{P}^{{\rm LMO}}$ rather than $\mathcal{P}_{v}$ is necessary for comparison with the Toda spectral curves, but of course it does not contain more information than $\mathcal{P}_{v}$.\\

The symmetry \eqref{Ysym} implies the palindromic symmetry:
\beq
\mathcal{P}^{{\rm LMO}}_{\mathcal{D}}(X,Y) = C\,X^{\bullet}Y^{\bullet}\,\mathcal{P}^{{\rm LMO}}_{\mathcal{D}}(1/X,1/Y)\,,
\label{eq:palin}
\eeq
where $\bullet$ are the degrees of $\mathcal{P}^{{\rm LMO}}_{\mathcal{D}}$ in the variables $X$ and $Y$, given in Table~\ref{proporPv}. $\mathcal{C}^{{\rm LMO}}_{\mathcal{D}}$ is a family of spectral curve with parameter $\lambda$, equipped with the $1$-form:
\beq
\tilde{\Omega} = \frac{a}{\chi_{{\rm orb}}\lambda}\ln(-Y/cX)\,\frac{\dd X}{X} = \frac{a}{\chi_{{\rm orb}}\lambda}\,\ln Y\,\frac{\dd X}{X} + \dd f(X)\,.
\eeq
Adding the differential of a rational function of $X$ does not change the free energies and correlators computed by the topological recursion, so we can equally choose the $1$-form:
$$
\Omega = \frac{a}{\chi_{{\rm orb}}\lambda}\,\ln Y\,\frac{\dd X}{X}
$$
Besides, the only effect of the rescaling by $a/\chi_{{\rm orb}}\lambda$ is that $W_{g,n}$ are multiplied by $(\chi_{{\rm orb}}\lambda/a)^{2g - 2 + n}$.

\begin{figure}[h!]
\begin{center}
\begin{tabular}{|c|c|c||c|c||c|c|}
\hline
{\rule{0pt}{3.2ex}}{\rule[-1.8ex]{0pt}{0pt}} fiber orders & $a$ & $\chi_{{\rm orb}}$ & $\mathcal{D}$ & $\mathcal{D'}$ & ${\rm deg}_{X}\,\mathcal{P}^{{\rm LMO}}_{\mathcal{D}}$ & ${\rm deg}_{Y}\,\mathcal{P}^{{\rm LMO}}_{\mathcal{D}}$ \\
\hline\
$(p)$ {\rule{0pt}{3.2ex}}{\rule[-1.8ex]{0pt}{0pt}} & $p$ & $1 + 1/p$ & $A_{p}$ & $A_p$ & $1$ & $p$ \\
\hline\ $(2,2,2p')$ {\rule{0pt}{3.2ex}}{\rule[-1.8ex]{0pt}{0pt}} & $2p'$ & $1/2p'$ & $D_{2p' + 2}$ & $A_{2p'}$ & $2$ & $2\cdot(2p' + 1)$ \\
\hline\
$(2,2,2p' + 1)$ {\rule{0pt}{3.2ex}}{\rule[-1.8ex]{0pt}{0pt}} & $2(2p' + 1)$ & $1/(2p' + 1)$ & $D_{2p' + 3}$ & $D_{2p' + 2}$ & $2$ & $4(p' + 1)$ \\
\hline\ $(2,3,3)$ {\rule{0pt}{3.2ex}}{\rule[-1.8ex]{0pt}{0pt}} & $6$ & $1/6$ & $E_6$ & $D_{4}$ & $4$ & $3\cdot 8$ \\
\hline\ $(2,3,4)$ {\rule{0pt}{3.2ex}}{\rule[-1.8ex]{0pt}{0pt}}  & $12$ & $1/12$ & $E_{7}$ & $E_6$ & $6$ & $2\cdot 27$  \\
\hline\ $(2,3,5)$   {\rule{0pt}{3.2ex}}{\rule[-1.8ex]{0pt}{0pt}}  & $30$ & $1/30$ & $E_8$ & $E_8$ & $18$ & $240$ \\
\hline
\end{tabular} 
\caption{\label{proporPv} $\mathcal{D}$ = Seifert geometry; $\mathcal{D}'$ = monodromy group of the spectral curve. $k\cdot d$ in the last column means that the reduced polynomial $\mathcal{P}_{v}$ has degree $d$ in $y$, and $\mathcal{P}^{{\rm LMO}}_{\mathcal{D}}$ contains $k$ factors of $\mathcal{P}_{v}$ differing by some rotations of $x$.}
\end{center}
\end{figure}

\subsection{The computation in practice}
\label{S42}
Since the details of the orbit analysis were presented in \cite{BESeifert}, we
focus here on the next step, i.e. the identification of the polynomial
equation for the spectral curve. Denote by $(w[i])_{i \in I}$ the list of
vectors in a chosen minimal orbit generated by $v$, and write 
\beq
\mathcal{P}_{v}(x,\mathcal{Y}_{v}(x)) = C\,\prod_{i \in I} (y - \mathcal{Y}_{w[i]}(x)) = 0
\eeq
for the equation of the spectral curve; the constant $C$ will be fixed later
on. Let 
$\overline{\mathcal{C}}_{v}$ be the compact Riemann surface which is a smooth model for $\overline{\big\{(x,y) \in \mathbb{C}^* \times \mathbb{C}^*\,\,:\,\, \mathcal{P}_{v}(x,y) = 0\big\}}$. It comes with a branched covering $x \,:\,\overline{\mathcal{C}}_{v} \rightarrow \mathbb{P}^1$, and $y$ defines a meromorphic function on $\overline{\mathcal{C}}_{v}$ whose value in the $i$-th sheet is $y^{(i)}(x) = \mathcal{Y}_{w[i]}(x)$.

\subsubsection*{Step A}

From \eqref{growth}, the functions $y^{(i)}(x)$
\bea
\label{slo1}x \rightarrow 0\,, & \qquad & y^{(i)}(x) \sim (-cx)^{n_0(w[i])}\,\zeta_{a}^{n_{1}(w[i])}\,,\\ 
\label{slo2} x \rightarrow \infty\,, & \qquad & y^{(i)}(x) \sim (-x/c)^{n_0(w[i])}\,\zeta_{a}^{n_{1}(w[i])}\,,
\eea
with:
\beq
n_0(w) = \sum_{j = 0}^{a - 1} w_{j},\qquad n_1(v) = \sum_{j = 0}^{a - 1} j w_{j}\,.
\eeq
This fixes the coefficients on the boundary of the Newton polygon of $\mathcal{P}_{v}$ up to the overall multiplicative constant $C$. \eqref{slo1}-\eqref{slo2} tell us that the slopes are $(\pm 1,n_{0}(w[i]))$, therefore there exists a single monomial of degree $0$ in $y$. We can fix $C$ by setting this coefficient to $1$. As we explained in the last paragraph, the symmetries observed by explicitly computing the orbits imply that $\mathcal{P}_{v}(x,y)$ is actually a polynomial in $x^{a'}$ with $a'$ given in Figure~\ref{Secondtab}.

In the $D$ geometries, at this stage there is a shortcut to the final solution, reviewed in Section~\ref{Dcases}. In the exceptional cases, we continue and proceed by necessary conditions.

\subsubsection*{Step B}

In the solution we look for, $y^{(i)}(x)$ derives from a single function $W(x)$ such that $y^{(i)}(x) = \mathcal{Y}_{w[i]}(x)$ given by formula \eqref{Ydefff}. If we write the Taylor expansion:
\beq
\label{mdef} x \rightarrow 0,\qquad \frac{\chi_{{\rm orb}} \lambda}{a}\,W(x) = \sum_{k \geq 1} \mu_k\,x^{k + 1}\,,
\eeq
we deduce that:
\beq
\label{expandYi}x \rightarrow 0,\qquad y^{(i)}(x) = (-cx)^{n_0(w[i])}\zeta_{a}^{n_1(w[i])}\,\exp\Big(\sum_{k \geq 1} \mu_{k - 1}\,\hat{w}[i]_{k\,\,{\rm mod}\,\,a}\,x^{k}\Big)\,,
\eeq
where we introduced the discrete Fourier transform:
\beq
\hat{w}_{k} \triangleq \sum_{j = 0}^{a - 1} \zeta_{a}^{jk}\,w_{j}\,.
\eeq
It turns out that many Fourier modes $k \in \mathbb{Z}_{a}$ are zero for all vectors in the orbit of $v$. The set of non-zero Fourier modes $K_{\mathcal{D}} = \bigcup_{i \in I} \{k \in \llbracket 0,a - 1 \rrbracket,\quad w[i]_{k} \neq 0\big\}$ is:
\begin{itemize}
\item[$\bullet$] $K_{E_{6}} = \{1,2,3,5\}$.
\item[$\bullet$] $K_{E_7} = \{3,4,6,8,9,11\}$.
\item[$\bullet$] $K_{E_8} = \{1,2,3,4,5,7,8,9,11,13,14,16,17,19,21,22,23,25,26,27,28,29\}$.
\end{itemize}
Only the $\mu_{k-1}$'s with $k \in K$ will appear in the Puiseux expansion of $y$. Then, the sought-for polynomial takes the form:
\beq
\label{expP} \mathcal{P}_{v}(x,y) = B(x)\,\prod_{i \in I} (y - y^{(i)}(x))
\eeq
where the monomial prefactor $B(x)$ is fixed by matching with the coefficient $1$ of the monomial $x^{\bullet}y^{0}$ in $\mathcal{P}_{v}$. By expanding the right-hand side of \eqref{expP} when $x \rightarrow 0$ using \eqref{expandYi}, we can express the coefficients of $\mathcal{P}_{v}$ in terms of a relatively small number of $\mu_k$ with $k \in K_{\mathcal{D}}$. Since we already know the Newton polygon and the symmetries of $\mathcal{P}_{v}$, we can impose those relations at the level of their expression in terms of $\mu_k$'s, which gives relations between the $\mu_k$'s and we can eliminate some of them. Doing so, we can express all coefficients of $\mathcal{P}_{v}$ only in terms of $c$ and:
\begin{itemize}
\item[$\bullet$] for $E_6$, $\mu_1$ and $\mu_2$.
\item[$\bullet$] for $E_7$, $\mu_2$, $\mu_3$, $\mu_5$ and $\mu_7$.
\end{itemize}
For $E_8$, we could not complete this computation: it requires expanding a product of $240$ factors to order $o(x^{540})$, and even if this would be achieved, it is still a formidable task to eliminate $\mu_k$'s.

\subsubsection*{Step C}

The ramification properties of the spectral curve we seek are easily described a priori. Call $d = \deg_{y} \mathcal{P}_{v}$ the size of the orbits, i.e. the degree of $x\,:\,\overline{\mathcal{C}}_{v} \rightarrow \mathbb{P}^{1}$, and $d' = \deg_{x} \mathcal{P}_{v}$ be the degree of $y\,\,:\,\,\overline{\mathcal{C}}_{v} \rightarrow \mathbb{P}^1$. By construction, the number of branchcuts of the function $y^{(i)}(x)$ in the $i$-th sheet is the number of non-zero components in $w[i]$, let us call it $b[i]$. The ramification points of $x\,:\,\overline{\mathcal{C}}_{v} \rightarrow \mathbb{P}^1$ are simple and correspond to the endpoint of the branchcuts, and since each branchcut is shared by two sheets, the total number of ramification points is $\sum_{i \in I} b[i]$. Riemann-Hurwitz formula then gives the genus of $\Sigma$:
\beq
{\rm genus}(\overline{\mathcal{C}}_{v}) = 1 - d + \frac{1}{2}\Big(\sum_{i \in I} b[i]\Big)\,.
\eeq
For $E_6$, we find that $\overline{\mathcal{C}}_{v}$ has genus $5$; for $E_7$, it has genus $46$, and for $E_8$, it has genus $1471$. We remark that the genus is much lower than the genus of a generic curve with same Newton polygon -- which is the number of interior points in the Newton polygon -- even if we take into account the symmetries. This means that the plane curve $\mathcal{P}_{v}(x,y) = 0$ must be singular. This puts a number of algebraic constraints on the coefficients inside the Newton polygon of $\mathcal{P}_{v}$. Taking into account symmetries, we can put an upper bound on the number of independent such constraints. From our experience with the $D$ and $E_6$ case, we expect that implementing these constraints gives only finitely many solutions for the sequence of $\mu_k$. The definition in \eqref{mdef} implies that $\mu_k = \cO(\lambda)$ for all $k \geq 1$ when $\lambda \rightarrow 0$ (i.e. $c \rightarrow 1$), and the $\mu_k$'s must have a power series expansion in $\lambda$ with rational coefficients. We expect that this extra piece of information singles out a unique solution (among the finitely many) to the algebraic constraints.\\

For $E_6$ this program is completed in Section~\ref{E6geom}. For $E_7$,
performing elimination in these algebraic constraints already seem
computationally hopeless, and we could not even solve them perturbatively in
$\lambda \rightarrow 0$. Therefore, our best result is the expression of
$\mathcal{P}_{v}$ in terms of the $\mu_2$, $\mu_3$, $\mu_5$ and $\mu_7$, which
should be considered as unknown (algebraic) functions of $c$. This expression
is given in Appendix~\ref{E7sp}. For $E_8$, as we have seen, our best result
is the Newton polygon and its boundary coefficients, given in
Appendix~\ref{E8bound}. Nonetheless, we will be able to compute the exact spectral curves at $c = 1$ in all cases (see Section~\ref{sec:LMOslice}).

\begin{table}[h!]
\begin{center}
\begin{tabular}{|c|c|c|c|c|c|}
\hline
{\rule{0pt}{3.2ex}}{\rule[-1.8ex]{0pt}{0pt}} $\mathcal{D}$ (geometry) & $D_{2p' + 2}$ & $D_{2p' + 3}$ & $E_6$ & $E_7$  & $E_8$ \\
\hline
$d = \deg_{y} \mathcal{P}_{v}$ {\rule{0pt}{3.2ex}}{\rule[-1.8ex]{0pt}{0pt}}  & $2p' + 1$ & $4(p' + 1)$& $8$ & $27$ & $240$ \\
\hline
$d' = \deg_{x} \mathcal{P}_{v}$ {\rule{0pt}{3.2ex}}{\rule[-1.8ex]{0pt}{0pt}}  & $4p'$ & $4(2p' + 1)$ & $8$ & $36$ & $540$ \\
\hline 
$a'$ {\rule{0pt}{3.2ex}}{\rule[-1.8ex]{0pt}{0pt}} &  $p'$ & $2p' + 1$ & $2$ & $6$ & $30$ \\
\hline
\end{tabular}
\end{center}
\caption{\label{Secondtab} Properties of $\mathcal{P}_{v}(x,y)$.}
\end{table}

Let us turn to the complete and explicit results that can be obtained for $D$ and $E_6$.

\subsection{$D_{p + 2}$ geometries}
\label{Dcases}
In that case, it is possible to guess a rational parametrization that has all
the required properties, and thus gives the correct solution bypassing Steps~$\mathbf{A}$-$\mathbf{B}$.

\subsubsection*{$p$ even}

$v = e_0 \triangleq w[0]$ is a minimal vector, and the other vectors in the orbit are $w[\ell] = (-1)^{\ell}e_{\ell}$ for $\ell \in \llbracket 1, p - 1\rrbracket$, and $w[p] = \sum_{l = 0}^{p - 1} (-1)^{l + 1}e_{l}$. We thus get:
\beq
\label{limiyd}\begin{array}{|c|c|c|}
\hline \sim {\rule{0pt}{3.2ex}}{\rule[-1.8ex]{0pt}{0pt}} & y^{(\ell)} & y^{(p)} \\
\hline
x \rightarrow 0 {\rule{0pt}{3.2ex}}{\rule[-1.8ex]{0pt}{0pt}} & \zeta_{p}^{\ell} (cx)^{(-1)^{\ell}} &  -1 \\
\hline
x \rightarrow \infty {\rule{0pt}{3.2ex}}{\rule[-1.8ex]{0pt}{0pt}}  &  \zeta_{p}^{\ell} (x/c)^{(-1)^{\ell}} & -1 \\
\hline
\end{array}
\eeq
The symmetries of the problem suggest to look for a parametrization of $\mathcal{P}_{v}(x,Y) = 0$ of the form:
\beq
\label{spcurveDeven} \left\{\begin{array}{rcl} x(z) & = & z\Big(\dfrac{z^{p} - \kappa^{2}}{\kappa^{2}z^{p} - 1}\Big) \\ Y(z) & = & -\dfrac{(z^{p/2} - \kappa)(\kappa z^{p/2} + 1)}{(\kappa z^{p/2} - 1)(z^{p/2} + \kappa)}\end{array}\right.
\eeq
where we impose that $z \rightarrow 0$ correspond to $x \rightarrow 0$ in the sheet of $w[p]$, and $z \rightarrow \zeta_{a}^{\ell}\kappa^{2/p}$ in the sheet of $w[\ell]$ for $\ell \in \llbracket 0,p - 1 \rrbracket$. Requiring \eqref{limiyd}, we must have:
\beq
\label{idenD}\frac{2\kappa^{1 + 1/p}}{1 + \kappa^2} = 1/c^2 = e^{-\lambda/4p^2}
\eeq
It can be checked that this satisfies all the desired properties of the spectral curve (including the positivity constraints), and by uniqueness, this is the solution we looked for. We can also write this parametrization with $\zeta = z^{p/2}$ and $X = x^{a} = x^{p}$:
\beq
\label{spcurveDeven2} \left\{\begin{array}{rcl}  X(\zeta)  & = & \zeta^2\,\Big(\dfrac{\zeta^2 - \kappa^{2}}{\kappa^{2}\zeta^{2} - 1}\Big)^{p} \\ Y(\zeta) & = & -\dfrac{(\zeta - \kappa)(\kappa \zeta + 1)}{(\kappa \zeta - 1)(\zeta + \kappa)}\end{array}\right.
\eeq
which is now a parametrization of $\mathcal{C}^{{\rm LMO}}_{D_{p + 2}}$, for $p$ even.

\subsubsection*{$p$ odd}

The minimal vector is $v \triangleq w[0] = e_0 + e_{p}$, and it generates the orbit consisting in $w[\ell] = e_{\ell} + e_{p + \ell}$ and $w[p + \ell] = -w[\ell]$ for $1 \leq \ell \leq p -1$, and $w[2p] = \sum_{\ell = 0}^{2p - 1} (-1)^{\ell}e_{\ell}$ and $w[2p + 1] = -w[2p]$.

\beq
\label{limiydodd}\begin{array}{|c|c|c|c|c|}
\hline \sim {\rule{0pt}{3.2ex}}{\rule[-1.8ex]{0pt}{0pt}} & y^{(\ell)} & y^{(p + \ell)} & y^{(2p)} & y^{(2p + 1)} \\
\hline
x \rightarrow 0 {\rule{0pt}{3.2ex}}{\rule[-1.8ex]{0pt}{0pt}} & -\zeta_{2p}^{\ell} (cx)^{2} &  -\zeta_{2p}^{-\ell} (cx)^{-2} & -1 & -1 \\
\hline
x \rightarrow \infty {\rule{0pt}{3.2ex}}{\rule[-1.8ex]{0pt}{0pt}}  &  -\zeta_{2p}^{\ell} (x/c)^{2} & -\zeta_{2p}^{-\ell} (x/c)^2 & -1 & -1 \\
\hline
\end{array}
\eeq
A similar guess leads to identification of the solution of $\mathcal{P}_{v}(x,Y) = 0$ in parametric form:
\beq
\label{spcurvDodd}\left\{\begin{array}{rcl} x^2(z) & = & z^{-2}\,\dfrac{z^{2p}\kappa^{2} - 1}{z^{2p} - \kappa^{2}} \\ Y(z) & = & - \dfrac{(z^{p} - \kappa)(\kappa z^{p} + 1)}{(z^{p}\kappa - 1)(z^{p} + \kappa)}\end{array}\right.
\eeq
Matching with the required behavior for $x \rightarrow 0$ or $\infty$ imposes:
\beq
\label{kappadef}\frac{2\kappa^{1 + 1/p}}{1 + \kappa^2} = 1/c = e^{-\lambda/4p^2}\,,
\eeq
which defines $\kappa$ as a function of $\lambda$ identically to \eqref{idenD}. The branch has to be chosen so that: 
\beq
\lambda \in [0,+\infty) \,\,\longleftrightarrow \,\,\kappa \in [1,0).
\eeq

We can also write in terms of $\zeta = z^{p}$ and $X = x^{a} = x^{2p}$:
\beq
\label{spcurvDodd2}\left\{\begin{array}{rcl} X(\zeta) & = & \zeta^{-2}\,\Big(\dfrac{\zeta^{2}\kappa^{2} - 1}{\zeta^{2} - \kappa^{2}}\Big)^{p} \\ Y(\zeta) & = & - \dfrac{(\zeta - \kappa)(\kappa \zeta + 1)}{(\zeta\kappa - 1)(\zeta + \kappa)}\end{array}\right.
\eeq
which is now a parametrization of $\mathcal{C}^{{\rm LMO}}_{D_{p + 2}}$, for $p$ odd.

\subsubsection*{Polynomial equation}

If we eliminate the variable $\zeta$ and keep only $X = x^{a}$ (which is equal to $x^{p}$ is $p$ is even and $x^{2p}$ if $p$ is odd) we obtain the polynomial equation $\mathcal{P}^{{\rm LMO}}_{D_{p + 2}}(X,Y) = 0$ for the spectral curve. We can factor a monomial in $\mathcal{P}^{{\rm LMO}}_{D_{p + 2}}(X,Y)$ to put it in the form:
\beq
\label{Lform}
(-1)^{p + 1}\,e^{-\lambda/2p}(X^2 + 1)(Y^2 + 1) + XY\,(\kappa^2 + 1)^{-(2p + 2)}\mathcal{Q}_{p}\big[(Y + 1/Y)(\kappa^2 + 1)^2\big] = 0,
\eeq
where:
\beq
\frac{2\kappa^{1 + 1/p}}{\kappa^2 + 1} = e^{-\lambda/4p^2}.
\eeq
and $\mathcal{Q}_{p}(\eta) = \eta^{p + 1} + \cdots$ is a polynomial in $\eta$ and $\kappa^2$, which does not have a uniform expression for all $p$'s. It is given for $p \leq 5$ in Appendix~\ref{Lpoly}. These results were obtained in \cite{BESeifert}, where A.~Wei\ss{}e also checked that the solutions \eqref{spcurveDeven} and \eqref{spcurvDodd} match the results of Monte-Carlo simulations of the matrix integral \eqref{zndef} \cite[Appendix]{BESeifert}.

\subsection{$E_6$ geometry}
\label{E6geom}

After Step~\textbf{A}, we arrive at a curve $\mathcal{P}_{v}(x,y) = \sum_{j = 0}^4 \sum_{k = 0}^8 \Pi_{j,k}\,x^{2j}\,y^{k}$ depending on the yet unknown parameters $\mu_1$ and $\mu_2$, with $c = \exp(\lambda/72)$.
$$
\begin{array}{|c|c|c|c|c|c|c|}
\hline
\Pi_{j,k} {\rule{0pt}{3.2ex}}{\rule[-1.8ex]{0pt}{0pt}} & j = 0 & 1 & 2 & 3 & 4 \\
\hline
k = 8 {\rule{0pt}{3.2ex}}{\rule[-1.8ex]{0pt}{0pt}} &             &                                          &    *                                                                                         &    &    \\
\hline
7       {\rule{0pt}{3.2ex}}{\rule[-1.8ex]{0pt}{0pt}}&             &  *                                      &   *                                                                                          &   * &  \\
\hline
6       {\rule{0pt}{3.2ex}}{\rule[-1.8ex]{0pt}{0pt}} &            &          *                               &   *                                                                                           &  * &  \\
\hline
5       {\rule{0pt}{3.2ex}}{\rule[-1.8ex]{0pt}{0pt}} &    *       &   *                                    &    *                                                                                           &  * &  * \\
\hline
4       {\rule{0pt}{3.2ex}}{\rule[-1.8ex]{0pt}{0pt}} & -c^{-4} &  -2c^{-2} - 4c^{-4}\mu_1  &   4(-1 + c^{-4} + 4c^{-2}\mu_1 + 8c^{-4}\mu_1^2 + 3c^{-3}\mu_2) &   * & *  \\
\hline
3      {\rule{0pt}{3.2ex}}{\rule[-1.8ex]{0pt}{0pt}}  & -c^{-4} &  2c^{-4}\mu_1                  &   -2 + c^{-4} - 6c^{-2}\mu_1 - 4c^{-4}\mu_1^2                             &   * &  * \\
\hline
2      {\rule{0pt}{3.2ex}}{\rule[-1.8ex]{0pt}{0pt}}  &              &        0                              &   2 + 12c^{-2}\mu_1 - 6c^{-3}\mu_2                                             &   * &  \\
\hline
1      {\rule{0pt}{3.2ex}}{\rule[-1.8ex]{0pt}{0pt}} &                &   c^{-2}                          &  1 + 2c^{-2}\mu_1                                                                    &    * &  \\
\hline
0      {\rule{0pt}{3.2ex}}{\rule[-1.8ex]{0pt}{0pt}} &                &                                      &  1                                                                                            &     &  \\
\hline
\end{array}
$$
and the * are the coefficients obtained by central symmetry, i.e $\Pi_{j,k} = \Pi_{j,8 - k} = \Pi_{4 - j,k}$. These expressions have been found in \cite{BESeifert}, and now we present a new computation.

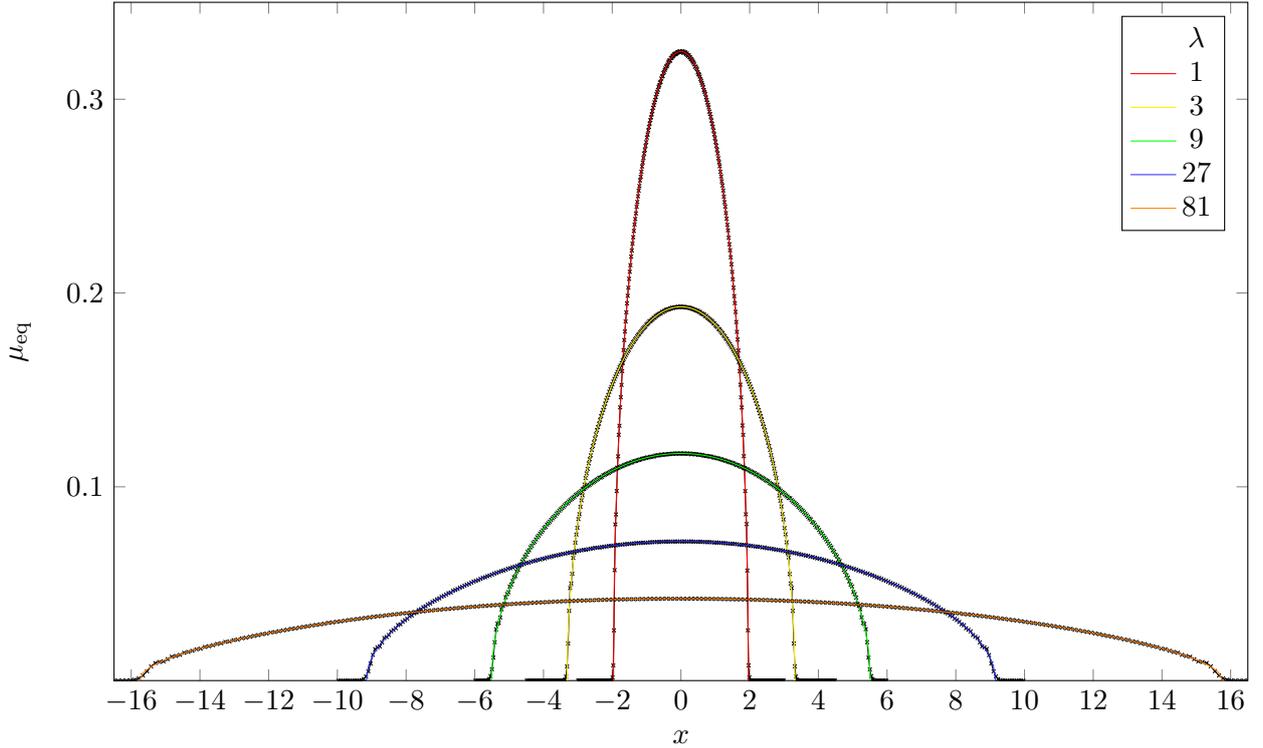
\begin{figure}[h!]
  \pgfplotsset{width=\linewidth,height=0.45\textheight}
  \centering 
  \begin{tikzpicture}
    \begin{axis}[clip marker paths=true,xlabel=$x$,ylabel=$\mu_{\text{eq}}$,
      xmin=-16.5,xmax=16.5,ymin=0,ymax=0.35,ytick={0.1,0.2,0.3}]
      \addplot[color=white] coordinates { (0,0) };
      \addlegendentry{$\lambda$}
      \addplot[color=black,mark=x,mark size=1pt,forget plot]
      file {histo_n200_02_03_03_u01.000_sc.dat};
      \addplot[color=red]
      file {newexpexact_02_03_03_u01.000.dat};
      \addlegendentry{$1$};
      \addplot[color=black,mark=x,mark size=1pt,forget plot]
      file {histo_n200_02_03_03_u03.000_sc.dat};
      \addplot[color=yellow]
      file {newexpexact_02_03_03_u03.000.dat};
      \addlegendentry{$3$};
      \addplot[color=black,mark=x,mark size=1pt,forget plot]
      file {histo_n200_02_03_03_u09.000_sc.dat};
      \addplot[color=green]
      file {newexpexact_02_03_03_u09.000.dat};
      \addlegendentry{$9$};
      \addplot[color=black,mark=x,mark size=1pt,forget plot]
      file {histo_n200_02_03_03_u27.000_sc.dat};
      \addplot[color=blue!80!white]
      file {newexpexact_02_03_03_u27.000.dat};
      \addlegendentry{$27$};
      \addplot[color=black,mark=x,mark size=1pt,forget plot]
      file {histo_n200_02_03_03_u81.000_sc.dat};
      \addplot[color=orange]
      file {newexpexact_02_03_03_u81.000.dat};
      \addlegendentry{$81$};
    \end{axis}
  \end{tikzpicture}
  \caption{$(a_1,a_2,a_3)=(2,3,3)$. 
    The dots display the Monte-Carlo simulation of the eigenvalue distribution for $N=200$ in the model \eqref{zndef}, with various choices of $\lambda$ given in the legend. The plain curves display the theoretical computation ensuing from the expression of $\mathcal{P}_{v}(x,y) = 0$ together with \eqref{these}.\label{fig:seifert_233}}
\end{figure}

Given the symmetry, let us define $\xi = x^2 + 1/x^2$ and $\eta = y + 1/y$, and eliminate $x$ and $y$ from the equation $\mathcal{P}_{v}(x,y) = 0$. We obtain an equation $\mathcal{Q}(\xi,\eta) = 0$ defining a curve of genus $2$. A birational transformation $(\xi,\eta) \mapsto (s,t)$ brings in in Weierstra\ss{} form $s^2 = \mathcal{R}(t)$ with a polynomial of degree $5$:
\bea
\mathcal{R}(t) & = & 12(1 - c^4 - 4c^2\mu_1 - 4\mu_1^2 + 8c\mu_2)t^5 + 3(-4 + c^4 + 4c^2\mu_1 + 4\mu_1^2 - 40c\mu_2)t^4 \nonumber \\
\label{Rt} & & + 24(c^3 + 2c\mu_1 + \mu_2)t^3 - 2c^2(11c^2 + \mu_2)t^2 + 8c^4t - c^4\,.
\eea
Since we are looking for a singular curve, $\mu_k$'s should be such that the
discriminant of $\mathcal{R}$ vanishes. This discriminant is a product of two
factors $\Delta_1$ and $\Delta_2$ given in Appendix~\ref{discri}, so that
gives us two equations for the two unknowns $\mu_1$ and $\mu_2$, that can be
solved explicitly. Among the finitely many solutions for $(\mu_1,\mu_2)$,
there is a unique branch in which $\mu_1 \rightarrow 0$ when $c \rightarrow
1$: that must be our solution. Then, $\mu_2$ is an explicit rational function of $\mu_1$ that we do not reproduce here, and $\mu_1$ itself is the branch of the solution of the degree $8$ equation:
\beq
\label{unifo}
\bary{rcl} 256\mu_1^8 + 4864 c^2\mu_1^7 + (-1024 + 
    35776 c^4)\mu_1^6 + (62112 c^2 + 125568 c^6)\mu_1^5 & & \\
    + (1536 - 81600 c^4 + 
    206064 c^8)\mu_1^4 + (4544 c^2 - 
    162576 c^6 + 128304 c^{10})\mu_1^3 & &  \\
    + (-1024 + 
    55116 c^4 - 78192 c^8 + 26244 c^{12})\mu_1^2 + (-5984 c^2 + 10332 c^6 - 4374 c^{10})\mu_1 &&  \\
    + 256 - 499 c^4 + 243 c^8  & = & 0\,,
    \eary \nonumber
    \eeq
which behaves like $\mu_1 = -2(c - 1) + \cO(c - 1)^2$ when $c \rightarrow 1$. As a matter of fact, the solution of \eqref{unifo} has a rational uniformization. We choose the uniformizing parameter $\kappa$ such that $\kappa \rightarrow 0$ corresponds to $c \rightarrow 1$, and our final result reads:
\bea
\label{these} \exp(\lambda/36) = c^2 & = & -\frac{(\kappa - 6)^4(\kappa - 2)^4}{16(\kappa - 3)(\kappa^2 - 6\kappa + 12)(\kappa^2 - 6\kappa + 6)^2}\,,  \\
\mu_1 & = & \frac{\kappa(\kappa - 4)(\kappa^2 - 6\kappa + 12)(\kappa^2 - 12)}{32(\kappa - 3)(\kappa^2 - 6\kappa + 6)}\,, \nonumber \\
c\cdot \mu_2 & = & -\frac{\kappa(\kappa - 2)^2(\kappa - 3)(\kappa - 4)(\kappa - 6)^2(\kappa^4 - 11\kappa^3 + 49\kappa^2  -108\kappa + 108)}{(\kappa^2 - 6\kappa + 6)^4(\kappa^2 - 6\kappa + 12)^{2}}\,.  \nonumber 
\eea
With these values, the spectral curve match perfectly the one computed numerically from Monte-Carlo simulations of the matrix model \eqref{zndef} by A.~Weisse (Figure~\ref{fig:seifert_233}). 

Remarkably, the smooth model of the curve $\mathcal{P}_{v}(x,y) = 0$ seen in variables $(x^2,y)$ has genus $1$, i.e. it defines an elliptic fibration over the base parameter $\kappa \in \mathbb{P}^1$, with discriminant:
$$
\Delta(\kappa) = 2^{80}\cdot 3^{4} \cdot \kappa(\kappa - 2)(\kappa - 3)^{16}
(\kappa - 4)(\kappa - 6)(\kappa^2 - 8\kappa + 18)(\kappa^2 - 4\kappa + 6)(\kappa^2 - 6\kappa + 6)^{14}(\kappa^2 - 6\kappa + 12)^{38}.
$$
Singular fibers occur at the critical values $c^2 \in \{0,\pm 1,7\epsilon_1/16 + {\rm i}\epsilon_2\sqrt{2}/4,\infty\}$ with $\epsilon_i = \pm 1$ independently. We have not found any striking feature in their Kodaira types, with are either $I_{1}$, $I_{2}$ or $I_{16}$.

\subsection{The $\lambda \rightarrow 0$ limit}
\label{eq:lto0}

This regime corresponds to $c\to 1$, where the monodromy along the fibers of the
flat connection tends to be deterministically equal to identity
under the Chern--Simons measure. This implies that $W(x) \in \cO(1)$, and
hence $\mu_k \rightarrow 0$ since it carries a prefactor of $\lambda$. As a result the spectral curve becomes easy to compute:
\beq
\mathcal{P}_{v}(x,y)|_{c = 1} = C(x)\prod_{i \in I} \big(y - (-x)^{n_0(w[i])}\zeta_{a}^{n_1(w[i])}\big)\,,
\eeq
i.e. $\mathcal{P}_{v}(x,y)|_{c = 1}$ is directly determined by the slope
polynomials of $P_{v}$, with no extra data. 
The description of the orbits leads to the following results.

\subsubsection*{$D_{p + 2}$, $p$ even}
\beq
\mathcal{P}_{v}(x,y)|_{c = 1} = (-1)^{p/2 + 1}(y + 1)\cdot ((-x)^{p/2}y^{p/2} + 1)(y^{p/2} - (-x)^{p/2})\,.
\label{eq:smthooft1}
\eeq
\subsubsection*{$D_{p + 2}$, $p$ odd}
\bea
\mathcal{P}_{v}(x,y)|_{c = 1} & = & (y + 1)^2 \prod_{\ell = 0}^{p - 1} (y + \zeta_{2p}^{\ell}x^2)(yx^2 + \zeta_{2p}^{-\ell}) \nonumber \\
& = & (y + 1)^2\cdot (y + x^2)(yx^2 + 1)\cdot \Big(\sum_{k = 0}^{p - 1}
(-1)^{k}\,y^{p - 1 - k}x^{k}\Big)\Big(\sum_{k = 0}^{p - 1} (-1)^k
(xy)^{k}\Big)\,. 
\eea
\subsubsection*{$E_6$ geometry}
\beq
\mathcal{P}_{v}(x,y)|_{c = 1} = (1 + y + y^2)\cdot (x + y)(1 + xy)\cdot (y^2 - x)(y^2x - 1)\,.
\eeq
\subsubsection*{$E_7$ geometry}
\beq
\mathcal{P}_{v}(x,y)|_{c = 1} = -(1 + y)(1 + y^2)^2 \cdot (y^2 - x^6)(y^2x^6 + 1) \cdot (x^6 + y^3)(x^6y^3 - 1)\cdot (y^6 + x^6)(x^6y^6 - 1)\,.
\eeq
Notice that the symmetry of the orbits implies $\mathcal{P}(x,y) = 0 \Leftrightarrow \mathcal{P}(\zeta_{12}x,1/y) = 0$, explaining how the factors come in pairs.
\subsubsection*{$E_8$ geometry}
\beq
\bary{rcl}
\mathcal{P}_{v}(x,y)|_{c = 1} & = & (y + 1)^2(y^3 - 1)^3(y^5 - 1)^5\cdot (y^{30} - x^{30})(y^{30}x^{30} - 1) \cdot (y^{15} + x^{30})^{2}(y^{15}x^{30} + 1)^2  \\
& &  \cdot (y^{10} - x^{30})^2(y^{10}x^{30} - 1)^2 \cdot (y^{15} - x^{60})(y^{15}x^{60} - 1) \cdot (y^6 - x^{30})(y^6x^{30} - 1)  \\
& & \cdot (y^5 + x^{30})(y^5x^{30} + 1) \,.
\eary
\label{eq:smthooft2}
\eeq

\section{Computations II: the Toda curves}
\label{Stringside}

\subsection{The computation in practice}
\label{Sstar}
Let us now construct explicitly the B-model geometries
$\cC_{\cG_\Gamma}^{\rm Toda}$ that are relevant for
Conjectures~\ref{conj:spec} and \ref{conj:gen}. Recall that we take $\rho_{\rm
  min}$ a minimal representation given in Table~\ref{tab:rmin}, and we denote $d_{\rm
  min} = \dim\rho_{\rm
  min}$. The
characteristic polynomial of the Lax operator  \eqref{eq:lax1} for the simple Lie group $\cG$,
\beq
\mathcal{P}_{\cG}(Y)
\triangleq\det\big[Y\mathbf{1} - \rho_{\rm min}(L^{\cG}_{\mathsf{w}})\big] =  \sum_{k=0}^{d_{\rm min}} (-1)^k\,Y^{d_{\rm min} - k}
\,\chi_{\Lambda^k \rho}(L^{\cG}_{\mathsf{w}}),
\eeq
can be regarded as a map
\beq
\cP_{\cG}: \cG_{\mathsf{w}} \longrightarrow \bbC[u,Y]
\eeq
that factors through a map $\cG_{\mathsf{w}} \longrightarrow \mathfrak{U}_\cG$ upon evaluation of
the antisymmetric characters $\chi_{\Lambda^k  \rho_{\rm min}}\in
\bbZ[\chi_{\omega_1}, \dots, \chi_{\omega_{R}}]$ in the representation ring ${\rm Rep}(\mathcal{G})$. Lifting this to the
co-extended affine situation amounts to turning on a spectral parameter as in
\eqref{eq:lax2}. Concretely, we are now looking at the loop space with a map:
\beq
\tilde{u}\,:\,{\rm Loop}(\mathfrak{U}_\cG) \longrightarrow \mathbb{C}[X^{\pm 1}],\qquad u_i(X) \triangleq \chi_{\omega_i}(L^{\cG^{\#}}_{\mathsf{w}})\,,
\eeq
whose constant term is given by $[X^0]\,\tilde u_i(X)=u_i$. The (affine
co-extended) Toda spectral curve \eqref{eq:sctoda} can then be computed in two
steps, for each $(\cG,\rho_{\rm min})$:
\begin{description}
\item[Step A1$'$] compute the decomposition of the exterior characters $\chi_{\Lambda^k
 \rho_{\rm min}}$ as polynomials in the fundamental characters $\chi_{\omega_i}$;
\item[Step A2$'$]
compute from \eqref{eq:lax2} the Casimir function $u_0$ in \eqref{eq:casimir}
and the dependence of $\tilde{u}(u_0, \dots, u_R\,; X) =
\chi_{\omega_i}[\rho_{\rm min}(L^{\cG^\#}_{\mathsf{w}})]$ on the Hamiltonians $u_i$
and the spectral parameter $X$.
\end{description}
Evaluating the result of Step~\textbf{A1}$'$ on a generic group element $g\in \cG$
expresses the characteristic polynomial of $\rho_{\rm min}(g)$ as
\bea
\det\big[Y \mathbf{1} -\rho_{\rm min}(g)\big] &=&
\sum_{k=0}^{d_{\rm min}} (-1)^k\,Y^{d_{\rm min}-k}\,\chi_{\Lambda^k \rho_{\rm min}}(g),\nn \\
&=&
\sum_{k=0}^{d_{\rm min}} Y^k \mathfrak{p}^{\cG}_k\l(g\,;\, u_1,\dots, u_{R}\r),
\eea
for some universal\footnote{These polynomials depend implicitly on $\rho_{\rm min}$, but
  we dropped the subscript $\rho_{\rm min}$ for the sake of readability.} polynomials $\mathfrak{p}^{\cG}_k \in \bbZ[u_1, \dots,
  u_{R}]$, while Step~\textbf{A2}$'$ amounts to plugging in the expression \eqref{eq:lax2}
of the Lax matrix and then expand the above in the spectral parameter,
\bea
\cP^{{\rm Toda}}_{\cG^\#}(X,Y;u_0,\ldots,u_{R})  &\triangleq & \det\big[Y\mathbf{1} -\rho(L^{\cG^\#}_{\mathsf{w}})\big] \nn \\
&=&
\sum_{k=0}^{d_{\rm min}} Y^k \mathfrak{p}^{\cG}_k\big[
\tilde{u}_1(X),\dots,
\tilde{u}_{R}(X)\big]\nn \\
&=&
\sum_{k=0}^{d_{\rm min}}\sum_{j=-d'_{{\rm min}}}^{d'_{{\rm min}}} Y^k X^j  \mathfrak{p}^{\cG}_{k,j}\big[u_0,u_1,\dots,
u_{R}\big]\,.
\label{eq:todascpol}
\eea
Here  $\mathfrak{p}^{\cG}_{k,j}$ denotes the result of the
expansion in the spectral parameter $X$ and
\beq
d'_{\rm
  min} \triangleq \mathrm{max}_{k,\sigma=\pm} \deg_{X^{\sigma}} \mathfrak{p}^{\cG}_{k}.
\eeq
The vanishing locus of  $\cP^{{\rm Toda}}_{\cG^\#} \in \bbC[u_1,\dots, u_{R}; X^{\pm 1},Y]$ in $\bbC^*_X \times \bbC^*_Y$ then returns \eqref{eq:sctoda}. \\

Once this is done, the naive expectation would be that, in light of the
discussion of the previous section, all
is left to do to prove Point~(a) in Conjecture~\ref{conj:spec} is just
to find a suitable restriction $u_i \leftarrow u_i(\lambda)$ of the Toda action variables such
that
\beq
\label{eq:P=P}
X^{d'_{\rm
  min}} \cP_{\cG^\#}^{{\rm Toda}}(X,Y;u(\lambda)) = \mathcal{P}_{\cG,v}(X,Y;\lambda),
\eeq
where $\mathcal{P}_{\cG,v}(X,Y;\lambda) = 0$ is the spectral curve found in Theorem~\ref{Lcal}, that we call here the ``naive LMO spectral curve''.  However, this is in general too much to ask.\\

First off, a rapid inspection of Tables~\ref{tab:rmin} and \ref{Secondtab} reveals that the $Y$-degrees in
\eqref{eq:P=P} will disagree in general. But more importantly, the qualitative
analysis of the naive LMO spectral curve $\overline{\mathcal{C}}_{v}$ given in
Section~\ref{S41} reveals that (a) its Galois group $\mathfrak{W}' =
\mathrm{Weyl}(\mathcal{D}') = {\rm Weyl}(\mathcal{G}')$ is a subgroup of the
Galois group $\mathfrak{W} = {\rm Weyl}(\cG) = {\rm Weyl}(\mathcal{D})$ of the
Toda spectral curve \eqref{eq:todascpol} with generic parameters $u$, and (b)
the branchcuts of $x = X^{1/a}\,:\,\overline{\mathcal{C}}_{v} \rightarrow
\mathbb{P}^1$ on the irreducible components of the LMO spectral curve must
necessarily be segments obtained from $x \in [1/\gamma,\gamma]$ by rotations
of angle multiple to $2\pi/a$, and the branching data of this curve is
completely determined by the analysis of orbits of $\mathfrak{W}'$ in
Section~\ref{S42}. \\
 
This actually suggests a way out of the conundrum: the sought-for subfamily of Toda curves should arise in the
sub-locus of the parameter space $\mathfrak{U}_\cG$ where the monodromy breaking
$\DD\to \DD'$ in Table~\ref{proporPv} is enforced. The simplest way to achieve this is to
consider an embedding of the subgroup $\iota\,:\,\cG' \hookrightarrow \cG$,
and the induced embedding $\iota\,:\,\cT'\hookrightarrow \cT$ of the maximal torus of $\cG'$
 into that of $\cG$. The restriction to $\iota(\cT')$ is cut out by homogeneous linear constraints on the Cartan subalgebra of $\mathrm{Lie}(\cG)$, and its image under the character map $\chi_\omega$ yields
an affine complete intersection $\tilde{\mathfrak{U}}_{(\cG,\cG')}$ in
$\mathfrak{U}_\cG$. Under the action of $\iota(\cG')$, the $\cG$-module
$\rho_{\rm min}$ decomposes into $\cG'$-modules:
\beq
\label{Gmode}\rho_{\rm min} = \bigoplus_{j \in J} \rho^{[j]}.
\eeq
For $u \in \mathfrak{U}_{\cG}$, the endomorphism $\rho_{{\rm min}}(L_{\mathsf{w}}^{\cG^{\#}})$ leaves
stable the direct sum in \eqref{Gmode}, and thus the characteristic polynomial
factors. Requiring that the latter are Laurent polynomials in $X$ yields an
additional constraint, i.e. $u$ must belong to a subvariety of higher codimension in $\tilde{\mathfrak{U}}_{(\cG,\cG')}$, that we denote
$\mathfrak{U}_{(\cG,\cG')}$. In both cases $(\cG,\cG') = (E_6,D_4)$ and
$(E_7,E_6)$ we will examine, $\mathfrak{U}_{(\cG,\cG')}$ will turn out to be a
subvariety of $\tilde{\mathfrak{U}}_{(\cG,\cG')}$ ruled along a distinguished direction $u_{{\rm rul}}$. Summing up:

\begin{description}
\item[Step B$'$] Consider the Lie group $\mathcal{G}'$ associated to $\mathcal{D}'$ as in Table~\ref{proporPv}, and the decomposition~\eqref{Gmode} as above. Restrict to $u \in \mathfrak{U}_{(\cG,\cG')}$ so that
\beq
\cP_{\cG^\#}^{{\rm Toda}}(X,Y) = \prod_{j \in J} \det\big[Y\mathbf{1} -\rho^{[j]}(L^{\cG^\#}_{\mathsf{w}})\big] \triangleq \prod_{j \in J} \mathcal{P}^{[j]}_{(\cG',\cG)}(X,Y)
\label{eq:Prestrict}
\eeq
for Laurent polynomials $\mathcal{P}^{[j]}_{(\cG,\cG')}(X,Y) \in
\bbC[X^{\pm 1},Y]$. 
\end{description}
This is the analogue of Step~$\textbf{B}$ on the LMO side (Section~\ref{S42}),
which consists in computing $\mathcal{P}_{\mathcal{D},v}(x,y)$ and thus $\mathcal{P}^{{\rm
    LMO}}_{\mathcal{D}}(X,Y)$ in terms of unknowns $\cM = (\mu_k)_{k \in K_0}$ for a
small set $K_0$. Let us denote $\mathcal{P}^{{\rm LMO}}_{\mathcal{D}}(X,Y;\cM,\lambda)$
this answer. In the following, we will check for $\mathcal{D} = E_6$ (resp. $\mathcal{D} = E_7$) that with $K_0 = \{1,2\}$ (resp. $K_0 = \{2,3,5,7\}$) the equality of polynomials\footnote{In \eqref{idM}, the $(Y - 1)^{\bullet}$ stands for $\bullet$ copies of the trivial representation appearing in \eqref{Gmode} and thus factoring out in the Toda spectral curve..} in $(X,Y)$:
\beq
\label{idM}(Y - 1)^{\bullet}\mathcal{P}^{{\rm LMO}}_{\mathcal{D}}(X,Y;\cM,\lambda) = X^{d'_{\rm
  min}}\mathcal{P}^{{\rm Toda}}_{\cG}(X,Y;u)
\eeq
gives an explicit isomorphism 
\beq
\bary{rccl}
\Upsilon: & \mathbb{C}^{|K_0|}_{\cM} & \stackrel{\sim}{\longrightarrow} & \mathfrak{U}_{\cG,\cG'}.
\eary 
\eeq
This $\Upsilon$ is given in \eqref{UpE6} for $E_6$ and \eqref{UpE7} for $E_7$. We expect the same property to hold for $\cG = E_8$ (here, $\cG' = E_8$ and $\mathfrak{U}_{(\cG,\cG')}$ is just equal to $\mathfrak{U}_{\cG}$) but this case was computationally out of reach.

\begin{description}
\item[Step C$'$] Determine the sublocus of $u \in \fU_{(\cG,\cG')}$ such that the Toda spectral curve has the ramification properties that were required for the LMO spectral curve (see Step \textbf{C}, Section~\ref{S42}). This should fix $u$ to live in a $1$-dimension variety $\mathfrak{U}^{{\rm LMO}}_{\cG}$ parametrized by $\lambda$.
\end{description}
In all cases, just by matching the coefficients on the boundary of the Newton polygon, we find that $u_0 = -1/c^{a}$, where we remind that $c = \exp(\chi_{{\rm orb}}\lambda/2a)$. Therefore, $\mathfrak{U}^{{\rm LMO}}_{\cG}$ is equivalently parametrized by $(u_i(u_0))_{i = 1}^{R}$.\\

By the previous remark, Step \textbf{C}$'$ is strictly equivalent to the
determination of $(\mu_k)_{k \in K_0}$ as functions of $\lambda$ on the LMO side
in Step \textbf{C}. Step {\bf A1}$'$-{\bf A2}$'$-{\bf B}$'$ and {\bf C}$'$
together give a complete derivation of the suitable restriction of the Toda
spectral curve, if we assume and verify the qualitative properties used in Step \textbf{B}$'$ and \textbf{C}$'$ that were dictated by the analysis of the matrix model.

\subsection{$A_{p-1}$ geometries}
\label{Anstring}
This is the case of lens spaces $\mathbb{S}^{\bbZ/p\bbZ}=L(p,1)$ already well-known in 
the literature, so we will only make a couple of passing remarks here to see
how it fits with the discussion above. Steps~\textbf{A1$'$} and \textbf{A2$'$} were
performed in \cite{Nekrasov:1996cz,Marshakov:2012kv} and
return\footnote{Choosing the minimal representation $\rho_{\rm min}$ to be the anti-fundamental
 has the sole effect of redefining $u_k \rightarrow (-1)^p u_{p-k}$.}
\eqref{eq:CA}, which is in exact agreement with the matrix model curve computed
by Halmagyi--Yasnov \cite{Halmagyi:2003ze} in a general flat background. This
proves Point~(a) of Conjecture~\ref{conj:gen} for the sphere and disk
potential; the rest of Point~(a) follows from the solution by
the topological recursion method of generalized loop equations
\cite{BEO}, which combined with the proof of the remodeling
conjecture \cite{EOBKMP} establishes Point~(b) as well. The restriction
relevant for Step~\textbf{B}$'$ and Conjecture~\ref{conj:spec} is simply $u_0=\re^{-t_{\rm B}/2}=c^{1/2}$, $u_i=0$ for $i \in \llbracket 1,p \rrbracket$: toric mirror 
symmetry shows that this amounts to setting to zero the insertion of twisted
classes in $H_{\rm orb}(Y^{\bbZ/p\bbZ})$, so that the resulting restricted
A-model theory is just the untwisted Gromov--Witten theory of the stack
$[\cO_{\mathbb{P}^1}(-1)^{\oplus 2}/(\bbZ/p\bbZ)]$.

\subsection{$D_{p + 2}$ geometries}

For $p + 2 = 4$, Steps~\textbf{A1}$'$-\textbf{A2}$'$ in this case can be extracted from
\cite{Kruglinskaya:2014pza} and found to be in agreement with 
\eqref{eq:CD}. 
\subsubsection*{Step A1$'$}
\begin{figure}[!h]
\centering
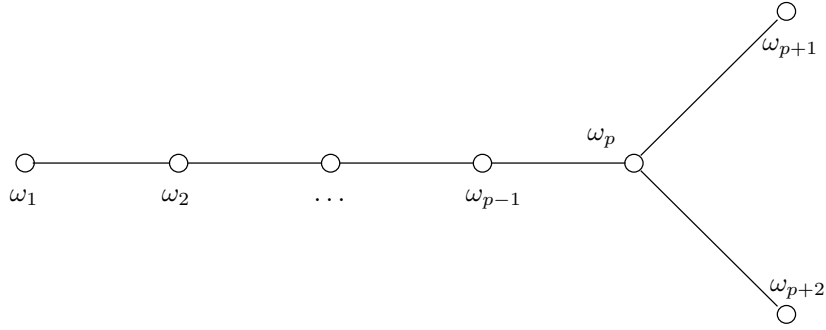
\caption{The Dynkin diagram of $\cG=D_{p + 2}$. Nodes in the diagram are
  labeled by the respective fundamental weights; we have $\rho_{\rm
    min}\triangleq \rho_{\omega_1}=\l(\mathbf{2p+4}\r)_{\bf v}$,
  $\rho_{\omega_i}=\Lambda^i \rho_{\omega_1}$ for $i\leq p$,
  $\rho_{\omega_{p+1}}=\mathbf{2}^{p+1}_{\bf c}$,
  $\rho_{\omega_{p+2}}=\mathbf{2}^{p+1}_{\bf s}$.}
\label{fig:dynkdn}
\end{figure}

Explicitly, the Dynkin diagram of $D_{p + 2}$ is represented in
Figure~\ref{fig:dynkdn}.

We write $\rho_{\rm min} \triangleq\mathbf{2(p + 2)}_{\bf v}$ for the defining module of $\mathrm{Spin}(2(p + 2))$.
In this specific case, we can slightly bypass Step~\textbf{A1}$'$ by
parametrizing $\mathfrak{U}_\cG$ using the exteriors characters $\epsilon_i = [X^0]\,\tilde{\epsilon}_i$ with:
\beq
\tilde{\epsilon}_{i} \triangleq\chi_{\Lambda^{i}\rho_{\omega_1}}(L^{D_{p + 2}^\#}_{\mathsf{w}}),\qquad  {\rm for}\,\, i \in \llbracket 1,p + 2 \rrbracket\,.
\eeq
We have $\tilde{\epsilon}_i = \tilde{u}_i$ for $i \leq  p$, and
\bea
\tilde{\epsilon}_{p + 1} &=& \tilde{u}_{p + 1}\tilde{u}_{p + 2}
-
\l\{
\bary{cc}
\sum_{k=0}^{p/2-1} \tilde{u}_{2k+1} & \quad \hbox{$p$ even}, \\
\sum_{k=0}^{p/2} \tilde{u}_{2k} & \quad \hbox{$p$ odd}.
\eary
\r. \\
\tilde{\epsilon}_{p + 2} &=& \tilde{u}_{p + 1}^2+ \tilde{u}_{p + 2}^2-2 
\l\{
\bary{cc}
\sum_{k=0}^{p/2} \tilde{u}_{2k} & \quad \hbox{$p$ even}, \\
\sum_{k=0}^{p/2} \tilde{u}_{2k+1} & \quad \hbox{$p$ odd}.
\eary
\r.
\eea
as a consequence of the decomposition rules of the tensor products $\cS_\pm\otimes
\cS_{\pm}$ of the chirality $\pm$ spin representations associated to the
fundamental weights $\omega_{p + 1}$ and $\omega_{p + 2}$. 
\subsubsection*{Step A2$'$} 
\begin{figure}[t]
\centering
\includegraphics[scale=0.4]{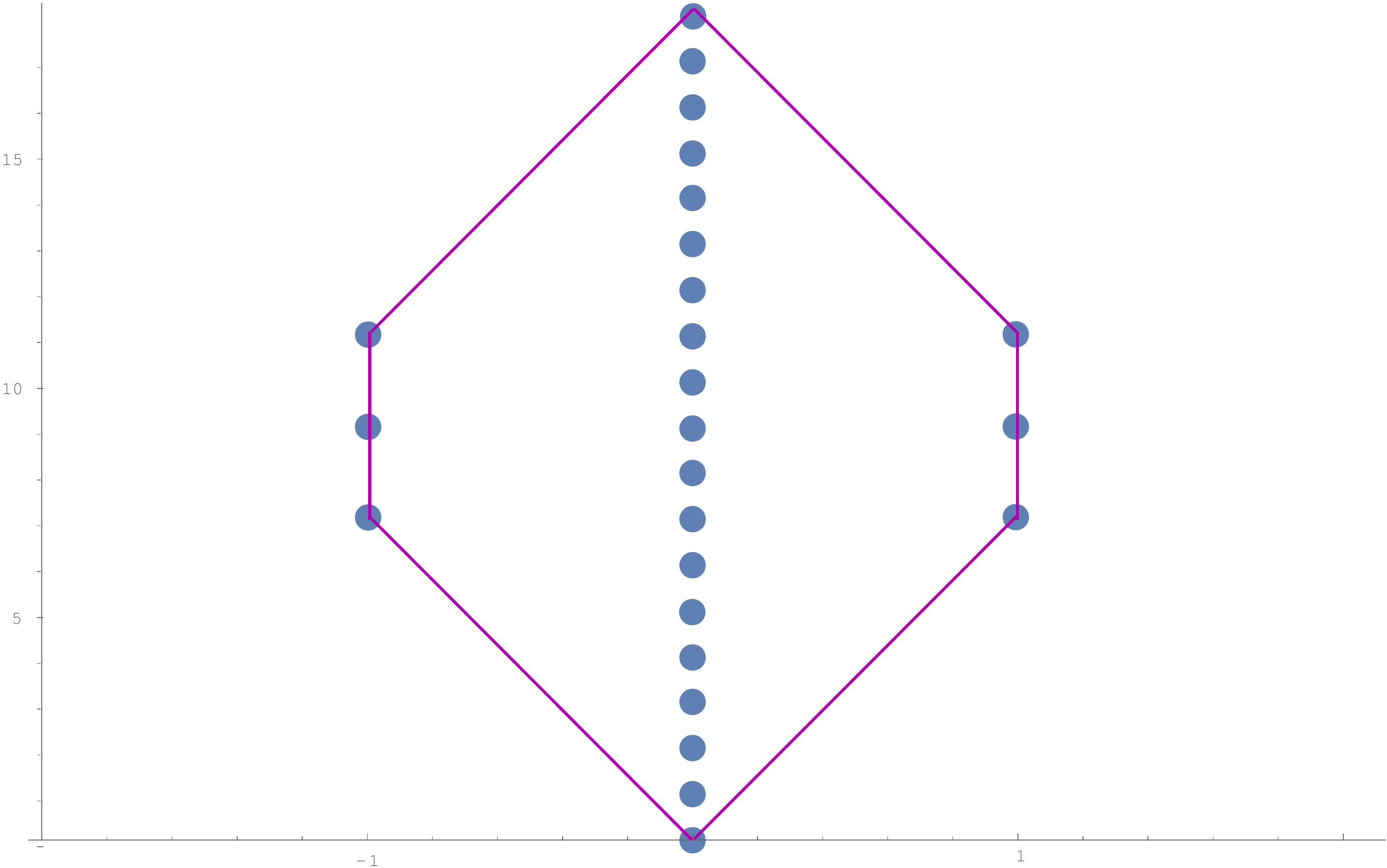}
\caption{\label{fig:DnToda} The Newton polygon of the Toda curve for $\cG=D_{p
    + 2}$, minimal representation $\rho_{\omega_1} =\mathbf{2(p + 2)}_{\bf v}$ for $p + 2 =9$.}
\end{figure}

The Casimir
function $u_0$ here reads
\beq
u_0^{-1}=\varkappa_0^{1/2} \varkappa_1 \varkappa_{p+1}\varkappa_{p+2} \prod_{1 < i < p+1}\varkappa_i^2,
\eeq
and the Laurent polynomials $\tilde{\epsilon}_i(X)$ can be computed straightforwardly from
\eqref{eq:lax2} using Newton identities. We
have 
\beq
\bary{rclc}
\tilde{\epsilon}_i(X) &=& \epsilon_i, \quad & i \neq p,p + 2,p + 4, \\
\tilde{\epsilon}_i(X) &=& \epsilon_i+ u_0(X+ 1/X), \quad & i=p, p+4, \\
\tilde{\epsilon}_{p + 2}(X) &=& \epsilon_{p + 2}- 2u_0(X+ 1/X).  \\
\eary
\eeq
The Newton polygon of the resulting plane curve is shown in
Figure~\ref{fig:DnToda}. In terms of exponentiated linear coordinates
$(r_j)_{j = 1}^{p + 2}$ on the
maximal torus $\cT_{D_{p + 2}}$, the resulting curve takes the form
\bea
X\,\cP_{D_{p + 2}^\#}^{{\rm Toda}}(X,Y) &=&
u_0(X^2 + 1)(Y - 1)^2(Y + 1)^2Y^{p} + \sum_{i=0}^{2(p + 2)} (-1)^i
\epsilon_i\,XY^i,\nn \\ 
&=&
u_0(X^2 + 1)(Y - 1)^2(Y + 1)^2Y^{p} + X\prod_{j = 1}^{p + 2} (Y - r_j)(Y - r_j^{-1}),\nn \\
\label{gauD}
\eea
which is just \eqref{eq:CD} with $u_0=\re^{-t_{\rm B}/2}(-1)^{p+1} 2^{2p}$.

\subsubsection*{Steps B$'$ and C$'$}

Comparing this Toda curve with the LMO curve \eqref{Lform}, we find agreement provided $r_{1} = 1$ and:
\beq
r_{j}^{\pm 1} = \frac{\tilde{r}_j}{2(\kappa^2 + 1)^2} \pm \sqrt{\frac{\tilde{r}_j^2}{4(\kappa^2 + 1)^4} - 1}
\eeq
where $\tilde{r}_{j}$ are the $(p + 1)$ roots of the polynomial $\mathcal{Q}_{p}$ given in Appendix~\ref{Lpoly}, and:
\beq
u_0 = (-1)^{p + 1}\,e^{-\lambda/2p}.
\eeq

\subsection{$E_6$ geometry}

\begin{figure}[!h]
\centering
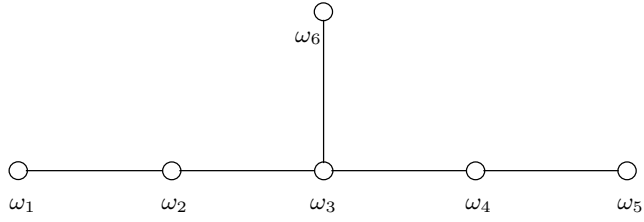
\caption{The Dynkin diagram of $\cG=E_6$. Nodes in the diagram are labeled by
  the corresponding fundamental weights; we have
  $\rho_{\rm min}\triangleq \rho_{\omega_1}=\mathbf{27}=\overline{\rho_{\omega_5}}$,
  $\rho_{\omega_2}=\Lambda^2\rho_{\omega_1}=\mathbf{351}=\overline{\rho_{\omega_4}}$,
  $\rho_{\omega_3}=\Lambda^3\rho_{\omega_1}=\mathbf{2925}$, $\rho_{\omega_6}=\mathrm{Adj}=\mathbf{78}$.}
\label{fig:dynke6}
\end{figure}

The Dynkin diagram of $\cG=E_6$ is represented in Figure~\ref{fig:dynke6}. We
write $\rho_{{\rm min}} \triangleq\mathbf{27}=\rho_{\omega_1}$ for the minimal irreducible
representation attached to the highest weight $\omega_1$; there is another
minimal $\cG$-module $\overline{\rho}_{\rm min} \triangleq \overline{\mathbf{27}}=\rho_{\omega_5}$,
which is complex-conjugate to $\rho_{{\rm min}}$.

\subsubsection*{Step \textbf{A1}$'$}

The fundamental representations of $E_6$ are antisymmetric powers of
$\rho_{{\rm min}}$ and
$\overline{\rho}_{\rm min}$ with the exception of $\rho_{\omega_6}$, which is the
78-dimensional adjoint representation. The antisymmetric characters
$\chi_{\Lambda^k \rho_{\omega_1}}(g):\cT_{E_6} \to \bbC$ of an element $g=\re^h$, which include the fundamental characters
$(\chi_{\omega_i}(g))_{i = 1}^5$, can be computed using the explicit representation
of the Chevalley generators in the representations $\rho_{{\rm min}}$ and $\overline{\rho}_{{\rm min}}$
\cite{Kruglinskaya:2014pza,MR1823074}. On the other hand, the regular character $\chi_{\omega_6}(g)$ is a trace in the
adjoint, which is computed straightforwardly from the root system of
$E_6$. We must have
\beq
\chi_{\Lambda^k \rho_{\omega_1}}(g) = \mathfrak{p}_k^{E_6}[\chi_{\omega_1}(g),\dots,\chi_{\omega_6}(g)]
\label{eq:lambdakve6}
\eeq
identically as functions on the Cartan torus. One possible brute-force way to compute
$\mathfrak{p}_k^{E_6}$ is to generate a finite-dimensional vector space of monomials
$\{\chi_{\omega_i}^{n_j}(g)\}_{i,j}$ satisfying a suitable dimensional upper bound on
$\chi_{\omega_i}^{n_j}(1)$, then evaluate \eqref{eq:lambdakve6} at a number of points $g\in
\cT$ equal to the dimension of this vector space, and then solve the linear
system ensuing from \eqref{eq:lambdakve6}. The resulting relations in ${\rm Rep}(\cG)$
read:
\beq
\bary{rcl}
 \mathfrak{p}_4^{E_6} &=& -u_5^2-u_2 u_5+u_1+u_4+u_4 u_6, \\
 \mathfrak{p}_5^{E_6} &=& u_1^2-2 u_5^2 u_1+2 u_4 u_1+u_4^2+u_5 u_6^2+u_2-2 u_3 u_5+u_5-u_2 u_6-u_5 u_6, \\
 \mathfrak{p}_6^{E_6} &=& -u_5^3-u_2 u_5^2+u_1 u_5+2 u_4 u_5-2 u_1 u_6 u_5+u_4 u_6 u_5+u_6^3+2 u_1 u_2-2 u_3+u_2 u_4-3 u_3 u_6, \\
 \mathfrak{p}_7^{E_6} &=& 2 u_2^2+u_5 u_2-2 u_5 u_6 u_2+u_3 u_5^2+u_1 u_6^2+u_4 u_6^2-3 u_1 u_3-2 u_3 u_4+u_4-u_5^2 u_6-u_1 u_6+u_4 u_6, \\
 \mathfrak{p}_8^{E_6} &=& u_2 u_5^3-u_1 u_6 u_5^2+u_6^2 u_5+u_1 u_2 u_5-2 u_3
 u_5-3 u_2 u_4 u_5+u_3 u_6 u_5-2 u_6 u_5+u_5-u_1^2-u_2 u_6^2 \\ &+& u_2 u_3+u_1 u_4+u_1^2 u_6-u_2
   u_6+2 u_1 u_4 u_6, \\
 \mathfrak{p}_9^{E_6} &=& u_1 u_5^4-u_6 u_5^3+u_2 u_5^2-4 u_1 u_4 u_5^2+u_2
 u_6 u_5^2-u_2^2 u_5-u_1 u_6^2 u_5-4 u_1 u_5+4 u_1 u_3 u_5+u_4 u_5 \\ &+& 3 u_4 u_6
   u_5+u_6^3+u_3^2+2 u_1 u_4^2+u_1 u_2-6 u_3+4 u_1^2 u_4-4 u_2 u_4-3 u_3 u_6-2 u_2 u_4 u_6+3, \\
 \mathfrak{p}_{10}^{E_6} &=& u_5^5-5 u_4 u_5^3+u_1 u_6 u_5^3-u_6^2 u_5^2-u_1
 u_2 u_5^2+5 u_3 u_5^2-u_5^2-2 u_1^2 u_5+5 u_4^2 u_5+u_2 u_5+u_2 u_3 u_5 
 \\ &+& 4 u_1 u_4
   u_5+u_1^2 u_6 u_5-2 u_2 u_6 u_5-3 u_1 u_4 u_6 u_5+u_1 u_6^2+2 u_4 u_6^2+2
   u_1+u_1^2 u_2-5 u_1 u_3  \\ &+& 2 u_1 u_2 u_4-5 u_3 u_4-u_2^2 u_6-2 u_1 u_6+u_1 u_3 u_6-u_4
   u_6, \\
 \mathfrak{p}_{11}^{E_6} &=& u_6 u_5^4-u_2 u_5^3+u_1 u_3 u_5^2-u_4 u_5^2+2 u_1
 u_6 u_5^2-4 u_4 u_6 u_5^2-2 u_6^2 u_5-u_1 u_2 u_5+3 u_3 u_5+3 u_2 u_4 u_5 
 \\ &-& 2 u_1 u_2 u_6
   u_5+3 u_3 u_6 u_5-u_6 u_5+u_1 u_2^2+2 u_4^2+2 u_1^2 u_6^2-2 u_2 u_6^2+2
   u_2-3 u_1^2 u_3+u_2 u_3  \\ &+& u_2^2 u_4+u_1 u_4-2 u_1 u_3 u_4-2 u_1^2 u_6+2 u_4^2 u_6+u_2 u_6,
   \\
 \mathfrak{p}_{12}^{E_6} &=& 2 u_6 u_1^3-u_1^3+u_5^2 u_1^2-u_4 u_1^2-2 u_2 u_5
 u_1^2-u_5^2 u_6 u_1^2+u_4 u_6 u_1^2+3 u_5 u_6^2 u_1+2 u_2 u_1-3 u_2 u_3
 u_1 \\ &+& u_2 u_4 u_5
   u_1+u_5 u_1-5 u_2 u_6 u_1-2 u_5 u_6 u_1+u_2^3+u_3 u_5^3-u_5^3-2 u_6^3+3
   u_3^2-u_3+u_2 u_4\\ &+& 3 u_2^2 u_5 -3 u_3 u_4 u_5+2 u_4 u_5+u_5^3 u_6-u_2 u_5^2 u_6+6 u_3
   u_6-3 u_4 u_5 u_6, \\
 \mathfrak{p}_{13}^{E_6} &=& u_1^4-2 u_5^2 u_1^3+2 u_4 u_1^3+u_4^2 u_1^2-3 u_2
 u_1^2-2 u_3 u_5 u_1^2-u_5 u_1^2-u_2 u_6 u_1^2+4 u_5 u_6 u_1^2+2 u_5^3
 u_1\\ &+& 2 u_2 u_5^2
   u_1-2 u_6^2 u_1+u_3 u_1-4 u_2 u_4 u_1+u_2^2 u_5 u_1-4 u_4 u_5 u_1-u_5^3 u_6
   u_1+3 u_3 u_6 u_1+u_4 u_5 u_6 u_1 \\ &-& u_6 u_1+2 u_1+u_2^2-2 u_2 u_4^2+u_3 u_5^2+u_2 u_4
   u_5^2+u_5^2 u_6^2-3 u_3 u_4+2 u_4+u_2 u_5-u_2 u_3 u_5 \\ &+& u_2^2 u_6-2 u_5^2
   u_6-2 u_2 u_5 u_6,
\eary
\label{eq:pke6}
\eeq
and $\mathfrak{p}_{27-k}^{E_6}(u_1, u_2,u_3,u_4,u_5,u_6)= \mathfrak{p}_{k}^{E_6}(u_5,u_4,u_3,u_2,u_1,u_6)$. This completes Step~\textbf{A1}$'$.
 
\subsubsection*{Step \textbf{A2}$'$}

\begin{figure}[t]
\centering
\includegraphics[scale=0.45]{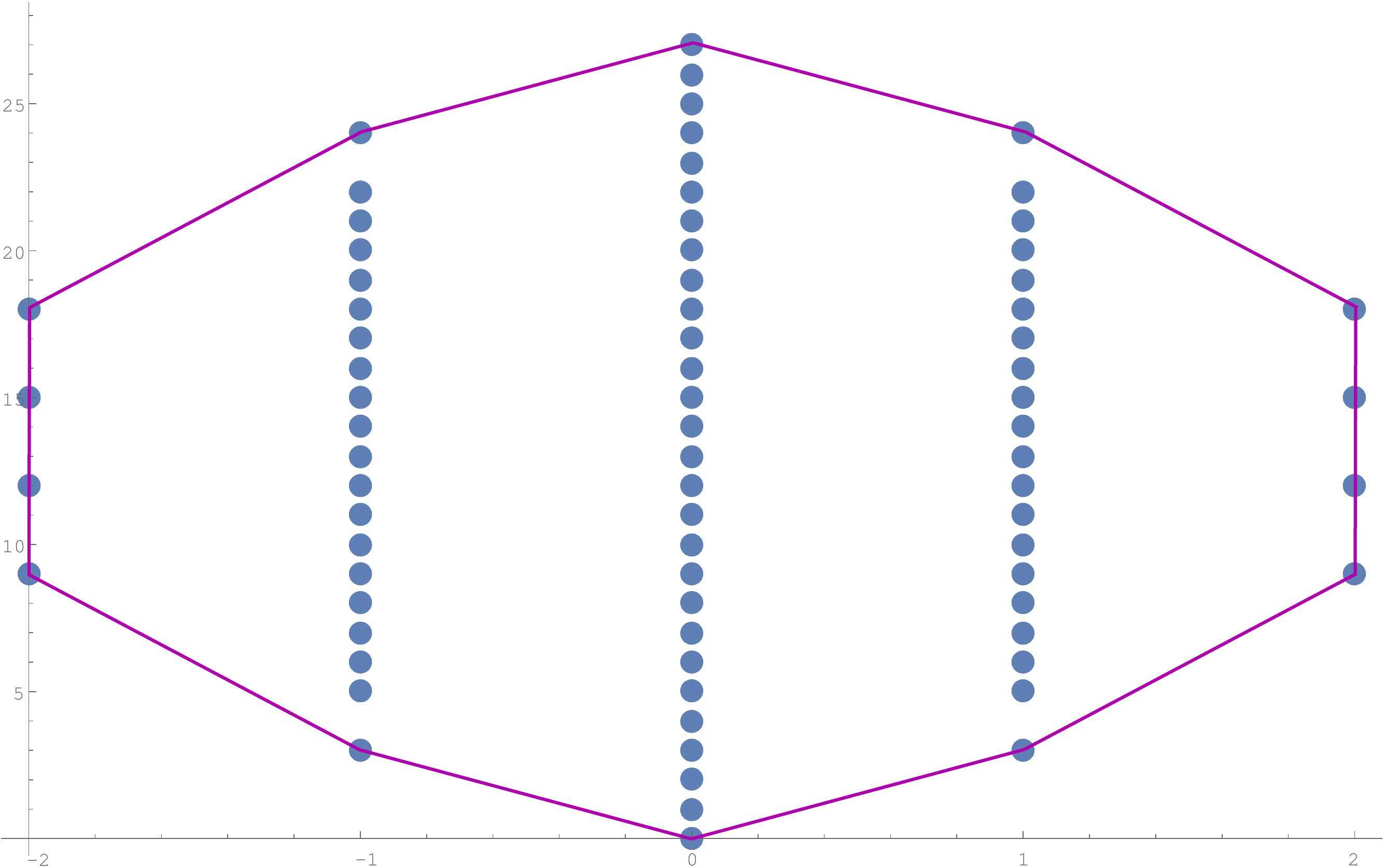}
\caption{\label{fig:npole6}The Newton polygon of the Toda spectral curve for
  $\cG^{\#}=E_6^{\#}$ and $\rho_{{\rm min}} =\mathbf{27}$.}
\end{figure}

As for the $D_{p + 2}$ case above, the spectral parameter dependence of
$(\widetilde{u}_i(X))_{i = 1}^5$ can be computed directly from
\eqref{eq:lax2} using
Newton identities. In particular, we obtain
\beq
\bary{rclr}
\tilde{u}_i(X) &=& u_i, \quad & \quad i \neq 3,\\
\tilde{u}_3(X) &=& u_3+ u_0(X+ 1/X),
\eary
\label{eq:uie6}
\eeq
in terms of the Casimir function $u_0^{-1}=\varkappa_0^{1/2}\varkappa_1 \varkappa_2^2 \varkappa_3 ^3 \varkappa_4^2 \varkappa_5
\varkappa_6^2$. The spectral parameter dependence of $\tilde{u}_6$ can be computed
from the first line of \eqref{eq:pke6}: the result is
\beq
\tilde{u}_6(X)=u_6.
\label{eq:u6e6}
\eeq
In other words, the $E_6$-Toda curve is computed as
\beq
0=\cP_{E_6^\#}^{{\rm Toda}}(X,Y) =
\sum_{k=0}^{27} \mathfrak{p}^{E_6}_k\big[u_1,u_2,u_3+u_0(X+1/X),u_4,u_5, u_6\big]\,Y^k.
\eeq
The resulting Newton polygon is depicted in Figure~\ref{fig:npole6}.

\subsubsection*{Step \textbf{B}$'$}

In this case, we have $\DD'= D_4$ while we started with $\DD= E_6$. There is an obvious embedding 
\beq
\bary{cccl}
\iota: & \cT_{D_4} & \longrightarrow & \cT_{E_6} \nn \\
& (Q_1,Q_2,Q_3,Q_4) & \longmapsto & (1,Q_1,Q_2,Q_3,1,Q_4)
\eary
\eeq
of the maximal tori, induced by the projection of the weight system of $E_6$
onto the sublattice of $\bbZ^6$ spanned by the unit lattice vectors
$\omega_2$, $\omega_3$, $\omega_4$ and $\omega_6$. Under this projection, the fundamental highest weight modules of $E_6$
decompose as $D_4$-modules in the following way: 
\bea
\label{eq:u1=u5}
\rho_{\omega_1}&=& \overline{\rho_{\omega_5}} =3(\irrep{1}) \oplus \irrepsub{8}{s} \oplus \irrepsub{8}{v} \oplus \irrepsub{8}{c},\\
\label{eq:u2=u4} 
\rho_{\omega_2}&=& \overline{\rho_{\omega_4}}=
3(\irrep{1}) \oplus 4(\irrepsub{8}{s}) \oplus 4(\irrepsub{8}{v}) \oplus 4(\irrepsub{8}{c}) \oplus 3(\irrep{28}) \oplus \irrepsub{56}{s} \oplus\irrepsub{56}{v} \oplus \irrepsub{56}{c},\\
\rho_{\omega_3}&=& 2(\irrep{1}) \oplus 8(\irrepsub{8}{s}) \oplus 8(\irrepsub{8}{v}) \oplus 8(\irrepsub{8}{c}) \oplus 
11(\irrep{28}) \oplus \irrepsub{35}{v} \oplus\irrepsub{35}{c} \oplus \irrepsub{35}{s} \nn \\
& & \oplus 6(\irrepsub{56}{s}) \oplus 6(\irrepsub{56}{v}) \oplus 6(\irrepsub{56}{c}) \oplus 2(\irrepsub{160}{s}) \oplus 2(\irrepsub{160}{v})\oplus 2(\irrepsub{160}{c}) \oplus \irrep{350},
\\
\rho_{\omega_6}&=&
2(\irrep{1}) \oplus 2(\irrepsub{8}{s}) \oplus 2(\irrepsub{8}{v}) \oplus 2(\irrepsub{8}{c})\oplus \irrep{28}.
\label{eq:E6D4fin}
\eea 
In particular, with \eqref{eq:u1=u5}:
\bea
\label{eq:PE6toD4}
\cP_{E_6^\#}^{{\rm Toda}}(X,Y)\big|_{u\in\tilde{\mathfrak{U}}_{(E_6,D_4)}} &=& (Y-1)^3
\prod_{\bullet={\bf c,v,s}}\det\big[Y\mathbf{1}-\rho_{8_\bullet}(L^{E_6^\#}_{\mathsf{w}})\big].
\eea
The resulting variety
$\tilde{\mathfrak{U}}_{(E_6,D_4)}=( u \circ \iota)(\cT')$ is a connected
codimension zero submanifold of the intersection of hyperplanes $u_1=u_5$,
$u_2=u_4$, hence it is locally ruled with respect to the $X$-dependent Casimir $u_{\rm rul}=u_3$. 
The degree of the factors in $Y$, leaving aside the trivial abelian component $(Y-1)^3$, now reproduces the structure of
$\mathcal{P}^{{\rm LMO}}_{E_6}(X,Y)$ as a product over the polynomials
associated to the $3$ minimal orbits
generated by $v$, $\varepsilon [v]$ and $\varepsilon^2 [v]$ as in
\eqref{eq:PmmE6}. In $\mathcal{P}^{{\rm LMO}}_{E_6}(X,Y)$, the $3$ factors are polynomials in $X^{1/3}$ that differ by order $3$ rotations $x \mapsto \zeta^j_3x$. However, this $\bbZ/3\bbZ$ rotational symmetry is
absent in \eqref{eq:PE6toD4} for generic values of $u\in \tilde{\mathfrak{U}}_{\cG,\cG'}$; more importantly, the individual factors
appearing in \eqref{eq:PE6toD4} are not guaranteed\footnote{This is an instance of the following, general problem: given a family of polynomials $P \in \mathbb{C}[x,y]$ depending on parameters $u = (u_i)_{i = 1}^R$, and given an integer $n \geq 2$, determine the locus of parameters for which there exists a factorization
\beq
\label{Odefu}P(x^n,y;u) = \prod_{j = 0}^{n - 1} Q(\zeta_{n}^{j}x,y;u)
\eeq
where $Q$ is also a polynomial in $x$ and $y$. It was communicated to us by
Don Zagier that there is no obvious strategy to solve this problem in general, but one can always try the naive approach consisting in writing down arbitrary coefficients for $Q$, expanding \eqref{Odefu} and solving for the parameters $(u_i)_{i = 1}^R$. For the example $(E_6,D_4)$ that we provided, the palindromic symmetry of the factors can be exploited to simplify a bit the derivation.}
to be polynomials.

This puts two constraints that are solved simultaneously as follows. Denote 
\beq
U_{\bullet} \triangleq \chi_{\bf 8_\bullet}\big[L^{E_6^\#}_{\mathsf{w}}\big],\qquad 
U_{\rm adj}\triangleq \chi_{\bf 28}\big[L^{E_6^\#}_{\mathsf{w}}\big].
\eeq
the evaluation of the $D_4$-fundamental characters on the reduced $L^{E_6^\#}_{\mathsf{w}}$ seen for $u \in \tilde{\mathfrak{U}}_{(E_6,D_4)}$ as a $D_4$ group element. The following character relations in ${\rm Rep}(D_4)$ are easily deduced from simple tensor multiplication rules:
\bea
\chi_{\bf 35_\bullet} &=& \chi_{\bf 8_\bullet}^2-\chi_{\bf 28}-1, \nn \\
\chi_{\bf 56_\bullet} &=& \chi_{\bf 8_\diamond}\chi_{8_\star}-\chi_{\bf 8_\bullet},
\nn \\
\chi_{\bf 160_\bullet} &=& \chi_{\bf 28}\chi_{\bf 8_\bullet}-\chi_{\bf
  8_\diamond}\chi_{8_\star}, \nn \\
\chi_{\bf 350} &=& \chi_{\bf 8_c}\chi_{\bf 8_v}\chi_{\bf 8_s}-\chi_{\bf
  8_c}^2-\chi_{\bf 8_v}^2-\chi_{\bf 8_s}^2+\chi_{\bf 28}+2,
\eea
where the formulas above should be intended as having the set equality $\{\bullet,\star,
\diamond\} = \{\mathbf{c},\mathbf{v},\mathbf{s}\}$. Also, $\Lambda^2 \mathbf{8}_\bullet=\mathbf{28}$, $\Lambda^3
\mathbf{8}_\bullet=\mathbf{56}_\bullet$, $\Lambda^4
\mathbf{8}_\bullet=\mathbf{35}_\star \oplus \mathbf{35}_\diamond$. Then, 
\bea
\det_{\bf 8_\bullet}\big[Y\mathbf{1} - \cR_{8_\bullet}(L^{E_6^\#}_{\mathsf{w}})\big] &=& Y^8-U_{\bullet}
Y^7+U_{\rm adj}
Y^6+\l(U_{\bullet}-U_{\star}U_{\diamond}\r)Y^5+\big(U_{\diamond}^2 + U_{\star}^2\nn \\ &-& 2
U_{\rm adj}-2\big) Y^4 + \l(U_{\bullet}-U_{\star}U_{\diamond}\r)Y^3+U_{\rm adj}Y^2-U_{\bullet}Y+1,
\eea
and it is immediate to see that restricting to
\beq
U_{\bf c}=\tilde{U}^{[2]}(\zeta_3^2 x)+ \tilde{U}^{[1]},\qquad U_{\bf
  s}=\tilde{U}^{[2]}(\zeta_3 x)+ \tilde{U}^{[1]},\qquad U_{\bf v}=\tilde{U}^{[2]}(x)+ \tilde{U}^{[1]}\,,
\eeq
with:
\beq \addtocounter{footnote}{-1}
\tilde{U}^{[2]}(x)=x+U_0/x,\qquad X=x^3,\qquad u_0=U_0^3
\eeq
is necessary and sufficient to attain the required cyclic symmetry with the spectral dependence dictated by
\eqref{eq:uie6}-\eqref{eq:u6e6}.

\subsubsection*{Comparison with the LMO spectral curve}

At this stage, we find by direct computation of the left-hand side of \eqref{idM} with \eqref{eq:PmmE6} and the table of coefficients at the beginning of Section~\ref{E6geom} that the equality:
\beq
\mathcal{P}^{{\rm Toda}}_{E_6^{\#}}(X,Y,u)\big|_{u \in \mathfrak{U}_{(E_6,D_4)}} = (Y - 1)^3 \mathcal{P}^{{\rm LMO}}_{E_6}(X,Y;\cM,\lambda),\qquad c = \re^{\lambda/72}\,,
\eeq
is realized if and only if:
\beq
\bary{rcl}
\tilde U^{[1]} &=& -1-2c^{-2}\mu_1 \\
U_{\rm adj} &=& 2+12c^{-2}\mu_1 -6c^{-1}\mu_2
\eary
\label{UpE6}
\eeq
Eliminating $\mu_1$ and $\mu_2$ from these equations, we retrieve exactly the
constraints defining $\mathfrak{U}_{(E_6,D_4)} \subset \mathfrak{U}_{E_6}$. This is the
equivalence between Step~\textbf{B}$'$ on the Toda side and Step~\textbf{B}
on the LMO side highlighted in Section~\ref{Sstar}. We can then insert in this
parametrization the determination of $\mu_k$'s in terms of $\lambda$ (or $c$)
performed at the end of Section~\ref{E6geom}. We obtain $u_0 = -1/c^{6}$ 
and the $(u_i)_{i = 1}^6$ as functions of the parameter $\kappa$ related to $c$ by \eqref{these}:
\bea
u_1 & = & u_5 \,\, = \frac{3\kappa(\kappa - 4)(\kappa^2 - 12)(\kappa^2 - 6\kappa + 12)^2}{(\kappa - 6)^4(\kappa - 2)^4}, \nonumber \\
u_2 & = & u_4 \,\, = \frac{3\kappa(\kappa - 4)(\kappa^2 - 6\kappa + 12)^2\,f_2(\kappa)}{(\kappa - 6)^8(\kappa - 2)^8}, \nonumber \\
u_3 & = & \frac{f_3(\kappa)}{(\kappa - 6)^{12}(\kappa - 2)^{12}}, \nonumber \\
u_6 & = & \frac{2f_6(\kappa)}{(\kappa - 6)^6(\kappa - 2)^6}, \label{fikappad}
\eea
where $f_i(\kappa)$ are polynomials given in Appendix~\ref{fkappaE6}.

\subsection{$E_7$ geometry}
\label{E7string}

\begin{figure}[!h]
\centering
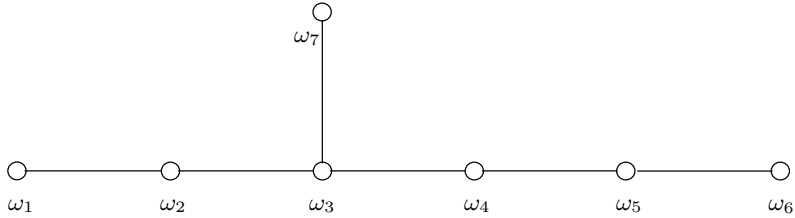
\caption{The Dynkin diagram of $\cG=E_7$. Nodes in the diagram are labeled
  here by the respective fundamental weights; we have that
  $\rho_{\omega_1}=\mathrm{Adj}=\mathbf{133}$,
  $\rho_{\omega_2}=\mathbf{8645}$, $\rho_{\omega_3}=\mathbf{365750}$,
  $\rho_{\omega_4}=\mathbf{27664}$, $\rho_{\omega_5}=\mathbf{1539}$,
  $\rho_{\omega_6}=\rho_{\rm min}=\mathbf{56}$, $\rho_{\omega_7}=\mathbf{912}$.}
\label{fig:dynke7}

\end{figure}

The Dynkin diagram of $\cG=E_7$ is represented in Figure~\ref{fig:dynke7}. We
write $\rho_{\rm min} \triangleq \mathbf{56}= \rho_{\omega_6}$ for the minimal irreducible
representation attached to the highest weight $\omega_6$, which is self-dual.

\subsubsection*{Step \textbf{A1}$'$}

The computation of $\mathfrak{p}_k^{E_7}$ can be performed exactly as for
the $E_6$ case. The anti-symmetric characters $\chi_{\Lambda^k\rho_{\omega_6}}(g)$ can be
computed e.g. from the explicit matrix representation of the exponentiated Cartan matrices of
$\rho$ \cite{MR1823074} via Newton identities. Also, as before, the fundamental
characters $(\chi_{\omega_k}(g))_{k = 1}^6$ are expressed via the characters of suitable
tensor powers of $\rho_{{\rm min}} = \rho_{\omega_{6}}$ and the character of the adjoint representation $\mathbf{133}=\mathrm{Adj}$:
\bea
\chi_{\omega_2} &=& \chi_{\Lambda^2 \mathrm{Adj}}- \chi_{\mathrm{Adj}}, \nn \\
\chi_{\omega_3} &=& \chi_{\Lambda^4 \omega_6}- \chi_{\Lambda^2 \omega_6}, \nn \\
\chi_{\omega_4} &=& \chi_{\Lambda^3 \omega_6}- \chi_{\omega_6}, \nn \\
\chi_{\omega_5} &=& \chi_{\Lambda^2 \omega_6}- 1.
\eea
The remaining fundamental character $\chi_{\omega_7}$ can be computed from the
following relation in ${\rm Rep}(E_7)$
\beq
\chi_{\omega_7}\chi_{\omega_2}= -(\chi_{\omega_5}+1)\chi_{\omega_6} -\chi_{\omega_4}+ \chi_{\omega_4} \chi_{\omega_1} + \chi_{\omega_6} \chi_{\mathrm{Sym}^2 \omega_6}\,,
\label{eq:u7}
\eeq
where the right-hand side can be computed again using Newton identities. Once this is done, the
relations $\mathfrak{p}_k^{E_7}=\mathfrak{p}_{56-k}^{E_7}$ for  $k \in \llbracket 5,28 \rrbracket$ in ${\rm Rep}(E_7)$
can be read off by specializing the identity
\beq
\chi_{\Lambda^k \rho_{\omega_6}}(g)=
\mathfrak{p}_k^{E_7}[u_1(g), \dots, u_7(g)]
\eeq
to a suitably large number of sample points $g$, and then solving for the coefficients of
$\mathfrak{p}_k^{E_7}$. We obtain for example that 
\beq
\bary{rcl}
 \mathfrak{p}_5^{E_7} &=& -\left(u_1-1\right) u_4+\left(-u_1^2+u_1+u_2+u_5+1\right) u_6+u_2 u_7, \\
 \mathfrak{p}_6^{E_7} &=& -2 u_1^3+\left(1-2 u_5\right)
 u_1^2+\left(u_6^2-u_7 u_6+u_7^2+4 u_2-2 u_3+2 u_5+2\right)
 u_1+u_2^2+u_5^2-u_3+2 u_5 \\ &+& 2 u_2 \left(u_5+1\right)+u_4 u_6-u_4 u_7-u_6
   u_7+1, \\
 \mathfrak{p}_7^{E_7} &=& u_4 \left(-u_1^2+u_1+u_2+2
 u_5+2\right)+\left(-u_1^3+\left(2 u_2+u_5+3\right) u_1+2 u_2-2
 u_3+u_5+1\right) u_6 \\ &+& u_7 \left(-2 u_1^2+\left(u_2-2 u_5+1\right)
   u_1+u_7^2+3 u_2-3 u_3\right), \\
 \mathfrak{p}_8^{E_7} &=&
\left(u_3-2 u_5-u_6 u_7+2\right) u_1^2+\left(2 u_6^2+u_4 u_6-2 u_7 u_6+u_7^2+4
u_2-4 u_3+2 u_5-2 u_4 u_7\right) u_1\\
&+& u_2^2+2 u_4^2+u_6^2+u_5 u_7^2+u_7^2-2
   u_3-3 u_3 u_5+3 u_4 u_6+u_4 u_7-u_5 u_6 u_7+u_6 u_7 \\ &+& u_2
   \left(u_6^2+u_7 u_6+u_7^2-2 u_3+2 u_5\right) -2 u_1^3, \\
\dots
\eary
\eeq
The expressions up to $k = 28$ are lengthy and are omitted here, but they are available upon request.
This completes Step~\textbf{A1}$'$.

\subsubsection*{Step \textbf{A2}$'$}

\begin{figure}[t]
\centering
\includegraphics[scale=0.4]{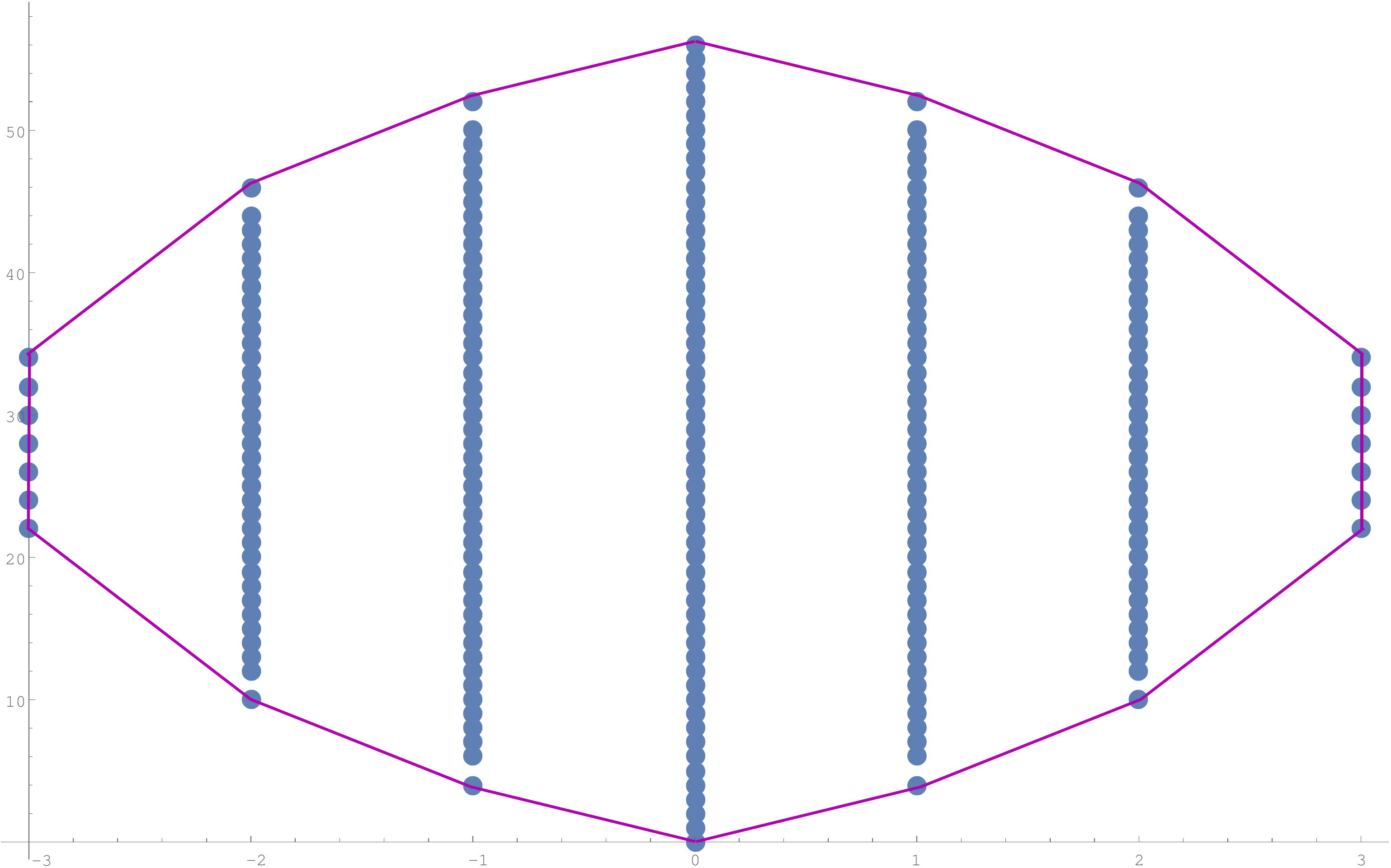}
\caption{The Newton polygon of the periodic relativistic Toda curve for
  $\cG^{\#}=E_7^{\#}$ and $\rho_{{\rm min}} =\mathbf{56}$.}
\label{fig:npole7}
\end{figure}

As before, the spectral parameter dependence of
$(\tilde{u}_i(X))_{i = 3}^6$ can be computed from
\eqref{eq:lax2} using
Newton identities applied to its explicit representation in terms of $56\times
56$ matrices. The same holds true for $(\tilde{u}_i(X))_{i = 1,2}$ and explicit
$133$-dimensional adjoint matrices.
Finally, $\tilde{u}_7(X)$ can be computed from
\eqref{eq:u7}. We obtain 
\beq
\bary{rclr}
\tilde{u}_i(X) &=& u_i, \quad & \quad i \neq 3,\\
\tilde{u}_3(X) &=& u_3+ u_0(X+ 1/X),
\eary
\eeq
in terms of the Casimir function $u_0^{-1}=\varkappa_0^{1/2}\varkappa_1^2 \varkappa_2^2 \varkappa_3 ^3 \varkappa_4^4 \varkappa_5^3
\varkappa_6^2 \varkappa_7$. The $E_7^{\#}$-Toda curve is then computed as:
\beq
\label{csaufn}0=\cP_{E_7^\#}^{{\rm Toda}}(X,Y) =
\sum_{k=0}^{56}\mathfrak{p}^{E_7}_k\big[u_1,u_2,u_3+u_0(X + 1/X),u_4,u_5, u_6,u_7\big]\,Y^k .
\eeq
The resulting Newton polygon is depicted in Figure~\ref{fig:npole7}.

\subsubsection*{Step \textbf{B}$'$}

Here we have $\DD'= E_6$ while we started with $\DD= E_7$. The embedding of the maximal tori

\beq
\bary{cccl}
\iota: & \cT_{E_6} & \longrightarrow & \cT_{E_7} \nn \\
& (Q_1,Q_2,Q_3,Q_4,Q_5,Q_6) & \longmapsto & (Q_1,Q_2,Q_3,Q_4,Q_5,1,Q_6)
\eary
\eeq
obtained upon projection onto the rank $6$ weight sublattice generated by $\omega_i$,
$i\neq 7$ gives rise to the following decomposition of the fundamental weight modules:
\bea
\label{eq:u1e7}
\rho_{\omega_1}&=& (\irrep{1})\oplus(\irrep{27})\oplus(\irrepbar{27})\oplus(\irrep{78}), \\
\rho_{\omega_2}&=& 
(\irrep{27})\oplus(\irrepbar{27})\oplus2(\irrep{78})\oplus2(\irrep{351})\oplus2(\irrepbar{351})\oplus(\irrep{\
650})\oplus(\irrep{1728})\oplus(\irrepbar{1728})\oplus(\irrep{2925}), \\
\rho_{\omega_3}&=& 
(\irrep{78})\oplus3(\irrep{351})\oplus3(\irrepbar{351})\oplus2(\irrep{650})\oplus3(\irrep{1728})\oplus(\irrep{2430})\oplus5(\irrep{2925})\oplus(\irrep{5824})\oplus(\irrepbar{5824})
\nn \\ &\oplus&
3(\irrep{7371})\oplus3(\irrepbar{7371})\oplus2(\irrep{17550})\oplus2(\irrepbar{17550})\oplus(\irrep{34749})\oplus2(\irrep{51975})\oplus(\irrep{70070}), \\
\rho_{\omega_4}&=&
(\irrep{27})\oplus(\irrepbar{27})\oplus2(\irrep{78})\oplus3(\irrep{351})\oplus3(\irrepbar{351})\oplus2(\irrep{650})\oplus
(\irrep{1728})\oplus(\irrepbar{1728})\nn \\ &\oplus&
2(\irrep{2925})\oplus(\irrep{7371})\oplus(\irrepbar{7371}) , \\
\rho_{\omega_5}&=&(\irrep{1})\oplus2(\irrep{27})\oplus2(\irrepbar{27})\oplus(\irrep{78})\oplus(\irrep{351})\oplus(\irrepbar{351})\oplus(\irrep{650}) , \\
\label{eq:u6e7}
\rho_{\omega_6}&=&
2(\irrep{1})\oplus(\irrep{27})\oplus(\irrepbar{27}), \\
\label{eq:u7e7}
\rho_{\omega_7}&=&
(\irrep{27})\oplus(\irrepbar{27})\oplus2(\irrep{78})\oplus(\irrep{351})\oplus(\irrepbar{351}).
\eea 
In particular, with \eqref{eq:u6e7}:
\bea
\label{eq:PE7toE6}
\cP_{E_7^\#}^{{\rm Toda}}(X,Y)\big|_{u\in\tilde{\mathfrak{U}}_{(E_7,E_6)}} &=& (Y-1)^2
\det\big[Y\mathbf{1}-\rho_{\mathbf{27}}(L^{E_7^\#}_{\mathsf{w}})\big]
\det\big[Y\mathbf{1}-\rho_{\overline{\mathbf{27}}}(L^{E_7^\#}_{\mathsf{w}})\big].
\eea
The variety $\tilde{\mathfrak{U}}_{(E_7,E_6)}=( u \circ \iota)(\cT')$, by
\eqref{eq:u1e7}-\eqref{eq:u7e7}, can be parametrized as the image of the
morphism $u :\bbC^6_U\to \mathfrak{U}_{E_7}$ given by 
\bea
 u_1 &=& U_1+U_5+U_6+1, \nn \\
 u_2 &=& U_2+U_3+U_4+U_1 U_5+U_1 U_6+U_5 U_6+U_6-1, \nn \\
 u_3 &=& -U_1^2+U_3 U_1+U_4 U_1-U_1-U_5^2+U_2+4 U_3+U_2 U_4 \nn \\ &+& U_4+U_2 U_5+U_3
 U_5-U_5+U_2 U_6+U_4 U_6-1, \nn \\
 u_4 &=& U_5 U_2+2 U_2+2 U_3+U_1 U_4+2 U_4+2 U_1 U_5-2, \nn \\
 u_5 &=& U_5 U_1+2 U_1+U_2+U_4+2 U_5, \nn \\
 u_6 &=& U_1+U_5+2, \nn \\
 u_7 &=& U_1+U_2+U_4+U_5+2 U_6.
\label{eq:uE7toE6}
\eea
This is not ruled however with respect to $u_{\rm rul}=u_3$, nor can it be
expected that the factorization \eqref{eq:PE7toE6} give polynomial factors with respect
to the spectral parameter $X$, let alone have the $\bbZ/2\bbZ$ symmetry of
\eqref{eq:PmmE7}. The first problem is solved as follows: introduce coordinates $(\tilde U_i)_{i = 1}^6$ and $U_{\rm rul}$ to
parametrize the maximal ruled subvariety $\tilde{\mathfrak{U}}_{(E_7,E_6)}\times \bbC_{U_{\rm rul}}$ of $\bbC^6_U$ w.r.t. to $u_{\rm
  rul}$; we are assuming at this stage the latter to be of dimension
higher than zero, with $U_{\rm rul}$ a curvilinear coordinate in the
distinguished ruling direction.
Imposing now that the factorization of $(Y-1)^2$ is preserved by
shifts along $u_{\rm rul}$ has the effect of restricting \eqref{eq:uE7toE6} to
$U_5=U_1=\tilde U_1$. Furthermore, the requirement that the functions
$u_i:\tilde{\mathfrak{U}}_{(E_7,E_6)}\to \bbC$ in \eqref{eq:uE7toE6} have vanishing
derivative along the distinguished direction $U_{\rm rul}$ is satisfied upon
setting, without loss of generality, $U_2=\tilde U_2+U_{\rm rul}$, $U_4=\tilde
U_2-U_{\rm rul}$, $U_3=\tilde U_3$, $U_6=\tilde U_4$. Equating now the third
line of \eqref{eq:uE7toE6} to the shifted Casimir $u_3+X+u_0/X$ as it appears in \eqref{csaufn}
sets $U_{\rm rul}=\pm\ri \sqrt{X+u_0/X}$ with no additional constraints on
$\mathfrak{U}_{E_7,E_6}$, which thus turns out to have dimension equal to
four. Restricting to $\mathfrak{U}_{E_7,E_6}$ therefore attains the
factorization \eqref{eq:PE7toE6}, which is {\it a fortiori} polynomial in
($X$, $X^{-1}$). The question of the $\bbZ/2\bbZ$ symmetry in $X$ is automatically solved by the fact
that replacing $\mathbf{27}$ with $\overline{\mathbf{27}}$ is tantamount to
switching $U_1 \leftrightarrow U_5$, $U_2 \leftrightarrow U_4$; on $\mathfrak{U}_{(E_7,E_6)}$ this reads
$U_{\rm rul}^2 \leftrightarrow -U_{\rm rul}^2$, which is just
$X\leftrightarrow -X$. Restricting to $\mathfrak{U}_{(E_7,E_6)}$ is thus
necessary and sufficient to have the factorization in polynomials of $X$ with
the desired $(\mathbb{Z}/2\mathbb{Z})$-symmetry.

\subsubsection*{Comparison with the LMO spectral curve}

By direct computation of the left-hand side of \eqref{idM} with \eqref{eq:PmmE7} and the table of coefficients given in Appendix~\ref{E7sp}, we find that the equality:
\beq
\mathcal{P}^{{\rm Toda}}_{E_7^{\#}}(X,Y,u)\big|_{u \in \mathfrak{U}_{(E_6,D_4)}} = (Y - 1)^2 \mathcal{P}^{{\rm LMO}}_{E_7}(X,Y;\cM,\lambda),\qquad c = e^{\lambda/72}\,,
\eeq
is realized by $u_0=-1/c^{12}$ and the morphism 
\bea
\tilde U_1 &=& -\frac{6 \mu _2}{c^3}-1,\nn \\
\tilde U_2 &=& \frac{6 \mu _5}{c^6}-\frac{12 \mu _3}{c^4}+\frac{6 \mu
  _2}{c^3}+2,\nn \\
\tilde U_3 &=& \frac{-2 c^8-24 c^5 \mu _2+24 c^4 \mu _3+36 c^2 \left(\mu
  _2^2-\mu _5\right)+144 c \mu _2 \mu _3+12 \left(2 \mu _3^2+\mu
  _7\right)}{c^8},\nn \\
\tilde U_4 &=& \frac{12 \mu _3}{c^4}+\frac{12 \mu _2}{c^3}-1.
\label{UpE7} 
\eea

\subsection{$E_8$}
\label{E8string}

The Dynkin diagram of $\cG=E_8$ is represented in Figure~\ref{fig:dynke7}. We
write $\rho_{{\rm min}} \triangleq {\rm Adj} = \mathbf{248}=\rho_{\omega_7}$ for the minimal irreducible
representation attached to the highest weight $\omega_7$: this is the adjoint
representation.

\begin{figure}[!h]
\centering
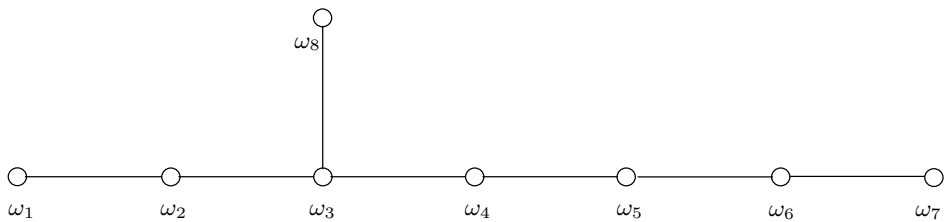
\caption{
The Dynkin diagram of $\cG=E_8$. We have $\rho_{\omega_1}=\mathbf{3875}$,
$\rho_{\omega_2}=\mathbf{6696000}$, $\rho_{\omega_3}=\mathbf{6899079264}$,
$\rho_{\omega_4}=\mathbf{146325270}$, $\rho_{\omega_5}=\mathbf{2450240}$,
$\rho_{\omega_6}=\mathbf{30380}$, $\rho_{\omega_7}=\mathbf{248}=\mathrm{Adj}$, $\rho_{\omega_8}=\mathbf{147250}$.}
\label{fig:dynke8}
\end{figure}

\subsubsection*{Step \textbf{A1}$'$}
Step~\textbf{A1}$'$ is the hardest bit here, and the one we could not complete entirely as the size
of the linear systems appearing in the calculation of $\mathfrak{p}_k^{E_8}$ grows uncontrollably all the way up to $k=124$. As a result, we do not
have a closed-form expression for $\mathfrak{p}_k^{E_8}$ but for the first few
orders, and in turn we could not find an explicit
expression for $\cP_{E_8^{\#}}$ for arbitrary values of $u_0, \dots,
u_8$. However, if we are interested in {\it any given} specific point $\overline{u}\in
\mathfrak{U}_{E_8}$, and in particular those with integer values for $\bar u_1, \dots,
\bar u_8$,
the value of $\mathfrak{p}_k^{E_8}$ at that point can be easily computed in
finite time, as
follows. Let 
 $(Q_i)_{i = 1}^8$ be exponential coordinates on
the maximal torus coming from linear coordinates on ${\rm Lie}(E_8)$. Then for a given group element $g$, $u_i=\chi_{\omega_i}(g)$ are
Laurent polynomials in the variables $Q_i$, and so is $\chi_{\Lambda^k({\rm Adj})}(g)$ for
any $k$: the latter in particular can be computed explicitly via Newton identities. For a given $\overline{u}$, let $\overline{Q}$ be any root of
the system of algebraic equations $\chi_{\omega_i}(g)=\overline{u}$. Plugging
$\overline{Q}$ into the expression of $\chi_{\Lambda^k({\rm Adj})}(g)$ then returns
$\mathfrak{p}_k^{E_8}\big|_{u=\overline{u}}$.\\

For generic $\overline{u}$, it is hopeless to find a manageable expression
of $\overline{Q}$ above that could yield a closed analytic expression for
$\mathfrak{p}_k^{E_8}\big|_{u=\overline{u}}$. However, if $\overline{u} \in
\bbZ^8$, a sensible thing to do is to find $\overline{Q}$ numerically to a
good accuracy, and then plug the result into the expression of $\chi_{\Lambda^k
  ({\rm Adj})}(g)$ as a Laurent polynomial in $\{Q_i\}_{i = 1}^8$: since the latter is on
general grounds a polynomial in $(u_i)_{i = 1}^8$ with integer coefficients, 
it follows that $\chi_{\Lambda^k({\rm Adj})}(g)|_{Q=\overline{Q}}\in \mathbb{Z}$. A reliable
integer rounding of the numerics gives then a prediction for the {\it exact} expression of $\mathfrak{p}_k^{E_8}\big|_{u=\overline{u}}$. We will provide an example of this procedure shortly.

\subsubsection*{Step \textbf{A2}$'$}

\begin{figure}[t]
\centering
\includegraphics[scale=0.4]{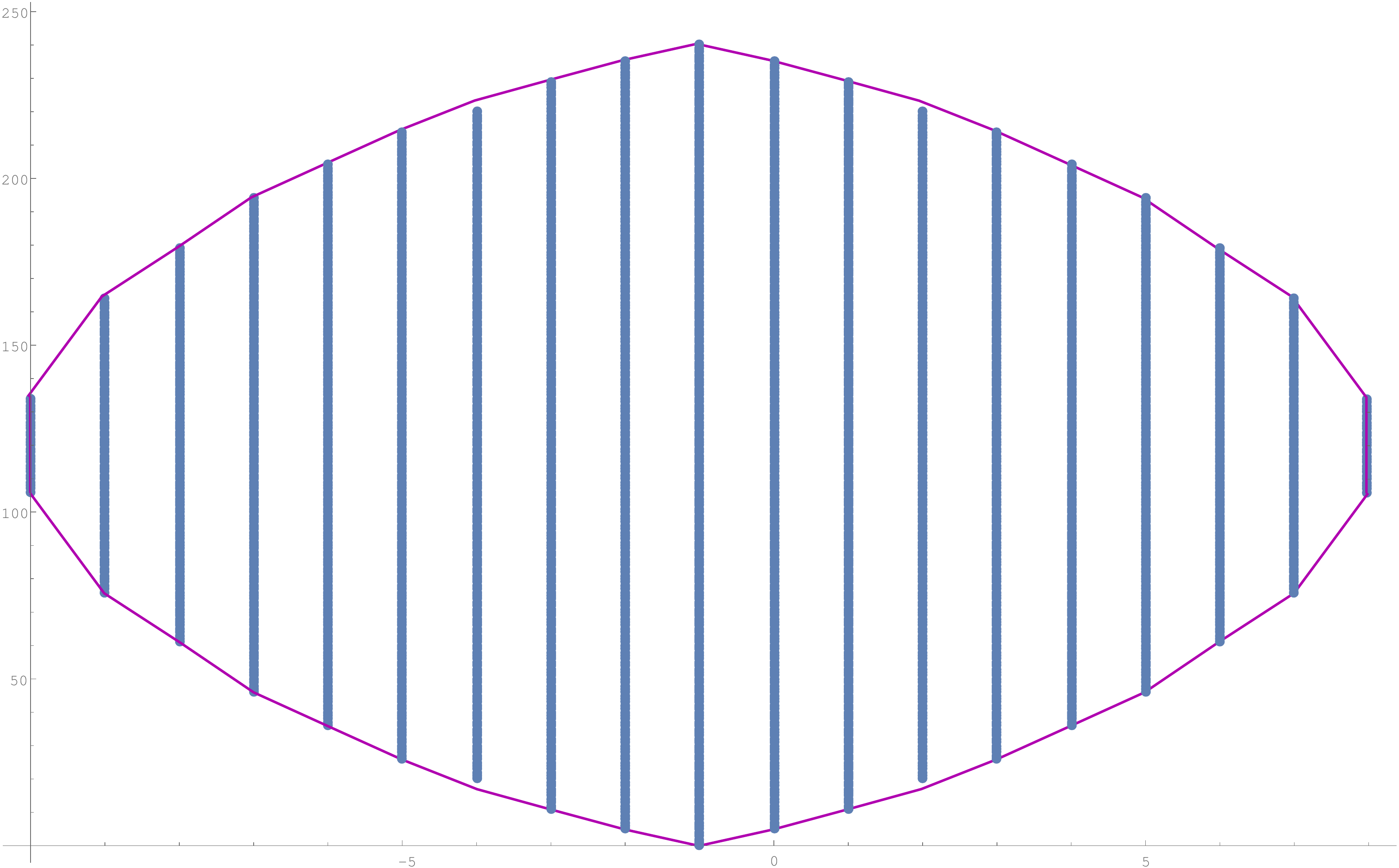}
\caption{The Newton polygon of the Toda spectral curve for
  $\cG^{\#}=E_8^{\#}$ and $\rho =\mathbf{248}$.}
\label{fig:npole8}
\end{figure}

The spectral parameter dependence of
$(\tilde{u}_i(X))_{i = 3}^7$ can be computed from
\eqref{eq:lax2} using Newton identities applied to its explicit representation in terms of $248\times
248$ matrices. We obtain
\beq
\bary{rclr}
\tilde{u}_i(X) &=& u_i, \quad & \quad i \in \llbracket 4,7 \rrbracket\\
\tilde{u}_3(X) &=& u_3+ u_0(X+ 1/X).
\eary
\eeq
in terms of the Casimir function $u_0^{-1}=\varkappa_0^{1/2}\varkappa_1^2 \varkappa_2^4 \varkappa_3 ^{6} \varkappa_4^{5} \varkappa_5^4
\varkappa_6^3 \varkappa_7^2 \varkappa_8^3$. Furthermore, a quick computation of the character relations $\mathfrak{p}_k^{E_8}$
for $k = 6,7,8$ reveals that $\tilde{u}_i(X)=u_i$ for $i=1,2,8$ as well. The $E_8^{\#}$-Toda curve is then computed as
\beq
0=\cP_{E_8^\#}^{{\rm Toda}}(X,Y; u) =
\sum_{k=0}^{248} \mathfrak{p}^{E_8}_k\big[u_1,u_2,u_3+u_0(X
+ 1/X),u_4,u_5, u_6,u_7, u_8\big]\,Y^k.
\label{eq:PE8v}
\eeq
The polynomials $\mathfrak{p}^{E_8}_k\big[\overline{u}_1,\overline{u}_2,\overline{u}_3+u_0(X
+ 1/X),\overline{u}_4,\overline{u}_5, \overline{u}_6,\overline{u}_7,
\overline{u}_8\big]$ at $u_i=\overline{u}_i$, $i \in \llbracket 1,8 \rrbracket$ can be computed by
interpolation of
$\mathfrak{p}^{E_8}_k\big[\overline{u}_1,\overline{u}_2,\overline{u}_3+n,\overline{u}_4,\overline{u}_5, \overline{u}_6,\overline{u}_7,
\overline{u}_8\big]$ for $n \in \llbracket 0,T \rrbracket$ with $T$ big enough. It turns out that the interpolation stabilizes at $T=9$. The resulting Newton polygon is depicted in Figure~\ref{fig:npole8}.

\subsubsection*{Step \textbf{B}$'$}

The computational strategy of Step~\textbf{A1}$'$ above only allows us to compute
$\cP_{E_8^\#}^{{\rm Toda}}$ at a fixed moduli point $u = \overline{u}$, leaving only $u_0$ unrestricted. This nevertheless leaves some limited space for
universal predictions, that are in particular relevant for comparison with
$\cP^{{\rm LMO}}_{E_8}$. \\

Firstly, as $\cP_{E_8^\#}^{{\rm Toda}}$ is a characteristic polynomial
in the adjoint representation, we automatically have a factor of
$(Y-1)^8$ pulling out. Factoring out this component leaves us
with a degree-240 polynomial $\tilde{\cP}_{E_8^{\#}}^{{\rm Toda}}$ in $Y$, as expected
from Table~\ref{Secondtab}. Secondly, notice that $T=9$ computed in Step~\textbf{A2}$'$ matches
with the fact that $\deg_{X} \mathcal{P}^{{\rm LMO}}_{E_8} =18$ in the same table. Thirdly, the
palindromic symmetry \eqref{eq:palin} is automatically enforced by
\eqref{eq:PE8v} and the reality of $\rho_{{\rm min}}$,
so that $\mathfrak{p}_k^{E_8}=\mathfrak{p}_{248-k}^{E_8}$.
Fourthly, in view of all preceding examples, it
is natural to assume that the coefficients of the monomials corresponding to
the vertical boundaries should depend on $u_0$ only. Setting $u_0=-1/c^{30}$, we get:
\beq
[X^9]\,\tilde{\cP}_{E_8^\#}^{{\rm Toda}} = [X^{-9}]\,\tilde{\cP}_{E_8^\#}^{{\rm Toda}}
=-c^{240} Y^{106} (Y+1)^2 \left(Y^2+Y+1\right)^3
\left(Y^4+Y^3+Y^2+Y+1\right)^5.
\eeq
This is precisely the vertical slope polynomial of the LMO curve given in Appendix~\ref{E8bound}.

\subsection{The LMO slice and the conifold point}

\label{sec:LMOslice}

The spirit of our calculations so far has been the following: we employed the orbit
analysis on the LMO side to enforce the Galois group reduction
 $\cG \to \cG'$ on $\cP_{\cG^\#}^{\rm Toda}$, as well as its compatibility with the affine deformation by the spectral parameter $X$ -- i.e.  we imposed that the factors of $\cP_{\cG^\#}^{\rm Toda}$ when this reduction are polynomials in $X$. It is quite remarkable that such limited piece of data, without any detailed input
from the matrix model, allowed us to establish Points (a)-(c) of
Conjecture~\ref{conj:spec} -- with the sole exception so far of
$\cG=E_8$. However, it would have been desirable to {\it predict} the restriction $u_i(\lambda)$
of the Toda action variables relevant for Conjecture~\ref{conj:spec}, based on
considerations purely within the dual A- and B-model, instead of deriving them
{\it a posteriori} from the comparison with the matrix model curve.\\

\subsubsection{Toric case}

For $\cG=A_{p-1}$, a complete interpretation of the
LMO restriction can be obtained from the Halmagyi--Yasnov solution of the Chern--Simons matrix model
in a generic Chern--Simons vacuum. Let
$[\widehat{X}^{\bbZ/p\mathbb{Z}}]=[\cO_{\mathbb{P}^1}(-1)^{\oplus 2}/(\bbZ/p\bbZ)]$ be
the $\bbZ/p \bbZ$ fiberwise orbifold of the resolved conifold: its coarse
moduli space is the GIT quotient arising from the stability conditions in the
maximally singular chamber of the secondary fan of $Y^{\bbZ/p\mathbb{Z}} \to
\widehat{X}^{\bbZ/p\mathbb{Z}}$, which contains \cite{Aganagic:2002wv} the
$\mathtt{t}=0$ point of Chern--Simons theory (see Section~\ref{conjs}). The space of marginal
deformations of the A-model chiral ring -- i.e. the degree-two Chen--Ruan cohomology of
$[\widehat{X}^{\bbZ/p\mathbb{Z}}]$ -- is parametrized by linear coordinates $t=\{t_{\rm B},
(\tau_{i/p})_{i=1}^{p-1})\}$: here $t_{\rm B}$ is dual to the K\"ahler class
$c_1(\cO_{\bbP^1}(1))$, and $\tau_{i/p}$ are dual to degree zero classes in
orbifold cohomology with fermionic age-shift \cite{Zaslow:1992rp} equal to one. A local analysis of the GKZ
system around $t=0$ then shows that the mirror map in the twisted sector
behaves asymptotically as
\beq
\tau_{i/p}=\cO\l(u_k^{i/(kp)}\r),
\eeq
for all $k$ such that $i k=p$. Therefore, the LMO restriction $u_i=0$, $u_0=-\re^{-t_{\rm B}/2}$
amounts to switching off the insertion of twisted classes, retaining only the
geometric modulus $t_{\rm B}=-2\log(c)=-\lambda/p(p+1)$. This cuts out a 1-dimensional slice of the orbifold
chamber of the 
K\"ahler moduli space of $[\widehat{X}^{\bbZ/p\bbZ}]$, containing two
distinguished boundary points: 
$c=0$ ($\Re (t_{\rm B})=+\infty$), corresponding geometrically to the large radius limit point in this orbifold phase
(that is, the decompactification $(\bbZ/p\bbZ)\backslash\bbC^2 \times \bbC
\hookrightarrow [\widehat{X}^{\bbZ/p\bbZ}]$), and the point $c=1$ ($t_{\rm
  B}=0$) to $X_{[0]}^{\bbZ/p\mathbb{Z}}$, the $\bbZ/p\mathbb{Z}$-orbifold of the conifold singularity.\\

\subsubsection{Non-toric cases}

\begin{figure}[t]
\centering
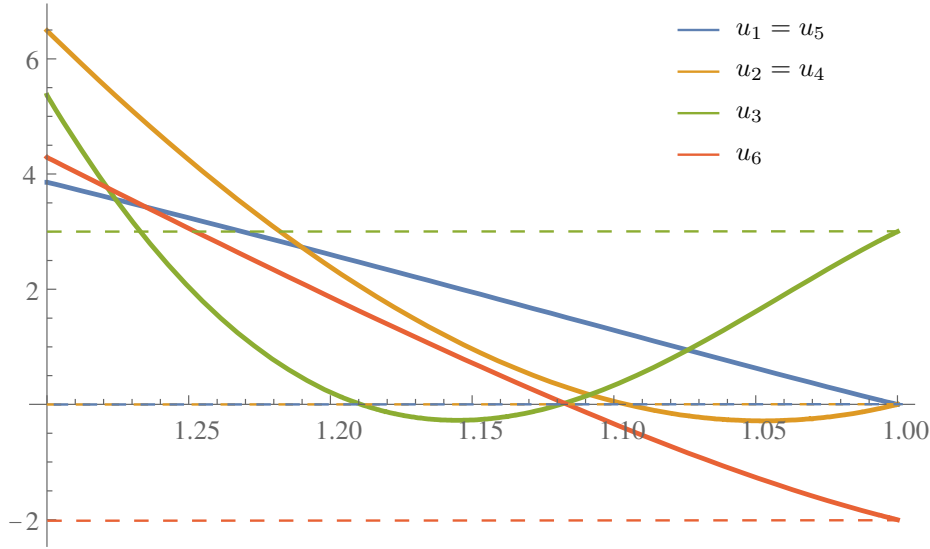
\caption{The LMO slice $u_i(\lambda)$ as a function of $c = \re^{\lambda/72}$ for $\cG=E_6$. Continuous lines depict the prediction
  for the Toda Hamiltonians upon restriction to the LMO slice. Dashed lines
  represent their value on the untwisted slice, where all orbifold moduli
  have been turned off. It is seen that the two disagree away from the
  conifold point $c=1$.}
\label{fig:LMO}
\end{figure}
A similar identification does not hold for $\cG=D, E$, and we are unable to
offer a poignant stringy interpretation of the LMO slice here. In this case $\de_c
u_i(c)\neq 0$ (Figure~\ref{fig:LMO}), 
and as a consequence the LMO slice does not correspond to the
slice where the twisted orbifold moduli are set to zero; it can also readily
be seen that the limit $c\to
0$
is a different limit point from the
(orbifold)
large radius point corresponding to $\Gamma\backslash\bbC^2 \times \bbC$.
However, there is at least one special moduli point where we must be able to offer a stringy
prediction of the LMO curve with no input from the matrix model: this is the weak 't Hooft limit $c\to 1$, which
should correspond to the point in the extended K\"ahler moduli space corresponding to the
$\Gamma$-orbifold of the singular conifold, $X_{[0]}^{\Gamma}$, where we have
contracted the exceptional curves on each $\widehat{\Gamma\backslash\bbC^2}\to \bbP^1$
fiber and we further blow-down the base $\bbP^1$.
For this point we do have a stringy prediction for the value of $u_i$: the
Bryan--Graber form of the Crepant Resolution Conjecture \cite{MR2483931} indeed
predicts that the condition of contracting the fibers takes the form, in
exponentiated linear coordinates $Q$ on the Cartan torus $\cT$,
\beq
Q_i=\exp\l(\frac{2\pi\ri l_i}{|\Gamma|}\r),
\eeq
where $l_i$ is the $i^{\rm th}$-component of the highest root of $\cG$ in the
$\omega$-basis (equivalently, the dimension of the corresponding irreducible
$\Gamma$-module). This sets the Toda actions $u_i$ to the values shown
in Table~\ref{tab:conifold}. As far as the K\"ahler modulus of the base $\bbP^1$ is concerned, this is
related to the Casimir as $u_0=-\re^{-t_{\rm B}/2}$, hence $u_0=-1$ is the conifold
limit. The conifold B-model curves can then be computed for
$\cG=A_{p-1},D_{p+2},E_6,E_7$ simply by restriction of the results of Section~\ref{Anstring}-\ref{E7string} to $u_0 = -1$. Furthermore, since we are sitting at a specific point in the
moduli space as per Table~\ref{tab:conifold}, we can fully compute
 the conifold Toda
spectral curve for $\cG=E_8$ upon employing the methods of Section~\ref{E8string}. The
results are given in the third column of Table~\ref{tab:conifold} below.\\

\begin{table}[!h]
\centering
\beq
\begin{array}{|c|c|c|}
\hline
\cG & {\rm character}\,\,{\rm values} & \cP^{\rm Toda}_{\cG^\#}=0  \\
\hline
A_{p-1} & u=\epsilon=(0,\dots, 0) & X+Y^p X^{-1}=Y^p+1 \\
\hline
D_4 & \epsilon=(0,-2,0,2) & X+X^{-1}=(Y^2+Y^{-2}) \\
\hline
D_{p+2} &  \epsilon=(0,(-1)^{p + 1}2,0,(-1)^p,0,\dots,0) & X+X^{-1}=(Y^p+Y^{-p})\\
\hline
E_6 & u=(0,0,3,0,0,-2) & \left(Y^3-1\right)^3 \left(X-Y^3\right)
\left(X Y^3-1\right) \left(X+Y^6\right) \left(X Y^6+1\right)=0 \\
\hline
E_7 & u=(-2,3,-3,0,1,0,0) & \bary{c}(X+1)^2 (X^2+1)^4 (X^4+Y) (X^6+Y) (X^{12}+Y)\\ (X^4 Y+1) (X^6 Y+1) (X^{12} Y+1)=0\eary\\
\hline
E_8 & u=(1,3,0,3,-3,3,-2,-2) & 
\bary{c} (Y+1)^2 \left(Y^2+Y+1\right)^3 \left(Y^4+Y^3+Y^2+Y+1\right)^5
\\ \left(X+Y^5\right) \left(X Y^5+1\right) \left(X-Y^6\right) \left(X Y^6-1\right)
   \left(X-Y^{10}\right)^2 \\ \left(X Y^{10}-1\right)^2  \left(X^2-Y^{15}\right)
   \left(X+Y^{15}\right)^2  \left(X Y^{15}+1\right)^2 \\ \left(X^2 Y^{15}-1\right)
   \left(X-Y^{30}\right) \left(X Y^{30}-1\right)(Y-1)^8=0
\eary
\\
\hline
\end{array}
\nn 
\eeq
\caption{The values of the B-model moduli at the $\Gamma$-orbifold of the
  conifold point and the corresponding spectral curves. It  corresponds to $u_0 = -1$. Here, $(u_i)_{i = 1}^R$ are the regular fundamental characters (for $\cG=A,E$), and
  $(\epsilon_i)_{i = 1}^R$  are the antisymmetric characters of the
  defining representation (for $\cG=A,D$).}
\label{tab:conifold}
\end{table}

As far as the LMO matrix model is concerned, the small 't~Hooft limit is the
$\lambda \rightarrow 0$ limit, for which the large $N$ spectral curve can be
fully determined as we have seen in Section~\ref{eq:lto0}. Using the definition of
$\mathcal{P}^{{\rm LMO}}_{\mathcal{D}}$ in terms of $\mathcal{P}_{v}$ given in
Section~\ref{S42}, we find exact agreement between
\eqref{eq:smthooft1}-\eqref{eq:smthooft2} and the Toda spectral curves in Table~\ref{tab:conifold}.

\section{Outlook}

We would like to point out a few directions that our findings suggest to explore in relation with existing works.

\subsubsection*{The full GOV correspondence}

Perhaps the most immediate question is how to extend the LMO/topological
strings correspondence of this paper to the full Chern--Simons partition
function, so as to give a proof of at least the B-side of the general
Conjecture~\ref{conj:gen}. For the string side, the relevant family of
spectral curves was constructed in Section~\ref{Stringside}; the only missing
ingredient is the full $E_8$-curve, whose computation is currently under way \cite{GWADE}.
Most of the burden of the comparison is borne by the matrix model side;
a good starting point here should be given by the large~$N$ analysis of the matrix integral
expression of \cite{MarinoCSM,BT2}. Establishing an explicit solution of the loop
equations for this matrix model in terms of the topological recursion applied
on the corresponding Toda spectral curve would give a full proof of the
B-side of the GOV correspondence. \\

\subsubsection*{Implications for GW theory}

The A-side of the correspondence requires substantially more work. While it
should be feasible to derive explicit all-genus results e.g. by degeneration
techniques \cite{MR2411404}, one should probably work harder to see the Toda
spectral setup and the topological recursion emerge. Perhaps the best route
to follow here will hinge on deriving the $S$- and $R$-calibrations of the
quantum cohomology of $Y^\Gamma$ emerge from the steepest descent asymptotics
of the Toda spectral data, as in \cite{Brini:2013zsa}, and then retrieve the
topological recursion from Givental's $R$-action on the associated
cohomological field theory \cite{DuninBarkowski:2012bw}. This would lead to a
proof of the remodeling conjecture beyond the toric case. Along the way it
would be interesting to prove a gluing property of ADE invariants analogous to
the one enjoyed by the topological vertex. A thorough study of mirror symmetry for the case at hand in the limit $u_0\to
\infty$ will appear in \cite{GWADE}, where the implications for the Crepant
Resolution Conjecture will be explored in detail.

\subsubsection*{Implications for gauge theory}

The topological recursion method applied to the Toda curves gives us a glimpse of one slice of the
$\Omega$-background for the associated gauge theory -- namely, the one with
$\epsilon_1=-\epsilon_2$. It would be very interesting to investigate how
the study of the stationary states of their quantized version -- which is itself an
open problem beyond the $A$- case -- is related to the twisted superpotential of the gauge theory
in the Nekrasov--Shatashvili limit, $\epsilon_2=0$. Our construction of
Section~\ref{Stringside} should also embody the solution to the associated
$K$-theoretic instanton counting problem \cite{Gottsche:2006bm} for the $A$-
and $D$-series; it is natural to imagine, for example, that the extrapolation of
the blow-up equation of \cite{Gottsche:2006bm} to exceptional root systems
will be solved by the Eynard--Orantin/Nekrasov--Shatashvili free energies of
our B-model setup in the respective limits.

\subsubsection*{Seifert matrix model and DAHA?}

Our results suggests that there should exist observables in the finite $N$ Seifert
matrix model -- expressible, moreover, in terms of fiber knot invariants in a
spherical Seifert manifold -- providing a basis of solutions for the matrix
$q$-difference equation $\Psi(qX) = \rho_{\rm
  min}(L_{\mathsf{w}}^{\cG^{\#}}(X))\Psi(X)$. Then,
Proposition~\ref{propsimple} would be the manifestation of this equation in
the $\hbar = \ln q \rightarrow 0$ limit. However, a point of caution must be
raised, as here we are considering only the contribution of the trivial flat
connection. Therefore, we rather expect $q$-difference equations related to affine Toda to
be found for observables in the Seifert matrix model with discrete eigenvalues
-- as opposed to the matrix integral considered here, where the eigenvalues
are integrated over the real Cartan subalgebra of ${\rm SU}(N)$. It is likely
that the difference between the continuous and the discrete model has no
impact on the large $N$ limit. 

Etingof, Gorsky and Losev established in \cite[Corollary 1.5]{ELG} an expression for the colored HOMFLY polynomial of $(p,q)$ torus knots in $\mathbb{S}^3$ in terms of characters of the rational double affine Hecke algebra (DAHA) of type $A_{p - 1}$, which are also related to characters of equivariant $D$-modules on the nilpotent cone of ${\rm SL}(p)$. A similar relation for the categorification of the HOMFLY of torus knots was also conjectured in \cite{OblomKnot}, and proved in the uncategorified case. From the present work, we are tempted to think that $D$ and $E$ version (instead of $A_{p - 1})$ of these results should be expressed in terms of fiber knot invariants in the $D$ and $E$ Seifert geometries. Even independently of the knot theory interpretation, establishing that certain observables in Seifert matrix model satisfy exact (for finite $N$) and explicit $q$-difference equations produced by DAHA of type $D$ and $E$ would be extremely interesting.

\subsubsection*{3d-3d correspondence}

From the point of view of the $3d-3d$ correspondence \cite{3d3d},
Chern--Simons theory on $M^3$ with simply-laced gauge group
$\exp(\mathfrak{g})$ is dual to a $3d$ gauge theory ${\rm
  T}_{\mathfrak{g}}[M^3]$ on $X^3$: they simultaneously appear in
compactification of the $(2,0)$ $6d$ SCFT with Lie algebra $\mathfrak{g}$ 
on $M^3 \times X^3$. The moduli space of classical vacua for ${\rm T}_{\mathfrak{su}}[L(p,1)]$ on $Y^2 \times \mathbb{S}^1$ is related to the Bethe states in the $N$ particle sector of the XXZ integrable spin chain on $p$ sites \cite{GukovPei}. One can wonder if a direct and thorough relation can be found between the theories ${\rm T}_{\mathfrak{sl}}(\mathbb{S}^{\Gamma})$ for $\Gamma$ of type $D$ and $E$, and the $\mathcal{G}_{\Gamma}^{\#}$ classical Toda integrable system, e.g. via spectral dualities in integrable systems.

\appendix

\section{$\pi^1$ and $H_1$ of Seifert spaces}
\label{pi1list}

The fundamental group is \cite{Seifertbook}:
\beq
\pi_1(M^3) = \left\langle h,c_0,\ldots,c_r \quad \left| \quad \begin{array}{l} c_0h^b = 1\,, \\
 c_m^{a_m}h^{b_m} = [c_m,h] = 1\,,\,\,\,\,m \in \llbracket 1, r \rrbracket  \\ c_0\cdots c_r = 1\,.\end{array}\right.\right\rangle\,,
\eeq
where $h$ is the generator of a regular fiber, and $c_1,\ldots,c_r$ project to loops around the orbifold points in the base $\mathbb{S}^2$. Denoting $\Sigma$ this orbifold $\mathbb{S}^{2}$, its orbifold fundamental group:
\beq
\pi_{1}^{{\rm orb}}(\Sigma) = \Big\langle c_1,\ldots,c_r\quad \big| \quad c_1^{a_1} = \cdots = c_r^{a_r} = \prod_{m = 1}^r c_{m} = 1\Big\rangle
\eeq
fits in the exact sequence:
\beq
\mathbb{Z} \mathop{\longrightarrow}^{\iota} \pi_1(M^3) \longrightarrow \pi_1^{{\rm orb}}(\Sigma) \longrightarrow 1\,,
\eeq
where $\iota(\mathbb{Z})$ is the central subgroup generated by $h$. One can show that $\pi_1(M^3)$ is finite iff $\chi_{{\rm orb}} > 0$ and $\sigma \neq 0$, which we now assume. Combining the relations in $\pi_1(M^3)$, one can show that $h^{a|\sigma|} = 1$, but it can happen that the order of $h$ is smaller than $a|\sigma|$. The complete description of the finite fundamental groups appearing here was derived in \cite{MilnorS,Orlik} (see also \cite{Tomoda} where three misprints of the final list of \cite{Orlik} were corrected), and is summarized below. By Hurewicz theorem, the abelianization of $\pi_1(M^3)$ gives $H_1(M^3;\mathbb{Z})$. 

The conditions $\chi_{{\rm orb}} > 0$ and $\sigma \neq 0$ are satisfied only for $r = 1,2$ (lens spaces), and for $r = 3$ with the orders of exceptional fibers among $(2,2,p)$, $(2,3,3)$, $(2,3,4)$ or $(2,3,5)$. We introduce the binary polyhedral groups:
\begin{itemize}
\item[$\bullet$] ${\rm Q}_{4p}$ the binary dihedral group of order $4p$ (abelianization $\mathbb{Z}/4\mathbb{Z}$ if $p$ is odd, $(\mathbb{Z}/2\mathbb{Z})^2$ if $p$ is even),
\item[$\bullet$] ${\rm P}_{24}$ for the symmetry group of the tetrahedron (abelianization $\mathbb{Z}/3\mathbb{Z}$),
\item[$\bullet$] ${\rm P}_{48}$ for that of the octahedron (abelianization $\mathbb{Z}/2\mathbb{Z}$),
\item[$\bullet$] ${\rm P}_{120}$ that of the icosahedron (trivial abelianization).
\end{itemize}
Introduce also the groups:
\bea
{\rm B}_{2^{k}\cdot (2k' + 1)} & = & \big\langle x,y\quad \big|\quad x^{2^{k}} = y^{2k' + 1} = xyx^{-1}y = 1\big\rangle \nonumber \\
& \simeq & (\mathbb{Z}/(2k' + 1)\mathbb{Z}) \rtimes (\mathbb{Z}/2^{k}\mathbb{Z})\,, \nonumber \\
{\rm P}'_{3^{k}\cdot 8} & = & \big\langle x,y,z \quad \big| \quad x^2 = (xy)^2 = y^2 = zxz^{-1}y^{-1} = zyz^{-1}x^{-1}y^{-1} = z^{3^{k}} = 1\big\rangle \nonumber \\
& \simeq & {\rm Q}_{8} \rtimes (\mathbb{Z}/3^{k}\mathbb{Z})\,, \nonumber
\eea
and for the latter we have ${\rm P}'_{3\cdot 8} \simeq {\rm P}_{24}$. Their abelianizations are:
\beq
[{\rm B}_{2^{k}\cdot (2k' + 1)}]_{{\rm ab}} = \mathbb{Z}/2^{k}\mathbb{Z},\qquad [{\rm P}'_{3^{k}\cdot 8}]_{{\rm ab}} = \mathbb{Z}/3^{k}\mathbb{Z}\,.
\eeq

\vspace{0.2cm}

\noindent $\mathbf{(2,2,p)\,-}$\, $\pi_1^{{\rm orb}}(\Sigma) = {\rm Q}_{4p}$. Denote $s = p|\sigma|$. If $s$ is odd, then $\pi_1(M^3)$ is the direct product $(\mathbb{Z}/s\mathbb{Z}) \times {\rm Q}_{4p}$. If $s$ is even, then $p$ is odd and $4$ divides $s$; decompose $s = 2^{k + 1}s'$ with $s'$ odd; then $\pi_1(M^3)$ is a non-trivial central extension of ${\rm Q}_{4p}$, namely $(\mathbb{Z}/s'\mathbb{Z}) \times {\rm B}_{2^{k + 3}\cdot p}$.

\vspace{0.2cm}

\noindent $\mathbf{(2,3,3)\,-}$\, $\pi_1^{{\rm orb}}(\Sigma) = {\rm P}_{24}$. Denote $s = a|\sigma|$, and decompose $s = 3^{k - 1}s'$ with $s'$ coprime to $3$. If $k = 1$, then $b_2 = b_3 = 1$, $s$ is coprime with $6$ and $\pi_1(M^3)$ is the direct product $(\mathbb{Z}/s\mathbb{Z}) \times {\rm P}_{24}$. If $k \geq 2$, then $(b_2,b_3) = (1,2)$, $s'$ is coprime to $6$, and  $\pi_1(M^3)$ is rather a non-trivial central extension of ${\rm P}_{24}$, namely $(\mathbb{Z}/s'\mathbb{Z}) \times {\rm P}'_{8\cdot 3^{k}}$.

\vspace{0.2cm}

\noindent $\mathbf{(2,3,4)\,-}$\, $\pi_1^{{\rm orb}}(\Sigma) = {\rm P}_{48}$, $a = 12$, and $\pi_1(M^3)$ is the direct product $(\mathbb{Z}/12|\sigma|\mathbb{Z})\times {\rm P}_{48}$.

\vspace{0.2cm}

\noindent $\mathbf{(2,3,5)\,-}$\, $\pi_1^{{\rm orb}}(\Sigma) = {\rm P}_{120}$, $a = 30$, and $\pi_1(M^3)$ is the direct product $(\mathbb{Z}/30|\sigma|\mathbb{Z}) \times {\rm P}_{120}$. In particular, for $b = -1$, $b_1 = b_2 = b_3 = 1$, we obtain $30\sigma = 1$, thus $\pi_1(M^3) = {\rm P}_{120}$ and $H_1(M,\mathbb{Z}) = 0$. This is the Poincar\'e sphere, i.e the unique integer homology sphere with finite fundamental group.

\vspace{0.2cm}

\noindent In all cases, the order of $H_1(M^3,\mathbb{Z})$ is $(\prod_{m = 1}^r a_m)|\sigma|$.

\section{Group actions on $\mathbb{S}^3$}
\label{gpact}
It is known that all groups acting smoothly and freely on $\mathbb{S}^3$ act (up to diffeomorphism) as subgroups of ${\rm SO}(4,\mathbb{R})$, see e.g. the account in \cite{Zimmermann}. We now review elementary facts to explain the classification of finite subgroups of ${\rm SO}(4,\mathbb{R})$. Consider $\mathbb{S}^3$ as the unit sphere of the quaternions. Thus it forms a Lie group:
\beq
\mathbb{S}^3 \simeq \mathrm{Sp}(1,\mathbb{H}) \simeq \mathrm{SU}(2,\mathbb{C}) \simeq \mathrm{Spin}(3,\mathbb{R}).
\eeq
We have a degree $2$ covering of Lie groups:
\bea
\label{A4}\Phi\,:\,\mathrm{SU}(2,\mathbb{C}) & \longrightarrow & \mathrm{SO}(3,\mathbb{R}) \\
q & \longmapsto & (x \mapsto qxq^{-1}) \nonumber
\eea
where in the right-hand side, we consider the linear map restricted to the set of purely imaginary quaternions (it is stable by conjugation since it is the subset of $\mathbb{H}$ orthogonal to $1$). The squared norm of a quaternion is $\det(q)$, so the conjugation by $q$ is an isometry of the $3$-space. Since ${\rm SU}(2,\mathbb{C})$ is connected, this isometry always preserve orientation. Since elements of $\mathrm{SO}(3,\mathbb{R})$ are a fortiori angle-preserving isomorphism of the sphere, they determine elements of the automorphism group of the Riemann sphere $\mathbb{S}^2$, i.e. we have a group homomorphism:
\beq
\psi\,:\,\mathrm{SO}(3,\mathbb{R}) \longrightarrow \mathrm{PSL}(2,\mathbb{C}).
\eeq
This is more easily understood starting directly from $\mathrm{SU}(2,\mathbb{C})$, since we have the canonical degree $2$ covering of Lie groups:
\beq
\Psi\,:\,\mathrm{SU}(2,\mathbb{C}) \rightarrow \mathrm{PSU}(2,\mathbb{C}) \subseteq \mathrm{PSL}(2,\mathbb{C}),
\eeq
which factors $\Psi = \psi\circ\Phi$. It is not difficult to see that any finite subgroup of $\mathrm{PSL}(2,\mathbb{C})$ must be conjugated to a finite subgroup of $\mathrm{PSU}(2,\mathbb{C})$. There are 3 ways to describe elements in those groups. First, as rotations in $\mathbb{R}^3$ of angle $\theta$ around the unit vector $\vec{x}$:
\beq
R(\vec{v}) = \cos \theta\,\vec{v} + (1 - \cos\theta)(\vec{x}\cdot\vec{v})\vec{x} + \sin\theta\,\vec{x}\times\vec{v}
\eeq
where $\times$ is the vector product. Second, as $2 \times 2$ unitary matrices up to a sign:
\beq
A = e^{\frac{{\rm i}\pi}{2}\vec{x}\cdot\vec{\sigma}} = \cos(\theta/2)\mathbf{1} + {\rm i}\sin(\theta/2)\vec{x}\cdot\vec{\sigma} = \left(\begin{array}{cc} a & -\overline{b} \\ b & \overline{a} \end{array}\right).
\eeq
where $\vec{\sigma}$ is the vector of Pauli matrices:
\beq
\sigma_{1} = \left(\begin{array}{cc} 0 & 1 \\ 1 & 0 \end{array}\right),\qquad \sigma_{2} = \left(\begin{array}{cc} 0 & -{\rm i} \\ {\rm i} & 0 \end{array}\right),\qquad \sigma_{3} = \left(\begin{array}{cc} 1 & 0 \\ 0 & -1 \end{array}\right).
\eeq
Third, as M\"obius transformations:
\beq
z \mapsto \frac{az - \overline{b}}{bz + \overline{a}}.
\eeq
These $3$ descriptions complement each other.

Then, $\mathrm{SU}(2,\mathbb{C})$ acts by right and left multiplications on $\mathbb{S}^3$ (which are isometries on $\mathbb{S}^3$). As a matter of fact, we have a degree $2$ covering of Lie groups:
\bea
\label{PhiA} \Phi_{4}\,:\,\mathrm{SU}(2,\mathbb{C})\times \mathrm{SU}(2,\mathbb{C}) & \longrightarrow & \mathrm{SO}(4) \nonumber \\
(q_1,q_2) & \longmapsto & (x \mapsto q_1xq_2^{-1})
\eea
which shows that $\mathrm{Spin}(4,\mathbb{R}) \simeq \mathbb{S}^3\times\mathbb{S}^3$. Therefore, the (finite) subgroups of $\mathrm{SO}(4)$ are of the form $\Phi_{4}(G_1\times G_2)$ where $G_1,G_2$ are (finite) subgroups of $\mathrm{SU}(2,\mathbb{C})$.

By the spin covering \eqref{A4}, the finite subgroups of ${\rm SU}(2,\mathbb{C})$ are either cyclic or obtained by adding the matrix $-\mathbf{1}$ to a finite subgroup of ${\rm PSL}(2,\mathbb{C})$. The finite subgroups of ${\rm PSL}(2,\mathbb{C})$ are the polyhedral groups. Their extension in ${\rm SU}(2,\mathbb{C})$ are the binary polyhedral groups. Let us give the $3$ descriptions of generators for those groups.

\begin{itemize}
\item[$\bullet$] Element $r_p$, order $p$:
\beq
r_p = \left(\begin{array}{cc} e^{2{\rm i}\pi/p} & 0 \\ 0 & e^{-2{\rm i}\pi/p} \end{array}\right).
\eeq
$\Phi(r_p)$ is the rotation of angle $4\pi/p$ around $\vec{e}_3$ (beware of the factor $2$), and $\Psi(r_p)$ is the M\"obius transformation $z \mapsto e^{4{\rm i}\pi/p}$.
\item[$\bullet$] Element $\iota$, order $2$:
\beq
\iota = \left(\begin{array}{cc} 0 & {\rm i} \\ {\rm i} & 0 \end{array}\right).
\label{eq:iota}
\eeq
$\Phi(\iota)$ is the symmetry of axis $\hat{e}_1$, and $\Psi(\iota)$ is the inversion $z \mapsto 1/z$.
\item[$\bullet$] Element $\jmath$ of order $3$:
\beq
\jmath = \frac{1}{2}\left(\begin{array}{cc} 1 + {\rm i} & 1 - {\rm i} \\ - 1 - {\rm i} & 1 - {\rm i}\end{array}\right).
\label{eq:j}
\eeq
$\Phi(\jmath)$ is the rotation of angle $2\pi/3$ around the vector $\frac{1}{\sqrt{3}}(-\vec{e}_1 + \vec{e}_2 + \vec{e}_3)$, and $\Psi(\jmath)$ is the M\"obius transformation $z \mapsto {\rm i}\frac{z - 1}{z + 1}$.
\item[$\bullet$] Element $\kappa$, order $2$:
\beq
\kappa = \frac{{\rm i}}{\sqrt{1 + c^2}}\left(\begin{array}{cc} 1 & k \\ k & -1 \end{array}\right),\qquad k = 2\cos(2\pi/5).
\label{eq:kappa}
\eeq
$\Phi(\kappa)$ is the symmetry of axis $\frac{1}{\sqrt{1 + k^2}}(k\vec{e}_1 + \vec{e}_3)$, and $\Phi(\kappa)$ is the M\"obius transformation $z \mapsto \frac{z + k}{kz - 1}$.
\end{itemize}
Coming back to the list of binary polyhedral groups: $\mathbb{Z}/p\mathbb{Z}$ is generated by $r_p$; ${\rm Q}_{4p}$ is generated by $r_{2p}$ and $\iota$; ${\rm P}_{24}$ is generated by $r_{4}$ and $\jmath$; ${\rm P}_{48}$ is generated by $r_8$ and $\jmath$; ${\rm P}_{120}$ is generated by $r_{4}$, $\jmath$ and $\kappa$.

Classifying the Seifert spaces that are finite quotients of the $3$-sphere amounts to classifying the pairs of binary polyhedral groups $(G_1,G_2)$ such that $\Phi_{4}(G_1 \times G_2)$ acts freely on $\mathbb{S}^{3}$ in \eqref{PhiA}. Up to isomorphism, this is the list given in Appendix~\ref{pi1list}.

\medskip

Notice that the action of ${\rm SU}(2)$ by left or right multiplication preserves the symplectic form
\beq
\omega = \ri\l(\dd w_0 \wedge \dd w_0^* + \dd \vec{w} \mathop{\wedge}^{\cdot} \dd\overline{\vec{w}}\r)
\eeq
on ${\rm Mat}(2,\mathbb{C}) = \{w_0 + {\rm i} \vec{w}\cdot \vec{\sigma},\quad w_0,\vec{w} \in \mathbb{C} \times \mathbb{C}^3\big\}$. It can be checked by direct computation, using $(\cos(\theta/2)\mathbf{1} + {\rm i}\sin(\theta/2) \vec{x}\cdot \vec{\sigma})(w_0 + {\rm i}\vec{w}\cdot\vec{\sigma}) = w_0' + {\rm i}\vec{w}'\cdot\vec{\sigma}$ with
\bea
w_0' & = & \cos(\theta/2) w_0 - \sin(\theta/2)\vec{x}\cdot\vec{w},\nonumber \\
\vec{w}' & = & \cos(\theta/2)\vec{w} + \sin(\theta/2)w_0\vec{x} - \sin(\theta/2)\vec{x}\times \vec{w},
\eea
for a unit vector $\vec{x}$, and the properties of the vector product:
\beq
\dd(\vec{x}\cdot\vec{w}) \wedge \dd(\vec{x}\cdot \vec{w}) + (\dd \vec{x} \times \vec{w})\mathop{\wedge}^{\cdot} (\dd \vec{x} \times \vec{w}^*) = \dd \vec{w} \mathop{\wedge}^{\cdot} \dd \vec{w}^*.
\eeq
Therefore, $\Gamma \subset {\rm SU}(2)$ acts by symplectomorphisms on the six-fold considered in Section~\ref{sec:amodel}.

\section{$D_{p + 2}$ geometry: LMO spectral curve}

\label{Lpoly}

For all $p \geq 2$, it takes the form \eqref{Lform}:
\beq
(-1)^{p + 1}\,e^{-\lambda/2p}(X^2 + 1)(Y + 1)^2 + XY\,(\kappa^2 +
1)^{-(2p + 2)}\,\mathcal{Q}_{p}\big[(Y + 1/Y)(\kappa^2 + 1)^2\big] = 0,
\eeq
with 
\beq
\qquad \qquad \frac{2\kappa^{1 + 1/p}}{\kappa^2 + 1} = e^{-\lambda/4p^2}\,.
\eeq
$\mathcal{Q}_{p}$ is a monic polynomial of degree $p + 1$. For $p \leq 5$, it reads:
\beq
\bary{rcl}
\mathcal{Q}_2(\eta) & = & \eta^3 + 2(\kappa^4 + 6\kappa^2 - 3)\eta^2 - 4(\kappa^8 - 4\kappa^6 + 2\kappa^4 + 12\kappa^2 - 3)\eta \nonumber \\
& & - 8(\kappa^2 + 1)^2(\kappa^8 + 14\kappa^4 - 8\kappa^2 + 1)\,, \\
\mathcal{Q}_3(\eta) & = & \eta^4 + 8(2\kappa^2 - 1)\eta^3 - 8(\kappa^8 - 4\kappa^4 + 12\kappa^2 - 3)\eta^2 - 32(2\kappa^{10} + 3\kappa^8 + 8\kappa^6 + 4\kappa^4 - 6\kappa^2 +1)\eta \\
& & + 16(\kappa^2 + 1)^2(\kappa^2 - 1)(\kappa^4 - 4\kappa^2 + 1)(\kappa^6 + 3\kappa^4 + 5\kappa^2 - 1)\,,\\
\mathcal{Q}_4(\eta) & = & \eta^5 - 2(\kappa^4 - 10\kappa^2 + 5)\eta^4 - 8(\kappa^8 + 4\kappa^6 - 12\kappa^4 + 20\kappa^2 - 5)\eta^3 \\
& &  + 16(\kappa^{12} - 6\kappa^{10} - 11\kappa^8 - 8 \kappa^6 - 33\kappa^4 + 30\kappa^2 - 5)\eta^2  \\
& & + 16(\kappa^{16} + 8\kappa^{14} - 8\kappa^{12} - 24\kappa^{10} + 64\kappa^4 - 30\kappa^8 + 56\kappa^6 - 40\kappa^2 + 5)\eta \\
& &  - 32(\kappa^2 + 1)^2(\kappa^{16} - 4\kappa^{14} - 4\kappa^{12} - 4\kappa^{10} - 10\kappa^8 - 44\kappa^6 + 44\kappa^4 - 12\kappa^2 + 1)\,,  \\
\mathcal{Q}_5(\eta) & = & \eta^6 - 4(\kappa^4 - 6\kappa^2 + 3)\eta^5 - 4(\kappa^8 + 20\kappa^6 - 46\kappa^4 + 60\kappa^2 - 15)\eta^4 \\
& &  + 32(\kappa^{12} - 2\kappa^{10} - 15\kappa^8 + 12\kappa^6 - 41\kappa^4 + 30\kappa^2 - 5)\eta^3  \\
& & - 16(\kappa^{16} - 24\kappa^{14} + 4\kappa^{12} + 40\kappa^{10} + 6\kappa^8 + 24\kappa^6 - 236\kappa^4 + 120\kappa^2 - 15)\eta^2 \\
& &  - 64(\kappa^{20} + 2\kappa^{18} - 17\kappa^{16} - 24\kappa^{14} - 30\kappa^{12} - 52\kappa^{10} - 98\kappa^8 + 8\kappa^6 + 77\kappa^4 - 30\kappa^2 + 3)\eta \\
& & + 64(\kappa^2 + 1)^2(\kappa^2 - 1)(\kappa^8 - 2\kappa^6 + 2\kappa^4 - 6\kappa^2 + 1)(\kappa^{10} - 3\kappa^8 - 12\kappa^6 - 8\kappa^4 + 7\kappa^2 - 1)\,. 
\eary\nonumber
\eeq

At $c = 0$, i.e. $\kappa = 0$, we always have $\mathcal{Q}_p(\eta)|_{\kappa = 0} = (\eta - 2)^{p + 1}$. At $c = 1$, i.e. $\kappa = 1$, the roots of those polynomials are:
\bea
p = 2 & & -2,\pm\sqrt{2} \nonumber \\
p = 3 & & 0,2,\pm\sqrt{3} \nonumber \\
p = 4 && -2, \epsilon_1\sqrt{2 + \epsilon_2\sqrt{2}} \nonumber \\
p = 5 & & 0,-2,\frac{\epsilon_1}{2}\sqrt{10 + 2\epsilon_2\sqrt{2}} \nonumber
\eea
where $\epsilon_i = \pm 1$.

\section{$E_6$ geometry}

\subsection{LMO side: discriminant of $\mathcal{R}$ in \eqref{Rt}}
\label{discri}
The discriminant has two factors:
\beq
\bary{rcl}
\Delta_{1} & = & -32+27c^2+68c^4+160\mu_1^2+32\mu_1-32\mu_1^3-144\mu_2^2-256\mu_1^4-128\mu_1^5  \\
& & +31c^{10}-36c^8-58c^6+408c^2\mu_1-408c\mu_2-864c\mu_2^3-144\mu_2^2\mu_1^2-288\mu_2^2\mu_1 \\
& & -1296c^2\mu_2^2-432c^4\mu_1+372c^3\mu_2-648c^2\mu_1^2-396c^7\mu_2+1832c^6\mu_1^2  \\
& & +432c^8\mu_1+3104c^4\mu_1^3+1404c^4\mu_2^2+1776c^2\mu_1^4-1440c^2\mu_1^3+648c\mu_1\mu_2  \\
& & -3240c^5\mu_1\mu_2-5136c^3\mu_2\mu_1^2+3888c^2\mu_2^2\mu_1+576c\mu_1^2\mu_2+2688c^3\mu_1\mu_2 \\
& & -480c\mu_1^3\mu_2+432c^5\mu_2-440c^6\mu_1-1392c^4\mu_1^2\,,  \\
\Delta_{2} & = & 32+27c^2-68c^4-160\mu_1^2+32\mu_1-32\mu_1^3-144\mu_2^2+256\mu_1^4-128\mu_1^5+31c^{10} \\
& & +36c^8-58c^6-408c^2\mu_1+408c\mu_2-864c\mu_2^3-144\mu_2^2\mu_1^2+288\mu_2^2\mu_1 \\
& & +1296c^2\mu_2^2 -432c^4\mu_1+372c^3\mu_2-648c^2\mu_1^2-396c^7\mu_2+1832c^6\mu_1^2+432c^8\mu_1 \\
& & +3104c^4\mu_1^3+1404c^4\mu_2^2+1776c^2\mu_1^4+1440c^2\mu_1^3+648c\mu_1\mu_2-3240c^5\mu_1\mu_2  \\
& & -5136c^3\mu_2\mu_1^2+3888c^2\mu_2^2\mu_1-576c\mu_1^2\mu_2-2688c^3\mu_1\mu_2-480c\mu_1^3\mu_2 \\
& & -432c^5\mu_2+440c^6\mu_1+1392c^4\mu_1^2\,.
\eary \nonumber
\eeq

\subsection{Toda side: the polynomials $f_i(\kappa)$ in \eqref{fikappad}}
\label{fkappaE6}
\beq
\bary{rcl}
f_2(\kappa) & = & 248832 - 912384 \kappa + 1119744 \kappa^2 - 617472 \kappa^3 + 115584 \kappa^4 + 
 43776 \kappa^5 - 32096 \kappa^6 + 8704 \kappa^7  \\
 & & - 1260 \kappa^8 + 96 \kappa^9 - 3 \kappa^{10}\,,  \\
f_3(\kappa) & = & 26748301344768 - 231818611654656 \kappa + 922816396394496 \kappa^2 - 
 2265845304655872 \kappa^3  \\
 & & + 3881228059017216 \kappa^4 - 4961293921419264 \kappa^5 + 
 4932729950699520 \kappa^6 \\
& & - 3918448994549760 \kappa^7  + 2531494971703296 \kappa^8 - 
 1345368215715840 \kappa^9 \\ 
 & & + 592090245808128 \kappa^{10} - 216319699795968 \kappa^{11} + 
 65493454344192 \kappa^{12} \\
 & & - 16315792478208 \kappa^{13} + 3293224915968 \kappa^{14} - 
 521639046144 \kappa^{15}  + 60092669952 \kappa^{16} \\
 & & - 3803240448 \kappa^{17} - 
 196007424 \kappa^{18} + 88858368 \kappa^{19} - 12725856 \kappa^{20} + 1122176 \kappa^{21} \\
 & &  -  64176 \kappa^{22} + 2208 \kappa^{23} - 35 \kappa^{24}\,. \\
 f_6(\kappa) & = & -2985984 + 13436928 \kappa - 24758784 \kappa^2 + 26085888 \kappa^3 - 17843328 \kappa^4 + 
 8404992 \kappa^5 \\
 & & - 2802048 \kappa^6 + 667008 \kappa^7 - 112752 \kappa^8 + 13248 \kappa^9 - 
 1032 \kappa^{10} + 48 \kappa^{11} - \kappa^{12}\,.
\eary \nonumber
\eeq

\section{$E_7$ geometry}

\subsection{LMO side: minimal orbit}

\label{appE7orb}

The orbit consists of $27$ $12$-dimensional vectors with entries $-1,0,1$. If $w$ is in the orbit, so does $-\varepsilon(w) = (-w_{j + 1\,\,{\rm mod}\,\,a})_j$. Below we give an element of $\{\pm w,-\varepsilon(w),\varepsilon^2(w),\ldots\}$ that has $n_0(w) = \sum_{k = 0}^{11} w_{k} \geq 0$, and indicate the size $l$ of this sub-orbit. We encode the vectors in $\underline{w}(t) = \sum_{k = 0}^{11} w_{k}t^{k}$.

\beq
\begin{array}{|c|c|l|}
\hline 
n_0 {\rule{0pt}{3.2ex}}{\rule[-1.8ex]{0pt}{0pt}} & l & \underline{w}(t) \\
\hline \pm 3 {\rule{0pt}{3.2ex}}{\rule[-1.8ex]{0pt}{0pt}}& 4 & 1 + t^4 + t^8 \\
\hline 
\pm 2 {\rule{0pt}{3.2ex}}{\rule[-1.8ex]{0pt}{0pt}}& 6 & t - t^2 + t^3 + t^7 - t^8 + t^9  \\
\hline
\pm 1 {\rule{0pt}{3.2ex}}{\rule[-1.8ex]{0pt}{0pt}} & 12 & 1 + t^6 - t^7 + t^9 - t^{11} \\
\hline
0 {\rule{0pt}{3.2ex}}{\rule[-1.8ex]{0pt}{0pt}}  & 4 & -1 + t^3 - t^4 + t^7 - t^8 + t^{11}  \\
0 {\rule{0pt}{3.2ex}}{\rule[-1.8ex]{0pt}{0pt}}  & 1 &  \sum_{k = 0}^{11} (-1)^{k}t^{k} \\
\hline 
\end{array} \nonumber
\eeq

\subsection{LMO spectral curve in terms of $\mu_k$'s}
\label{E7sp}
We find $\mathcal{P}_{v}(x,y) = \sum_{j = 0}^{6} \sum_{k = 0}^{27} \Pi_{j,k}\,x^{6j}\,y^{k}$ with the symmetries $\Pi_{j,k} = (-1)^{j + 1}\Pi_{6 - j,k} = (-1)^j\Pi_{j,27 - k}$. All non-zero coefficients are deduced by symmetry from the following list, and depend on the $4$ parameters $\mu_2$, $\mu_3$, $\mu_5$ and $\mu_7$ which are unknown functions of $c$:

\beq
\bary{rcl}
\Pi_{0,11} & = & \Pi_{0,12} = -c^{-18}  \\
\Pi_{0,13} & = & -2c^{-18} \\
\Pi_{1,5} & = & \Pi_{1,6} = -c^{-12} \\
\Pi_{1,7} & = &-2c^{-12} \\
\Pi_{1,8} & = & -3c^{-15}(c^3 + 6\mu_2) \\
\Pi_{1,9} & = & -c^{-16}(c^4 + 18c\mu_2 + 24\mu_3) \\
\Pi_{1,10} & = &- 2c^{-16}(c^4 + 15c\mu_2 + 6\mu_3)  \\
\Pi_{1,11} & = &-6c^{-18}(5c^3\mu_2+6c^2\mu_3 - \mu_5)  \\
\Pi_{1,12} & = & c^{-18}(c^6 - 12c^3\mu_2  - 36\mu_2^2 - 36c^2\mu_3 + 18\mu_5) \\
\Pi_{1,13} & = & -c^{-18}(c^6 + 12c^3\mu_2 + 36\mu_2^2 - 12c^2\mu_3 + 12\mu_5) \\
\Pi_{2,2} & = &  - c^{-6} \\
\Pi_{2,3} & = & 0 \\
\Pi_{2,4} & = & c^{-10}(-c^4 + 6c\mu_2 + 12\mu_3)  \\
\Pi_{2,5} & = & -12c^{-12}(c^3\mu_2 - c^2\mu_3 + \mu_5)  \\
\Pi_{2,6} & = & 2c^{-13}(c^3 + 6\mu_2)(c^4 - 3c\mu_2 - 6\mu_3) \\
\Pi_{2,7} & = &  -6c^{-14}\big(c^5\mu_2 - 6c^4\mu_3 - 24c\mu_2\mu_3 + 16\mu_3^2 + c^2(-12\mu_2^2 + 8\mu_5) - 4\mu_7\big)  \\
\Pi_{2,8} & = & 3c^{-15}\big(c^9 + 72\mu_2^3 - 20c^5\mu_3 + 12c^3(\mu_2^2 + \mu_5) + c(40\mu_3^2 - 4\mu_7)\big) \\
\Pi_{2,9} & = & c^{-16}(c^3 + 6\mu_2)\big(c^7 + 12c^4\mu_2 - 24c^3\mu_3 + 72\mu_2\mu_3 + 36c(-2\mu_2^2 + \mu_5)\big)  \\
\Pi_{2,10} & = & -6c^{-17}\big(3c^8\mu_2 + 4c^7\mu_3 - 24c^4\mu_2\mu_3 + 12c^5(2\mu_2^2 - \mu_5) + 24c\mu_3(3\mu_2^2 - \mu_5)  \\
& & + 36c^2(\mu_2^3 + 2\mu_2\mu_5) + 8c^3(2\mu_3^2 + \mu_7) - 12\mu_2(2\mu_3^2 + \mu_7)\big) \\
\Pi_{2,11} & = & c^{-18}\big(3 + c^{12} + 24c^9\mu_2 - 60c^8\mu_3 - 432c^5\mu_2\mu_3 + 36\mu_5^2 - 432c^2\mu_3(2\mu_2^2 + \mu_5)  \\
& &  + c^6(-36\mu_2^2 + 48\mu_5) + 72c^3(9\mu_2^3 + 2\mu_2\mu_5) + 12c^4(34\mu_3^2 - \mu_7)\big)  \\
& & - 144c\mu_2(2\mu_3^2 + \mu_7)\big)  \\
\Pi_{2,12} & = & -3c^{-18}\big(-1 + c^{12} + 8 c^9 \mu_2 - 144 c^3 \mu_2^3 - 24 c^8 \mu_3 - 
    48 c^5 \mu_2 \mu_3 + 144 c^4 \mu_3^2 \\
    & &  + 36 \mu_5^2 + 
    12 c^6 (2 \mu_2^2 + \mu_5) - 144 c^2 \mu_3 (2 \mu_2^2 + \mu_5)\big)  \\
\Pi_{2,13} & = &2c^{-18}\big(3 - c^{12} + 3 c^9 \mu_2 + 24 c^8 \mu_3 + 72 c^5 \mu_2 \mu_3 + 
   6 c^6 (15 \mu_2^2 - 8 \mu_5)  \\
   & & - 72 c^2 \mu_3 \mu_5 + 36 \mu_5^2 - 
   36 c^3 (9 \mu_2^3 - \mu_2 \mu_5) + 24 c^4 (2 \mu_3^2 + \mu_7)  \\
   & & + 
   36 c \mu_2 (2 \mu_3^2 + \mu_7)\big)  \\
 \Pi_{3,0} & = & 1  \\
 \Pi_{3,1} & = & 1 + 6c^{-3}\mu_2  \\
 \Pi_{3,2} & = & 2 + 6c^{-3}\mu_2 - 12c^{-4}\mu_3 + 6c^{-6}\mu_5  \\
 \Pi_{3,3} & = & 2c^{-8}\big(c^8 + 12 c^5 \mu_2 - 12 c^4 \mu_3 - 72 c \mu_2 \mu_3 - 
  18 c^2 (\mu_2^2 - \mu_5) - 6 (2 \mu_3^2 + \mu_7)\big)  \\
  \Pi_{3,4} & = & 6c^{-10}\big(4 c^7 \mu_2 + 2 c^6 \mu_3 - 24 c^3\mu_2 \mu_3 - 24 c^2 \mu_3^2 + 
 18 c \mu_2 \mu_5 + 12 \mu_3 \mu_5  \\
 & &  + c^4 (12 \mu_2^2 + \mu_5)\big)  \\
 \Pi_{3,5} & = & 2c^{-12}\big(1 + 15 c^9\mu_2 - 6 c^8 \mu_3 - 288 c^5 \mu_2 \mu_3 + 24 c^6 \mu_5 + 
 252 c^3 \mu_2 \mu_5  \\
 & & + 108 c^2 \mu_3 \mu_5 - 18 \mu_5^2 + 
 72 c \mu_2 (4 \mu_3^2 - \mu_7) - 12 c^4 (8 \mu_3^2 + \mu_7)\big)  \\
 \Pi_{3,6} & = & c^{-13}\big(-3 c^{13} + 30 c^{10} \mu_2 + 84 c^9 \mu_3 - 864 c^6 \mu_2 \mu_3 - 
 216 c^3 \mu_3 (6 \mu_2^2 - 5 \mu_5)  \\
 & & - 432 \mu_2 \mu_3 \mu_5 - 
 36 c^7 (\mu_2^2 + \mu_5) - 36 c^4 (24 \mu_2^3 - 31 \mu_2 \mu_5)  \\
 & & - 
 432 c^2 \mu_2 \mu_7 + 12 c^5 (-82 \mu_3^2 + \mu_7)  \\
 & & +  
 c (2 + 864 \mu_3^3 - 648 \mu_2^2 \mu_5 + 36 \mu_5^2 - 432 \mu_3 \mu_7)\big) \\
 \Pi_{3,7} & = & -c^{-14}\big(3 c^{14} + 12 c^{11} \mu_2 - 72 c^{10} \mu_3 + 360 c^7 \mu_2 \mu_3 + 
 864 c^6 \mu_3^2  \\
 & & + 72 c^4 \mu_3(18 \mu_2^2 - 19 \mu_5) + 
 864 c \mu_2 \mu_3 (6 \mu_2^2 + \mu_5) + 12 c^8 (6 \mu_2^2 + 5 \mu_5)  \\
 & & + 
 36 c^5 (6 \mu_2^3 - 7 \mu_2 \mu_5) - 216 c^3 \mu_2 (6 \mu_3^2 - \mu_7)  \\
 & & + 
 4 c^2 (-1 + 324 \mu_2^4 - 288 \mu_3^3 + 126 \mu_5^2 + 72 \mu_3 \mu_7)  \\
 & & + 
 144 \big(\mu_5 (4 \mu_3^2 - \mu_7) + 3 \mu_2^2 (2 \mu_3^2 + \mu_7)\big)\big) 
 \eary \nonumber 
 \eeq
  
 \beq
 \bary{rcl}
 \Pi_{3,8} & = & -3c^{-15}\big(c^{15} + 4 c^{12} \mu_2 - 288 c^6 \mu_2^3 - 20 c^{11} \mu_3 - 
 168 c^5 \mu_3 \mu_5  \\
 & & + 288 c^2 \mu_2 \mu_3 (9 \mu_2^2 + 2 \mu_5) + 
 c^9 (-36 \mu_2^2 + 38 \mu_5) + 4 c^7 (26 \mu_3^2 - 5 \mu_7)  \\
 & & - 
 48 c^4 \mu_2 (8 \mu_3^2 + \mu_7) + 
 12 \mu_2 (-1 + 48 \mu_3^3 + 36 \mu_2^2 \mu_5 - 18 \mu_5^2 + 
    24 \mu_3 \mu_7) 
  \\
 & & + c^3 (-2 + 432 \mu_2^4 - 384 \mu_3^3 - 648 \mu_2^2 \mu_5 + 36 \mu_5^2 + 
    96 \mu_3 \mu_7)  \\
    & & + 
 24 c \big(\mu_5 (10 \mu_3^2 - \mu_7) + 12 \mu_2^2 (14 \mu_3^2 + \mu_7)\big)\big) \\
 \Pi_{3,9} & = & c^{-16}\big(-3 c^{16} - 78 c^{13} \mu_2 + 36 c^{12} \mu_3 + 1296 c^9 \mu_2 \mu_3 - 
 432 c^3 \mu_2 \mu_3 (24 \mu_2^2 - 19 \mu_5)  \\
 & & + 
 18 c^{10} (2 \mu_2^2 - 5 \mu_5) - 144 c^6 \mu_3 (3 \mu_2^2 - 5 \mu_5) + 
 108 c^7 (4 \mu_2^3 - 9 \mu_2 \mu_5)  \\
 & & - 432 c^5 \mu_2 (10 \mu_3^2 - \mu_7) + 
 36 c^8 (-2 \mu_3^2 + \mu_7)  \\
 & & + 
 36 c \mu_2 (1 + 216 \mu_2^4 - 192 \mu_3^3 - 216 \mu_2^2 \mu_5 + 
    30 \mu_5^2 - 96 \mu_3 \mu_7)
 \\
& & - 2 c^4 (-1 + 2592 \mu_2^4 - 720 \mu_3^3 - 3456 \mu_2^2 \mu_5 + 
    594 \mu_5^2 + 72 \mu_3 \mu_7)  \\
    & & - 
 48 (12 \mu_3^4 + \mu_3 (-1 + 54 \mu_2^2 \mu_5 - 18 \mu_5^2) + 
    12 \mu_3^2 \mu_7 + 3 \mu_7^2)  \\
    & & - 
 864 c^2 \big(\mu_5 (2 \mu_3^2 - \mu_7) + 3 \mu_2^2 (6 \mu_3^2 + \mu_7)\big)\big) \\
 \Pi_{3,10} & = & -2c^{-17}\big(39 c^{14} \mu_2 + 72 c^{13} \mu_3 - 648 c^{10} \mu_2 \mu_3 - 
 72 c^7 \mu_3 (9 \mu_2^2 - 20 \mu_5)  \\
 & & - 1296 c^4 \mu_2 \mu_3 \mu_5 - 
 6 c^{11} (3 \mu_2^2 + 4 \mu_5) - 54 c^8 (4 \mu_2^3 - 13 \mu_2 \mu_5)  \\
 & & - 
 6 c^2 \mu_2 (5 + 648 \mu_2^4 - 108 \mu_2^2 \mu_5 - 54 \mu_5^2) - 
 12 c \mu_3 (1 + 648 \mu_2^4  \\
 & & - 540 \mu_2^2 \mu_5 - 18 \mu_5^2) + 
 216 c^6 \mu_2 (2 \mu_3^2 - \mu_7) + 6 c^9 (-190 \mu_3^2 + \mu_7)  \\
 & & + 
 216 \mu_2 \mu_5 (2 \mu_3^2 + \mu_7) + 
 2 c^5 (-1 + 648 \mu_2^4 + 936 \mu_3^3 + 324 \mu_2^2 \mu_5  \\
 & & - 270 \mu_5^2 - 
    180 \mu_3 \mu_7) - 
 216 c^3 \big(\mu_5 (6 \mu_3^2 - \mu_7) + 2 \mu_2^2 (14 \mu_3^2 + \mu_7)\big)\big) \\
 \Pi_{3,11} & = & -6c^{-18}\big(14 c^{15} \mu_2 + 26 c^{14} \mu_3 - 288 c^{11} \mu_2 \mu_3 - 
 108 c^8 \mu_3 (8 \mu_2^2 - 5 \mu_5) - 
  \\
 & & 1296 c^5 \mu_2 \mu_3 (4 \mu_2^2 - \mu_5) - 19 c^{12} \mu_5 + 
 c^9 (-72 \mu_2^3 + 264 \mu_2 \mu_5)  \\
 & & - 
 12 c^2 \mu_3 (1 + 216 \mu_2^4 + 72 \mu_2^2 \mu_5 + 6 \mu_5^2) + 
 2 (\mu_5 + 18 \mu_5^3)  \\
 & & - 
 144 c \mu_2 (3 \mu_2^2 - \mu_5) (2 \mu_3^2 + \mu_7) - 
 48 c^7 \mu_2 (\mu_3^2 + 2 \mu_7)  \\
 & & + c^{10} (-400 \mu_3^2 + 4 \mu_7) + 
 2 c^3 \mu_2 (-5 + 648 \mu_2^4 - 720 \mu_3^3 - 324 \mu_2^2 \mu_5  \\
 & & - 
    162 \mu_5^2 - 360 \mu_3 \mu_7) - 
 12 c^6 (108 \mu_2^4 - 48 \mu_3^3 - 99 \mu_2^2 \mu_5 + 8 \mu_5^2  \\
 & &  + 
    12 \mu_3 \mu_7) - 
 36 c^4 \big(\mu_5 (10 \mu_3^2 - \mu_7) + 4 \mu_2^2 (22 \mu_3^2 + 5 \mu_7)\big)\big) \\
 \Pi_{3,12} & = & c^{-18}\big(c^{18} - 84 c^{15} \mu_2 - 180 c^{14} \mu_3 + 1368 c^{11} \mu_2 \mu_3 + 
 144 c^8 \mu_3 (21 \mu_2^2 - 31 \mu_5)  \\
 & & + 
 864 c^5 \mu_2 \mu_3 (33 \mu_2^2 - 4 \mu_5) + 
 c^{12} (-288 \mu_2^2 + 198 \mu_5) - 72 c^9 (9 \mu_2^3 + 20 \mu_2 \mu_5)  \\
 & & + 
 36 (\mu_5 (-1 + 6 \mu_5^2) + \mu_2^2 (2 + 36 \mu_5^2)) + 
 96 c^{10} (25 \mu_3^2 - \mu_7)  \\
 & & + 216 c^7 \mu_2 (2 \mu_3^2 + \mu_7) - 
 2592 c \mu_2 (\mu_2^2 - \mu_5) (2 \mu_3^2 + \mu_7)  \\
 & & - 
 24 c^3 \mu_2 (-1 + 972 \mu_2^4 - 432 \mu_3^3 - 864 \mu_2^2 \mu_5 + 
    306 \mu_5^2 - 216 \mu_3 \mu_7)
  \\
 & & + 2 c^6 (-1 + 8424 \mu_2^4 - 864 \mu_3^3 - 4968 \mu_2^2 \mu_5 + 
    1386 \mu_5^2 + 864 \mu_3 \mu_7) 
 \\
& & + 72 c^2 (24 \mu_3^4 + 
    \mu_3 (1 - 648 \mu_2^4 + 360 \mu_2^2 \mu_5 - 18 \mu_5^2) + 
    24 \mu_3^2 \mu_7 + 6 \mu_7^2)  \\
    & & + 
 432 c^4 \big(-\mu_5 (4 \mu_3^2 + 5 \mu_7) + \mu_2^2 (66 \mu_3^2 + 9 \mu_7)\big)\big) \\
 \Pi_{3,13} & = & c^{-18}\big(c^{18} - 96 c^{14} \mu_3 - 288 c^{11} \mu_2 \mu_3 + 
 864 c^5 \mu_2 \mu_3 (6 \mu_2^2 - 7 \mu_5)  \\
 & & - 
 1296 c^8 \mu_3 (3 \mu_2^2 + \mu_5) + c^{12} (-36 \mu_2^2 + 60 \mu_5) - 
 36 c^9 (24 \mu_2^3 - 5 \mu_2 \mu_5)  \\
 & & - 
 24 c^2 \mu_3 (1 + 1944 \mu_2^4 - 432 \mu_2^2 \mu_5 + 126 \mu_5^2) + 
 24 (\mu_5 + 18 \mu_5^3  \\
 & & + \mu_2^2 (3 - 162 \mu_5^2)) + 
 288 c^7 \mu_2 (10 \mu_3^2 - \mu_7) + 12 c^{10} (118 \mu_3^2 - \mu_7)  \\
 & & - 
 432 c \mu_2 (12 \mu_2^2 + \mu_5) (2 \mu_3^2 + \mu_7) - 
 24 c^3 \mu_2 (-1 + 648 \mu_2^4 - 288 \mu_3^3  \\
 & &  - 486 \mu_2^2 \mu_5 - 
    144 \mu_3 \mu_7) + 
 2 c^6 (1 + 3240 \mu_2^4 - 2448 \mu_3^3 + 324 \mu_2^2 \mu_5  \\
 & &  - 
    126 \mu_5^2 + 72 \mu_3 \mu_7) + 
 144 c^4 \big(6 \mu_2^2 (22 \mu_3^2 - \mu_7) + \mu_5 (50 \mu_3^2 + \mu_7)\big)\big) 
\eary\nonumber
 \eeq

\section{$E_8$ geometry}

\subsection{LMO side: minimal orbit}
\label{appE8orb}

The orbit consists of $240$ $30$-dimensional vectors with entries $-2,-1,0,1,2$. If $w$ is in the orbit, so does $-w$ and its shift $\varepsilon(w) = (w_{j + 1\,\,{\rm mod}\,\,a})_j$. Below we give an element of $\{\pm w,\pm \varepsilon(w),\ldots\}$ that has $w_{0} \neq 0$ and $n_0(w) \geq 0$, and indicate the size $l$ of this sub-orbit. The vectors are compactly encoded in $\underline{w}(t) = \sum_{k = 0}^{29} w_{k}t^{k}$.

\beq
\begin{array}{|c|c|l|}
\hline 
n_0 {\rule{0pt}{3.2ex}}{\rule[-1.8ex]{0pt}{0pt}} & l & \underline{w}(t) \\
\hline \pm 6 {\rule{0pt}{3.2ex}}{\rule[-1.8ex]{0pt}{0pt}}& 2\cdot 5 & 1 + t^5 + t^{10} + t^{15} + t^{20} + t^{25} \\
\hline 
\pm 5 {\rule{0pt}{3.2ex}}{\rule[-1.8ex]{0pt}{0pt}}& 2 \cdot 6 & -1 + t + t^{5} - t^{6} + t^{7} + t^{11} - t^{12} + t^{13} + t^{17} - t^{18} + t^{19} +
t^{23} - t^{24} + t^{25} + t^{29} \\
\hline
\pm 4 {\rule{0pt}{3.2ex}}{\rule[-1.8ex]{0pt}{0pt}} & 2\cdot 15 & -1 + t + t^4 - t^5 + t^7 + t^{13} - t^{15} + t^{16} + t^{19} - t^{20} + t^{22}
+ t^{28} \\
\hline
\pm 3 {\rule{0pt}{3.2ex}}{\rule[-1.8ex]{0pt}{0pt}}  & 2\cdot 10 & -1 + t + t^3 - t^4 + t^7 - t^{10} + t^{11} + t^{13} - t^{14} + t^{17} - t^{20} + t^{21} + t^{23} - t^{24} + t^{27} \\
\pm 3 {\rule{0pt}{3.2ex}}{\rule[-1.8ex]{0pt}{0pt}}  & 2\cdot 10 &  1 - t + t^2 - t^3 + t^4 + t^{10} - t^{11} + t^{12} - t^{13} + t^{14} + t^{20}
- t^{21} + t^{22} - t^{23} + t^{24}  \\
\hline 
\pm 2 {\rule{0pt}{3.2ex}}{\rule[-1.8ex]{0pt}{0pt}}  & 2 \cdot 15 &  -1 + t + t^2 - t^3 + t^7 - t^9 + t^{11} - t^{15} + t^{16} + t^{17} - 
t^{18} + t^{22} - t^{24} + t^{26} \\
\pm 2 {\rule{0pt}{3.2ex}}{\rule[-1.8ex]{0pt}{0pt}}  & 2 \cdot 15 & 1 - t + t^3 - t^4 + t^5 - t^6 + t^8 + t^{15} - t^{16} + t^{18} - t^{19}
+ t^{20} - t^{22} + t^{23} \\
\hline 
\pm 1 {\rule{0pt}{3.2ex}}{\rule[-1.8ex]{0pt}{0pt}}  & 2 \cdot 30 & 2 - t + t^{6} - t^7 + t^{10} - t^{11} + t^{12} - t^{13} + t^{15} - 
t^{17} + t^{18} - t^{19} + t^{20} - t^{23} + t^{24} - t^{29} \\
\hline 
0 {\rule{0pt}{3.2ex}}{\rule[-1.8ex]{0pt}{0pt}}  & 2 \cdot 5 & -1 + t^4 - t^5 + t^9 - t^{10} + t^{14} - t^{15} + t^{19} - t^{20} + t^{24} - t^{25}
+ t^{29} \\
0 {\rule{0pt}{3.2ex}}{\rule[-1.8ex]{0pt}{0pt}}  & 2 \cdot 5 & 1 - t^2 + t^5 - t^7 + t^{10} - t^{12} + t^{15} - t^{17} + t^{20} - t^{22} +
t^{25} - t^{27} \\
0 {\rule{0pt}{3.2ex}}{\rule[-1.8ex]{0pt}{0pt}}  & 2 \cdot 3 & \bary{l}1 - t + t^3 -
t^4 + t^6 - t^7 + t^{9} - t^{10} + t^{12} - t^{13} + t^{15}  - 
t^{16} + t^{18} - t^{19} + t^{21} \\ - t^{22} + t^{24} - t^{25} + t^{26} - t^{28}
\eary\\
0 {\rule{0pt}{3.2ex}}{\rule[-1.8ex]{0pt}{0pt}}  & 2 & \sum_{j = 0}^{29} (-1)^{j}t^{j} \\
\hline
\end{array} \nonumber
\eeq

\subsection{LMO side: Newton polygon}
\label{E8bound}

The boundary (and coefficients therein) of the Newton polygon of the full curve $\mathcal{P}_{E_8}^{{\rm LMO}}(X,Y)$ is the same as the one of the polynomial computed in terms of the minimal orbit data:
\beq
C\,\prod_{i = 1}^{240} \big(Y - (-cx)^{n_0(w[i])} \zeta_{30}^{n_1(w[i])}\big),\qquad X = x^{30},\qquad c = e^{\lambda/1800}.
\eeq
where the constant $C$ is fixed so that the monomial $X^{9}Y^0$ appears with coefficient $1$. The result is:
\beq
\bary{rcl}
\mathcal{P}_{E_8}^{{\rm LMO}}(X,Y) & = &  -c^{-240}(1 + X^{18})Y^{106}(Y + 1)^2(Y^2 + Y + 1)^3(Y^4 + Y^3 + Y^2 + Y + 1) \\
& & + c^{210}(X + X^{17})(Y^{76} + \cdots + Y^{164}) + 2c^{180}(X^2 + X^{16})(Y^{61} + \cdots + Y^{179}) \\
& & +c^{150}(X^3 + X^{15})(Y^{46} + \cdots + Y^{194})   - 2c^{120}(X^4 + X^{14})(Y^{36} + \cdots + Y^{204}) \\
& & + c^{90}(X^5 + X^{13})(Y^{26} + \cdots + Y^{214}) - c^{60}(X^7 + X^{11})(Y^{11} + \cdots + Y^{229}) \\
& & + c^{30}(X^8 + X^{10})(Y^5 + \cdots + Y^{235}) + X^9(1 + Y^{240})\,,
\eary \nonumber
\eeq
where the $\cdots$ lie in the interior of the polygon.

\newpage

\bibliographystyle{amsalpha}
\bibliography{SDbibli2.bib}

\end{document}